\documentclass[aps, 
twocolumn,showpacs,floatfix,nofootinbib, reprint,showkeys,longbibliography]{revtex4-1}
\usepackage{graphicx}
\usepackage{amssymb}
\usepackage{amsmath}
\usepackage{color}
\usepackage[yyyymmdd,hhmmss]{datetime}
\usepackage{MnSymbol}
\usepackage[english]{babel}
\usepackage{manfnt}
\usepackage{rotating}
\begin{document}

\newcommand{\tr}{\mbox{$\,{\rm tr\,}$}}
\newcommand{\Tr}{\mbox{$\,{\rm Tr\,}$}}
\newcommand{\vol}{\mbox{$\,{\rm vol\,}$}}
\newcommand{\tor}{\mbox{$\,{\rm tor\,}$}}
\newcommand{\volg}{\mbox{$\,{\rm vol}\, g$}}
\newcommand{\voltg}{\mbox{$\,{\rm vol}\, \tilde g$}}
\newcommand{\torg}{\mbox{$\,{\rm tor}\, g$}}
\newcommand{\mathbfg}{\mbox{${\textit{\textbf g}}$}}
\newcommand{\mm}{\mbox{$\mathfrak{m}$}}
\def\myfrac#1{\mbox{\Large $#1$}}
\def\myfracc#1{\mbox{\large $#1$}}
\newcommand{{\superstar}}{\mbox{{${{\otimes}}$}\llap{${\oplus}$}}}
\newcommand{\insulator}{{\mbox{{\color[cmyk]{1,0,1,0}{{\superstar}}}}}}
\newcommand{{{\attractor}}}{{\mbox{${\color[cmyk]{1,0,1,0.3}{{\oplus}}}$}}}
\newcommand{\repulsor}{{\mbox{${\color{red}{{\ominus}}}$}}}
\newcommand{{\smallsuperstar}}{\mbox{{${{_\otimes}}$}\llap{${_\oplus}$}}}
\newcommand{\smallinsulator}{{\mbox{{\color[cmyk]{1,0,1,0}{{\smallsuperstar}}}}}}
\newcommand{{{\smallattractor}}}{{\mbox{${\color[cmyk]{1,0,1,0.3}{{_\oplus}}}$}}}
\newcommand{\smallrepulsor}{{\mbox{${\color{red}{{_\ominus}}}$}}}
\newcommand{\connector}{{\mbox{ $\stackrel{{\color{blue}{{{\otimes}}}}}{\longleftrightarrow}$ }}}
\newcommand{\qcp}{{\mbox{${\color{blue}{{\otimes}}}$}}}
\newcommand{\smallqcp}{{\mbox{${\color{blue}{{_\otimes}}}$}}}
\newcommand{\qcpT}{\mbox{$\qcp_{_T}$}}
\newcommand{\qcpR}{\mbox{$\qcp_{_R}$}}
\newcommand{\qcpS}{\mbox{$\qcp_{_S}$}}
\newcommand{\xqcp}{{\mbox{${\color{purple}{{{\qcp}}}}$}}}
\newcommand{{\mmapdown}}{\mm\,\raisebox{-1.3ex}{\rotatebox{90}{$\longleftarrow$}}}
\newcommand{{\longdownarrow}}{\raisebox{-1ex}{\rotatebox{90}{$\longleftarrow$}}}
\newcommand{\papillon}{\mbox{\scalebox{1}{$\bowtie$}}}

\title[Elliptic mirror $\dots$]{Elliptic mirror of the quantum Hall effect}

\author{C.A.  L\"utken}
\affiliation{Physics Dept., University of Oslo, NO-0316 Oslo, Norway}
\date{\today~@~\currenttime}
\begin{abstract}
Toroidal sigma models of magneto-transport are analyzed, 
in which integer and fractional quantum Hall effects automatically are unified by a 
{holomorphic modular symmetry}, whose group structure is determined by the 
{spin structure} of the toroidal target space (an \emph{elliptic curve}).
Hall quantization is protected by the topology of stable holomorphic vector bundles $\mathcal V$ on this space, 
and plateaux values $\sigma_H^{\smallattractor} = \mu\in\mathbb Q$ of the Hall conductivity are rational because 
such bundles are classified by their \emph{slope} $\mu(\mathcal V) = {\rm deg}(\mathcal V)/{\rm rk}(\mathcal V)$,
where ${\rm deg}(\mathcal V)$ is the degree and ${\rm rk}(\mathcal V)$ is the rank of $\mathcal V$.
By exploiting a quantum equivalence called \emph{mirror symmetry}, these models are mapped to
tractable mirror models (also elliptic), in which topological protection is provided by more familiar
winding numbers. 

Phase diagrams and scaling properties of elliptic models are 
compared to some of the experimental and numerical data accumulated over the past three decades.
The geometry of scaling flows extracted from quantum Hall experiments is in good agreement
with modular predictions, including the location of many quantum critical points.
One conspicuous model 
has a critical delocalization exponent
$\nu_{\rm tor}  =  18 \ln 2 /(\pi^2 G^4) = 2.6051\dots$ ($G$ is Gauss' constant)
that is in excellent agreement with the value $\nu_{\rm num} = 2.607\pm\,.004$ calculated in the 
numerical Chalker-Coddington model,
suggesting that these models are in the same universality class.
The real delocalization exponent may be disentangled from other scaling exponents
in finite size scaling experiments, giving an experimental value $\nu_{\rm exp} = 2.3\pm 0.2$.
The modular model suggests how these theoretical and experimental results may be reconciled,
but in order to determine if these theoretical models really are in the quantum Hall universality 
class, improved finite size scaling experiments are urgently needed.
\end{abstract}


\maketitle

\section{Introduction and summary}\label{sec:intro}
 
The quantum Hall effect (QHE) is a corner stone of modern metrology, but almost four decades after its 
discovery it continues to challenge our understanding of quantum physics. It is an emergent property of
$10^{10}$ strongly interacting electrons in two dirty dimensions that cannot be analyzed using only conventional
perturbative techniques.  While special cases are well understood, we do not have a comprehensive 
effective field theory (EFT) that exhibits all the universal properties observed in low temperature 
magneto-transport experiments. It is our purpose here to try to remedy this situation.

All else being equal, we would prefer to derive such a model from the well understood microphysics of
quantum electrodynamics and localization in a disordered medium, but since we are unable to do so, 
we must instead rely on experiments to tell us what the essential (universal) features of such an emergent 
model are. The most important information characterizing a universality class is the symmetry 
representation of the low energy (long wavelength) modes.
Since continuous symmetries, which includes exact symmetries like Lorentz and gauge invariance,
are infinite, they are difficult to accommodate, and therefore especially useful.
Discrete symmetries, on the other hand, are usually finite and therefore much less constraining. 

We shall here exploit a new type of symmetry (infinite and discrete), which is called \emph{modular}, 
that appears to be relevant for the QHE \cite{LR:1992, *LR:1993, *Lutken:1993a, *Lutken:1993b}.
This rigid emergent order is encoded in a fractal phase diagram tightly harnessed by the modular 
symmetry, which may be compared directly with scaling data
(cf.\;Ref.\;\cite{OLLutken:2018} for a recent review of the experimental situation).
Although these are finitely generated approximate (emergent) discrete symmetries, 
because they are nonabelian and infinite they provide unusually strong constraints on model building.

This section tries to motivate the construction of a family of ``toroidal" (``elliptic") models that 
automatically contain modular symmetries.  Toroidal sigma models have been extensively 
studied in string theory \cite{Mirror:2003}, and we make use of some of the technology discovered in that 
context (including modular and mirror symmetry, conformal field theory, Zamolodchilov's $C$-theorem, 
and topological properties of gauge fields carried by ``open strings"), which previously have seen little or 
no use in condensed matter physics.  
The family of toroidal models  studied here, which are conformal only at isolated quantum critical points,
have not previously been used. There is at present no deep understanding of how 
such a model can emerge from the messy microphysics of the QHE.  
However, since the model is extremely rigid it is easy to falsify, and its usefulness may therefore 
be evaluated by confronting its predictions directly with experimental data.

In order not to clutter this introduction, which serves only to motivate the choice of ansatz 
that will be used to model the QHE, but at the risk of making the narrative less self-contained, 
only basic definitions and results are included in this section. These are expanded on in the main body of
the paper, and a more complete review of modular mathematics may be found in a technical appendix.

The rest of this section may be regarded as a simplified preview of coming attractions, and also 
introduces some of the novel non-perturbative field theoretic techniques gifted us by strings.
Its main objective, however, is to explain the rationale behind the choice of EFT discussed here.
Since the narrative relies heavily on intuition and results from string theory, readers unfamiliar with this
enterprise may not find this introduction very useful. Whether they think the reasoning unconvincing, 
or just prefer to skip unfamiliar mathematics, they can simply take the scaling function $\varphi$ obtained 
in Sec.\;\ref{subsec:partitions} [cf.\;Eq.\;(\ref{eq:Tpot})] as an ad hoc ansatz, and proceed to analyze its 
properties. It is explained in the Appendix how this ansatz, apart from the normalization, follows directly 
from the modular symmetry and general properties of RG flows. It is therefore logically independent of 
the model from which it happens to be derived here. This EFT is itself an ansatz, arguably better 
motivated and perhaps more useful, but nevertheless (so far) disconnected from our microphysical 
understanding of the QHE. 

The potential $\varphi$ is a sharply defined, alternative, stand-alone starting point, 
but without pedigree its usefulness is not so obvious. This ansatz is, however, far from vacuous, since
it was pointed out a long time ago that universal quantum Hall data are encoded in this extremely 
rigid function, including phase and scaling diagrams, and the exact location of 
quantum critical points \cite{LR:1992, *LR:1993, Lutken:1993a, *Lutken:1993b, Lutken:2006, 
BLutken:1997, *BLutken:1999, LR:2007,LR:2006, *LR:2009, *LR:2011, LR:2010, LR:2014, 
NLutken:2012a, *NLutken:2012b,  Lutken:2014a, *Lutken:2015, OLLutken:2018}.  
Its properties have been extensively investigated over the years (cf.\;Ref.\;\cite{OLLutken:2018}, 
and references therein), so a brief summary comparison with data in Sec.\;\ref{sec:data} will suffice here.

An important reason for trying to derive this function from an EFT is that this can give us access to critical 
exponents, whithout which we cannot identify the universality class of the QHE.
These depend on the normalization of $\varphi$, which is not fixed by symmetry arguments alone. 
Sections \ref{sec:toromodels} and \ref{sec:RGflow} are devoted to extracting observable 
predictions from the toroidal ansatz, by exploiting both the modular symmetry and a quantum equivalence 
called \emph{mirror symmetry}, so that a comparison with data can be carried out in Sec.\;\ref{sec:data}.

Another reason for investigating toroidal models is that we shall ultimately need an EFT that is \emph{derived} 
from the microphysics of the QHE, and it would be helpful to have something consistent with data to aim for.
Our goal here is much more modest.  The question we attempt to answer is if a more 
or less plausible (depending on your point of view) EFT \emph{exists} that is consistent with 
currently available data. The conclusion that the guiding principles (discussed below) extracted from these
experiments appear to lead naturally to a toroidal sigma model, should be tempered by the fact that 
the argument is quite heuristic.   Nevertheless, predictions extracted from this guess are encouraging,
and most importantly, non-negotiable.   Provided that enough experimental evidence 
accumulates to convince us that a useful effective model has been identified, 
it is not unreasonable to expect that this will help us to ``reverse engineer" a derivation of 
that model from microphysics. Alas, this is a daunting task that
appears to be well beyond our current theoretical grasp.

Before we can construct the model we must first recall a number of powerful field theoretic ideas 
that have appeared in the wake of string theory.

\subsection{Emergent sigma model}\label{subsec:sigmamodels}

This subsection recalls how low energy physics sometimes may be encoded in the 
geometry and topology of an effective field theory that is a non-linear sigma model.
This is an old idea that has been explored since the earliest days of the QHE.
Our story is framed in a language that is suited for toroidal generalizations, not previously 
considered in this context, that will be explored below.

The main idea to be investigated here is that universal properties of the QHE 
are captured by an EFT that is a sigma model whose target space 
geometry is determined by two bosonic fields $\varphi^1$ and $\varphi^2$.
To leading order in the derivative expansion the most general form of such a model is
\begin{equation*}
\mathcal L_{\rm EFT} 
= \gamma_{\mu\nu} \partial^\mu \varphi^a \partial^\nu \varphi^b K_{ab} + \dots,
\end{equation*}
where $\gamma$ is a 2-tensor on a two-dimensional manifold $\Sigma$, and
$K(t)$ is a matrix of running coupling constants, i.e., parameters that renormalize 
or ``flow" with the dominant scale parameter $t$ (usually $t = \ln T$, where $T$ is temperature).
The ellipses represent gauge and matter fields to be added later.
If we allow parity violation, as we must in any model of the QHE, 
then these matrices have both symmetric and antisymmetric parts:
$\eta_{\mu\nu} = \gamma_{(\mu\nu)}$, $\epsilon_{\mu\nu} = \gamma_{[\mu\nu]}$,  
$g_{ab} = K_{(ab)}$, and $h_{ab} = K_{[ab]}$. 

The fields $\varphi$ take values in a target manifold  ${\rm M}$, and should be viewed as 
coordinates on this space.  In other words, they are maps 
$\Sigma\stackrel{\varphi}{\longrightarrow} {\rm M}$.
This allows us to convert low energy physics into geometry, 
which is intrinsically non-linear, and the powerful machinery of 
differential geometry and topology becomes available for analyzing non-perturbative 
properties of the model.  The most relevant low energy field configurations are
deformations of ${\rm M}$, which we can think of as changing its shape and size.
The matrix $g$ of effective couplings gives a metric on ${\rm M}$, 
and if $h \neq 0$ the manifold has torsion. 

The running coupling constants in an EFT of the QHE are the magneto-conductivity $\sigma_D(t)$
and the Hall conductivity $\sigma_H(t)$.  $K(t)$ is identified  as the scale dependent transport tensor,
\begin{eqnarray*}
K_{ab}(t) = \sigma_{ab}(t)
 = \left[\begin{array}{cc}
 \phantom{-}\sigma_D(t) & \sigma_H(t)\\
 - \sigma_H(t)& \sigma_D(t)  
\end{array}\right] =  \sigma_D \delta_{ab} +  \sigma_H\varepsilon_{ab},
\end{eqnarray*}
which is allowed to be asymmetric by the generalized Onsager relation,
because parity is broken by the external magnetic field.  
Our starting point is therefore what appears to be the simplest possible (parity violating) effective action,
\begin{equation}
\mathcal L_{\rm EFT} 
=   \sigma_D \, \partial_\mu \varphi_a \partial^\mu \varphi^a 
+ \sigma_H \, \epsilon_{\mu\nu} \partial^\mu \varphi^a \partial^\nu \varphi^b \varepsilon_{ab} + \dots,
\label{eq:LEFT}
\end{equation}
but this is delusive, since most of the nontrivial information about the physical system will be encoded in 
constraints on the fields and suppressed ``decorations".

The prototypical sigma model in condensed matter physics is the spherical model 
of quantum magnetism, often called the $O(3)$-model \cite{footnote:AltmanS2006}.
In this case the origin of the compact curved target sphere is well understood,
as it belongs to a fertile family of models that derive from the geometry of 
spontaneous symmetry breaking.
Spin waves, whose quanta are magnons, are collective excitations
of a spin chain that are modeled by fluctuations in the spherical target metric.
The target geometry parametrizes Goldstone modes that appear 
because an $SO(3)$ symmetry is broken to $SO(2)$ when 
the spins ``spontaneously" pick some direction.  
To lowest order in a derivative expansion the EFT
is unambiguously determined by the symmetries, and the leading local term is parametrized 
by the spherical metric on the target space ${\rm M} =  SO(3)/SO(2) = {\rm S}^2$.
This is typical of spontaneous symmetry breaking:
if the system has a global symmetry $G$ that breaks to a subgroup $H$, then
the low energy modes are fluctuations of the coset geometry $G/H$.
In other words, the target space coordinates are Goldstone bosons.  

In some cases a topological term may be added to the effective action, provided that it does 
not break any symmetry respected by the physical system.  
The $O(3)$-model does admit a topological term, which is not invariant under time reversal.
This term can not be included in the antiferromagnetic case, 
but it is allowed in the ferromagnetic case \cite{Burgess:2000}, and it must be included since there
is no symmetry that forbids it.  This is fortunate, since the topological piece of the effective action 
is the only thing at leading order that distinguishes these two models, so without it sigma models 
of quantum magnetism would make no sense.  
This explains why ferromagnetic magnons have linear dispersion, while antiferromagnetic magnons 
have quadratic dispersion.  It is hard to find a better illustration of the importance of 
topological terms in the low energy effective action.

An effective field theory of magneto-transport, obtained by adding a topological term parametrized by 
$\sigma_H$ to a non-linear tensor sigma model of localization that predates the QHE \cite{EfetovLK:1980},
was proposed soon after discovery of the integer QHE (IQHE) \cite{LevineLP:1983, *LevineLP:1984a, 
*LevineLP:1984b,*LevineLP:1984c, *Pruisken:1984, *LevineL:1985}.
This is sufficient to emulate some qualitative features of the integer phase and renormalization group 
(RG) flow (scaling) diagram proposed in Ref.\;\cite{Khmelnitskii:1983}, but reliable non-perturbative 
results that may be confronted with critical data (location of saddle points, and flow rates near these points) 
have not been extracted from this model. Furthermore, this model does not include the interacting case, 
which exhibits the perplexing fractional QHE (FQHE), and therefore does not contain the 
modular symmetry observed in the data (cf.\;Ref.\;\cite{OLLutken:2018} and references therein).

This is in sharp contrast to the toroidal sigma model proposed in 1991 \cite{LR:1992,LR:1993}.
Not only does it automatically unify the integer and fractional QHE in a cohesive and attractive 
geometric framework, it also provides strong constraints on critical behavior that are only now, 
almost three decades later, becoming experimentally accessible \cite{LR:2007, LR:2014}.  
We shall here extend and explore this model, and argue that it accounts for most, if not all, 
universal data observed in the QHE, in a profoundly geometrical and topological manner \cite{footnote:openstring}.

Since we are not yet able to derive this model from microphysics, we rely instead on two  guiding principles  
distilled directly from empirical evidence (``phenomenology"), which impose severe constraints 
on the construction of an effective (emergent) theory:\\
\emph{(i) Symmetry:} The scaling of the model 
should match the observed modular scaling of $\sigma(t) = \sigma_H(t) + i \sigma_D(t)$.
\emph{(ii) Topology:} Hall quantization should be protected by topological properties of the 
effective quantum fields relevant at low energy.

\subsection{Modular symmetry}\label{subsec:modsym}

This subsection recalls how the geometry of an EFT is intertwined with the geometry of the parameter space
of the model. Quantum field theories are best regarded as families of models, one for each point in parameter 
space. Conventional renormalization theory only explores small regions of parameter space close to scale invariant 
fixed points.  The full renormalization group acts non-perturbatively on the whole parameter space, 
connecting fixed points by flow lines whose global geometry typically is beyond our perturbative reach.
Modular symmetries are discrete symmetries acting directly on the whole parameter space.
They are intrinsically non-perturbative since they connect widely separated models, and may be regarded as 
non-abelian generalizations of the more familiar Kramers-Wannier duality in Ising models.

Our starting point is the assumption that the two scale dependent conductivities $\sigma_H$ and $\sigma_D$ parametrize an EFT that captures all universal transport properties of the QHE.  Our first task is to discuss
the parameter space $\overline{\mathcal M}$ of this model, which in mathematics and string theory usually 
is called a \emph{moduli space}, especially its topology and any symmetries it 
may have. Since $\sigma_D \geq 0$, the conductivities take values in a compactification of 
the upper half of the complex conductivity plane \cite{footnote:overline},
$\overline{\mathcal M}[\sigma(t; \dots)] = \overline{\mathbb C^+}(\sigma = \sigma_H + i\sigma_D)$,
where $t$ is the dominant scale parameter, and the ellipsis represent non-universal quantities that 
depend on the type of material used and other ``evironmental" (non-scaling) parameters 
that can be dialled at will (e.g., the magnetic field $B$). We first give a brief glossary of the 
vernacular used in the theory of scaling and renormalization, as it is used throughout this article. 
Most of it is borrowed from hydrodynamics.

All universal data are believed to be encoded in a scalar function $C$ (the \emph{RG potential})
that can be visualized as a potential landscape over the parameter space $\overline{\mathcal M}$
(cf.\,Figs.\;\ref{fig:Landscape} - \ref{fig:NewLandscape}).  
Since the properties of $C$ are inextricably intertwined with the topology and geometry of 
$\overline{\mathcal M}$, it is important to define this space properly. 
 
As long as the inelastic scattering length is smaller than the size of the Hall bar, the dominant scale parameter 
is the temperature $T$, and $t = \ln T$.  The \emph{phase and flow diagrams} on display in this article are 
obtained by studying families $\sigma(t;\,B,\dots)$ of quantum Hall data.

A \emph{flow line} tracks how the effective (renormalized) values of the transport coefficients change when the scale parameter (temperature) is changed.  We are free to dial any starting point for a flowline by changing 
non-scale parameters, like the magnetic field $B$, which must subsequently remain fixed while only 
the scale parameter $t$ is changed. 

A \emph{flow diagram} is a collection of such flow lines, which if possible are 
chosen so that they ``spread out" and probe as much of parameter space as possible (in practice 
experimental limitations severely constrains access to initial values).  Since flow lines cannot
cross phase boundaries they map out the phase diagram. The geometry of any flow diagram is 
controlled by \emph{fixed points} of the flow.  These are points in parameter space where the flow 
(scaling) stops, so they are by definition scale invariant.  

Sources for the flow are called \emph{repulsive ultraviolet (UV) fixed points}, which we represent by the 
icon $\repulsor$.   Experiments reveal that these fixed points lie on the boundary of parameter space.
Sinks for the flow are called \emph{attractive infrared (IR) fixed points}, which we represent by the 
icon ${\attractor}$. In the QHE these are the plateaux, where $\sigma_H$ is a rational number ($\in \mathbb Q$) 
in units with $e^2/h = 1$, and $\sigma_D$ vanishes.  Rational points should therefore be 
included in the physical parameter space for the QHE. In mathematics this is called a 
\emph{compactification} of $\mathcal M = \mathbb C^+$ to the set
${\overline{\mathcal M}}(\sigma) = \mathcal M(\sigma) \cup \{\attractor\}$  (cf.\;Fig.\;\ref{fig:TRSgrid} 
in the Appendix). For topological reasons only rational points may be added (including the 
``fraction" $\infty = 1/0$), and this set is called the \emph{boundary} of $\overline{\mathcal M}$.
Notice that each phase is ``attached to" (compactified by) a single, unique attractor ${\attractor} = p/q$, 
whence the pair of integers $(p, q)$ labels that phase (cf.\;Fig.\;\ref{fig:DyonGrid}).

Saddle points for the flow (one attractive and one repulsive direction) are called \emph{semi-stable fixed points}, which we represent by the icon $\qcp$.  In the QHE these are the quantum critical points that control
quantum phase transitions, sometimes called the localization-delocalization transition, between phases 
attached to different plateaux.  Physical critical points are vanishing points (simple zeros) of 
the vector field $\beta = (\beta_H, \beta_D)$ of scaling functions,
which belong to the interior of parameter space.  

This fixed point structure can be extracted directly from the geometry of the data 
without any theoretical bias.  If they reveal a hidden order (symmetry), then they are the 
DNA of this symmetry from which all else will follow.  Our main assertion is that quantum Hall data 
does reveal such an order, encoded in the nested hierarchical structure of phase portraits 
(cf.\;Figs.\;\ref{fig:7Sisters} - \ref{fig:NewLandscape}).  
This is the signature of an approximate global discrete symmetry, which, given some familiarity 
with infinite discrete groups, is surprisingly easy to identify by finding some of the fixed points.   
The symmetry in question is called \emph{modular}.

We have previously shown that there is substantial experimental evidence that the scaling of transport 
coefficients in the QHE is harnessed by a modular symmetry \cite{LR:1992, 
*LR:1993, Lutken:1993a, *Lutken:1993b, Lutken:2006, BLutken:1997, *BLutken:1999, 
LR:2007,LR:2006, *LR:2009, *LR:2011, LR:2010, LR:2014, NLutken:2012a, *NLutken:2012b,  
Lutken:2014a, *Lutken:2015, OLLutken:2018}.  The symmetry constraint \emph{(i)} is therefore 
imposed by experimental data, unadulterated by theoretical assumptions, and is therefore non-negotiable.

A modular transformation  $\gamma$ is simply a special type of M\"obius transformation
\begin{equation*}
\sigma \in \overline{\mathcal M} \longrightarrow \gamma(\sigma) = \frac{a\sigma + b}{c\sigma + d} 
\in\overline{\mathcal M},
\end{equation*}
where the coefficients are integers satisfying $ad - bc = 1$.
These transformations preserve the compactified moduli space, and the set of all transformations 
restricted in this way (there are infinitely many) form a group that is
called the \emph{modular group} $\Gamma(1) = {\rm PSL}(2,\mathbb Z)$.
It is generated by a (horizontal) \emph{translation} $T(\sigma) = \sigma +1$, and a \emph{duality} transformation,
which in this case is the inversion (``reflection") $S(\sigma) = -1/\sigma$ in the unit circle.
Since the generators $T$ and $S$ do not commute, 
e.g., $TS(\sigma) = (\sigma - 1)/\sigma \neq -1/(\sigma + 1) = ST(\sigma)$, 
this is an infinite, discrete, non-abelian group that we sometimes write as $\langle T, S\rangle$.
The symmetry observed in experiments is always slightly smaller, usually one of the maximal 
subgroups $\Gamma_{\rm X}\subset \Gamma(1)$ \mbox{(X = R, S, or T)} \cite{footnote:GammaRST},
in which case the integer coefficients satisfy additional parity constraints (cf.\;Sec.\;\ref{subsec:targets}
and the Appendix).

This is a target space symmetry, or more precisely, a symmetry of the space of target spaces.  
Unlike in string theory, we are not demanding modular invariance on 
the ``world-sheet" $\Sigma$, which would be tantamount to requiring the model to be critical (conformal) 
for all target space geometries, and there would be no RG flow. 
This is what happens in string theory, where all deformations of the moduli of a given 
space-time vacuum [which includes a compact Calabi-Yau (CY) space of fixed topology] are truly marginal, 
and the central charge of the conformal algebra is fixed at a value given only by the dimension of the target 
space, independent of its shape and size.

Since modular mathematics derives from the theory of \emph{elliptic curves} (tori), 
this is a strong indication that a toroidal model (a sigma model with a toroidal target space 
${\rm M} = {\rm T}^2$) may be useful.
Furthermore, the mathematical structure of a toroidal model forces the plateaux values
of the Hall conductivity to be rational. This follows from the fact that
any toroidal model automatically is endowed with a modular symmetry, and for this symmetry to 
act ``properly" (in a strict mathematical sense) on the mathematical moduli space, 
this space cannot contain all real points, only rational ones \cite{DiamondS:2005}.    
Therefore, since a plateau by definition belongs to the boundary of moduli space ($\sigma_D^\smallattractor = 0$), 
it must be rational ($\sigma_H^{\smallattractor} \in\mathbb Q$), as is observed in all quantum Hall experiments.

This is the first example (more will follow) of a surprising and remarkable confluence of toroidal mathematics and 
quantum Hall physics: \emph{in any toroidal model modular symmetry, rational Hall quantization,
and  unification of integer and fractional quantum Hall effects, is automatic and inescapable.}

This is, however, just the first hurdle, as it only explores the boundary of the moduli space 
$\overline{\mathcal M}$ of the model. The real test is what happens in the interior of 
$\overline{\mathcal M}$, which is where all quantum critical points reside.
The most important (experimentally accessible) information is the location of critical points, 
and the principal flow rates $y^\pm$ near these semi-stable RG fixed points.
The relevant ($\nu^+>0$) and irrelevant ($\nu^-<0$) critical (delocalization) exponents 
are the inverse flow rates, $\nu^\pm = 1/y^\pm$.

\subsection{Renormalization}\label{subsec:renorm}

This subsection recalls how properties of RG flows are encoded in the geometry of sigma model
parameter spaces. We also recall the concept of ``geometrostasis", which will be used to model 
quantum critical points in the QHE, and how infinite (modular) symmetries under favorable circumstances
conspire to pin down the non-perturbative geometry of renormalization.

We shall ultimately construct what is arguably the simplest and most natural viable toroidal model,
whose scaling predictions are in good agreement with all experiments to date, 
including reasonable values of the critical exponents (the meaning of ``reasonable" in this context is 
discussed at length in Sec.\;\ref{subsec:flowrates}).  To prepare for our investigation of scaling properties 
of these \emph{elliptic models} (a class of toroidal sigma models constructed below), especially the 
calculation of critical exponents, we first collect here some general (model independent) 
results about two-dimensional RG flows that have been obtained in the wake of the conformal field theory 
approach to string theory. Although our models only are conformal (scale invariant) at isolated (quantum critical)
points in parameter (moduli) space, conformal symmetry in low dimensions severely constrains scaling 
near these points, as described by the $C$-theorem \cite{Zamo:1986}.

Renormalization of a two-dimensional sigma model is \emph{geometrical}, i.e., given by 
``quantum deformations" of the target geometry (shape, size and torsion), with beta functions
\begin{equation}
\beta_{ab} = \dot K_{ab} = \frac{d K_{ab}}{dt} 
= \frac{1}{2\pi} \hat R_{ab} + \dots\propto \frac{\delta C}{\delta K_{ab}}.
\label{eq:Ricci}
\end{equation}
$\hat R$ is the Ricci curvature of the generalized connection $\hat\Gamma = \Gamma - H$, where
$\Gamma$ is the Christoffel connection compatible with the metric $g$, 
and $H = dh$ is the torsion  \cite{CurtrightZ:1984, BraathenCZ:1985, Zamo:1986, footnote:torsion}.

$C\in \mathbb R$ is an ``effective action" or ``RG potential" (on the moduli space 
of the target manifold) that gives a gradient RG flow \cite{Zamo:1986}.
A critical point is an inflection point of $C$, whose value at this scale invariant point is 
the central charge of the enhanced conformal symmetry.
$C$ decreases monotonically along RG flow lines, which suggests that it is a kind of 
``vacuum entropy" that measures the degree of entanglement of the ground state of the effective 
quantum field theory, as a function of scale
\cite{Preskill:2000, *VLatorreRK:2003, *LatorreRV:2004, *LLutkenRV:2005, *Nishioka:2018, 
*HollandsS:2018, *Witten:2018a, *Witten:2018b}.

The beta functions define a vector field tangent to RG flow lines in the parameter space $\overline{\mathcal M}$.
By definition a critical point is a point in ${\mathcal M}$ where all beta functions vanish.
The set of these critical geometries may be a finite dimensionalal sub-manifold of $\mathcal M$ (as in string theory),
a line of fixed points (as in the Baxter model), or isolated fixed points, as in the QHE.
If we only have massless toroidal bosons nothing much can happen, and classical conformal symmetry is 
preserved at the quantum level (no RG flow). Models with interesting flows are obtained by adding other fields. 
Typically these are charged matter fields coupled to gauge potentials, but beware that  
the distinction between fermions and bosons can be blurred in two dimensions.

If the RG flow, for exceptional (fine-tuned) initial conditions, stops at a nontrivial fixed point 
in the interior of the moduli space of target geometries (a saddle point $\qcp\in\mathcal M$,
cf.\;Fig.\;\ref{fig:TRSgrid}), then the target space is frozen (albeit unstable) at this quantum critical point, 
and scaling properties of the model are determined by the shape, size and torsion of the critical 
target space geometry. It is this phenomenon, sometimes called \mbox{\emph{geometrostasis}} 
\cite{BraathenCZ:1985}, that will be employed here to model quantum critical points in the QHE.

To leading order a sigma model is critical when the target space is \^Ricci flat 
[$\hat R = 0$, cf.\;Eq.\;(\ref{eq:Ricci})], but this perturbative result does not provide enough information 
for us to locate critical points, much less calculate critical exponents.  However, if the model also has some 
degree of modular symmetry, then this is so constraining that the beta functions in simple cases are all but fixed. 
This follows from the explicit classification of modular forms of the required type: 
so-called \emph{weight two forms}, if they exist, transform like vectors, as do beta functions.
It is the fortunate fact that modular beta functions are essentially unique (up to an overall normalization) 
if the symmetry is large enough, but not too large, and uniquely associated with modular (RG) potentials, 
which will allow us to calculate critical exponents. In order not to interrupt this introductory narrative, 
a summary of previous work where this surprising result was obtained is deferred to the Appendix.
This is arguably another example of the compelling convergence of toroidal mathematics and quantum Hall physics.

In short, a sufficiently large modular symmetry (which are the ones appearing in the QHE), 
combined with geometric properties of RG flows (embedded in the $C$-theorem), 
renders the beta function all but unique, up to an overall normalization that may be taken to be real 
(as explained in the Appendix).  It is remarkable that these symmetries reduce our ignorance about 
the flow to a single number.  It is even more remarkable  that three decades of experiments have failed 
to falsify this statement. 

Modular symmetry is not by itself sufficient to give the values of critical exponents, since it does not
fix the normalization of the beta function, so this is as far as we can go with symmetry arguments. 
Our ``only" remaining task is to identify a properly (physically) normalized modular 2-form (beta function), 
and to expand this form near a convenient quantum critical point in order to obtain 
values for the delocalization exponents $\nu^\pm$ that can be compared with numerical 
and real experiments.  In order to advance we need the dynamical information encoded in an EFT,
and to aid in the identification of a suitable model we now appeal to constraint \emph{(ii)}.

\subsection{Topology}\label{subsec:topology}

This subsection recalls topological ideas that will be used to construct the EFT.
Some of these are familar from first quantized (wave function) models of the QHE,
but appear here in an entirely different way.  Our second quantized (field theory) models
are ``derived" or inferred from some rudimentary phenomenological and mathematical observations.
This automatically includes the fractional case, where the topological ``slope"
of a vector bundle on target space generalizes the Chern class that suffices in the non-interacting
(integer) case.

We have argued above that the topology and geometry of modular symmetry ``explains" the remarkable 
precision of rational Hall quantization. The elegance of this argument notwithstanding, it relies on a 
delicate mathematical property of an exact (infinite and discrete) global symmetry, which at first sight 
seems unlikely to reflect reality. 
Nevertheless, Hall quantization is as close to simple rational numbers as we can measure (parts per billion), 
so there must be some property of the effective low energy fields that ``protects" or ``rigidifies" the
modular symmetry, and therefore Hall quantization, rendering both essentially exact 
for all practical purposes.  A plausible candidate would be some kind of ``topological kink" in 
the emergent quantum fields that cannot be altered by small fluctuations.

The topological constraint \emph{(ii)} is more theoretical and therefore less obvious than \emph{(i)}, 
but this constraint also derives from empirical observations.  
The remarkable precision of Hall quantization is one motivation;
another is the remarkable robustness of the QHE, observed in many different materials 
over many decades in temperature. Topological charges of a quantum 
field are by definition robust, i.e.,  insensitive to continuous deformations.  

Unlike \emph{(i)}, which is a sharp statement about an unfamiliar but unique group 
(easily extracted from each experiment), by itself \emph{(ii)} is quite vague and unpromising. 
All it does is to instruct us to look for some kind of suitable topological property of the low energy 
effective fields.  However, in conjunction with \emph{(i)} it is surprisingly effective.  
Given that we are looking for a topological structure on elliptic curves, as inferred from \emph{(i)},
the possibilities are greatly reduced.  Furthermore, since we expect gauge invariance to 
play a central role in this model \cite{Laughlin:1981}, a natural candidate presents itself. 

To see this we recall a truly serendipitous mathematical fact, which appears to be another example of 
the convergence of toroidal mathematics and quantum Hall physics:
\emph{the topological classification of vector bundles (gauge fields) on elliptic 
curves provides precisely the kind of framework required by the topological constraint (ii)}.
We have no compelling physical argument forcing us to decorate the model with topologically
interesting gauge potentials, but rather than looking this gift horse in the mouth we will see how far it can carry us. 

This kind of ``pattern matching" of experimental data to an unfamiliar mathematical structure, whose
microscopic origin is obscured by our inability to solve strongly coupled disordered QED, leaves a lot to be desired.
At this point it is at best an educated guess, but this is not inconsistent with the ethos of EFT.
Furthermore, since it is abundantly clear that emergent phenomena may look absolutely nothing like their
provenance, there are no a priori constraints on the mathematical structures that may be employed in the
construction of a model. The most fruitful question is not where the model came from, but where it is going:
does it make falsifiable predictions about the outcome of experiments?  If not, it is metaphysics 
and of no use to physics. Fortunately, the predictions made by the model we are in the process 
of ensnaring here are very rigid indeed, and that is its main virtue.

Consider first (holomorphic) vector bundles $\mathcal V_{r,d}$ (complex analytic 
gauge theories) over any ``nice" curve, i.e.,  any Riemann surface ${\rm M}_g$. 
They have integer valued topological invariants $\{ g, r, d\}$, 
where $g = {\rm genus}({\rm M}_g)\geq 0$ is the number of holes in ${\rm M}_g$, 
$r = {\rm rk}(\mathcal V_{r,d}\rightarrow{\rm M}_g)\geq 1$ is the rank of  the bundle, 
and $d = {\rm deg}(\mathcal V_{r,d}\rightarrow{\rm M}_g)\in\mathbb Z$ is its degree.  
We do not have a classification of these objects, except when the base space 
(the target space of the sigma model) is spherical 
(${\rm M}_{g = 0} = {\rm S}^2$), or toroidal (${\rm M}_{g = 1} = {\rm T}^2$).  
Since every holomorphic bundle on a sphere is a direct sum of line bundles $\mathcal L_d\rightarrow{\rm S}^2$
($r = 1,\; d\in\mathbb Z$) \cite{Grothendieck:1957}, a sigma model with this target geometry could 
conceivably provide a topological model of the IQHE [$\sigma^{\smallattractor}_H = c_1(\mathcal L_d) = d$], 
but it is not rich enough to model the FQHE. 

The first nontrivial case is the elliptic curve, in which case indecomposable holomorphic vector 
bundles $\mathcal V_{r,d}$ that are not line bundles do exist ($r\geq 1,\;d\in\mathbb Z$).  
They are classified by the topological ratio 
\begin{equation*}
\mu(\mathcal V_{r,d}\rightarrow {\rm T}^2) = \frac{c_1(\mathcal V)}{{\rm rk}(\mathcal V)} 
= \frac{d}{r} \in \mathbb Q , \quad\gcd(d, r) = 1,
\end{equation*}
which is called the \emph{slope} of the bundle, 
together with a complex number that may be identified with a point on the target torus \cite{Atiyah:1957}.
An indecomposable vector bundle is \emph{stable}, which means that all sub-bundles 
have smaller slope.  

Given this classification, it is natural to conjecture that each plateau (and the universality class to 
which it belongs) is paired with a stable holomorphic vector bundle $\mathcal V_{r,d}\rightarrow{\rm T}^2$, 
in such a way that the value of the Hall conductivity is fixed by the topological slope of this bundle: 
\begin{equation*}
\sigma^{\smallattractor}_H = \mu(\mathcal V) = d/r\in\mathbb Q.
\end{equation*}
So the numerator of $\sigma^{\smallattractor}_H$ is identified as the degree of a stable bundle 
(gauge theory) on the target torus ${\rm T}^2$, and the denominator is the dimension of the fibers in this bundle.
Furthermore, these gauge invariant objects appear naturally in the field theory limit of toroidal strings. 

In short, the paucity of suitable mathematical structures that can satisfy both of the phenomenological
constraints $(i)$ and $(ii)$ singles out gauged toroidal sigma models for closer inspection.  
We do not attack this head on by trying to investigate explicit dynamical details of these models, 
e.g., how the gauge symmetry changes during a phase transition \cite{footnote:tachyons}. 
Instead we bypass this difficult problem by drawing on information from the corresponding string theory,
which contains the toroidal sigma model we are studying in a field theory limit.
As it is reasonable to assume that topological features of the toroidal string survive in the field theory limit, 
this gives us a relatively painless way to identify candidate topological degrees of freedom in the low energy 
gauge theory. Furthermore, since the toroidal geometry and the infinite modular symmetry severely constrains 
the energy function for these modes, we can calculate topological partition functions that contain enough
non-perturbative information to give us the location of quantum critical points in $\overline{\mathcal M}$, 
as well as critical exponents at these points.

This is analogous to how symmetries are exploited in order to analyze and classify two-dimensional 
conformal field theories \cite{DiFMS:1997}. This example shows that infinite symmetry can be so constraining 
that sometimes the degrees of freedom, the operator algebra, and ultimately the 
partition function, can be deduced without exhibiting any effective action (Lagrangian or Hamiltonian).  
This has the virtue that although an explicit action may belong to the universality class of interest, 
it is always an approximate representation of the real system.  Unless nonuniversal microscopic 
details are of interest, the least biased approach is therefore to work directly with symmetry representations, 
since that is where universal information resides. Unfortunately, this ``bootstrapping" appears to be possible 
only if the symmetry is infinite.
 
It is intriguing that this reasoning [essentially a process of mathematical elimination guided by $(i)$ and $(ii)$]
converges on an EFT containing the same type of mathematical machinery (Chern classes) that was found 
to be useful in the first-quantized approach using approximate wave functions of non-interacting particles
\cite{ThoulessKNdN:1982, NThouless:1984, NThoulessW:1985, AvronSS:1983, Simon:1983, 
AvronS:1985, AvronSSS:1989}, but since the field theory argument makes no reference to multi-particle states
it is mysterious why this should be so.
Another intriguing fact is that the symmetry will force us to 
consider so-called \emph{quasimodular forms}, whose Fourier expansions may be identified as instanton 
expansions, allowing us to compare some aspects of toroidal models more or less directly with the 
dilute instanton gas analysis of the original tensor 
sigma model \cite{LevineLP:1983, *LevineLP:1984a, *LevineLP:1984b,*LevineLP:1984c, *Pruisken:1984, 
*LevineL:1985}  (cf.\;the Appendix).

\subsection{Mirror symmetry}\label{subsec:mirrorsymmetry}

This subsection recalls that some sigma models are constrained by a quantum equivalence 
that is called ``mirror symmetry".  It is the serendipitous conjunction of mirror and modular symmetry in 
toroidal models that allows us to analyse these models in some detail.

\begin{figure}[tbp] 
\includegraphics[scale = .38]{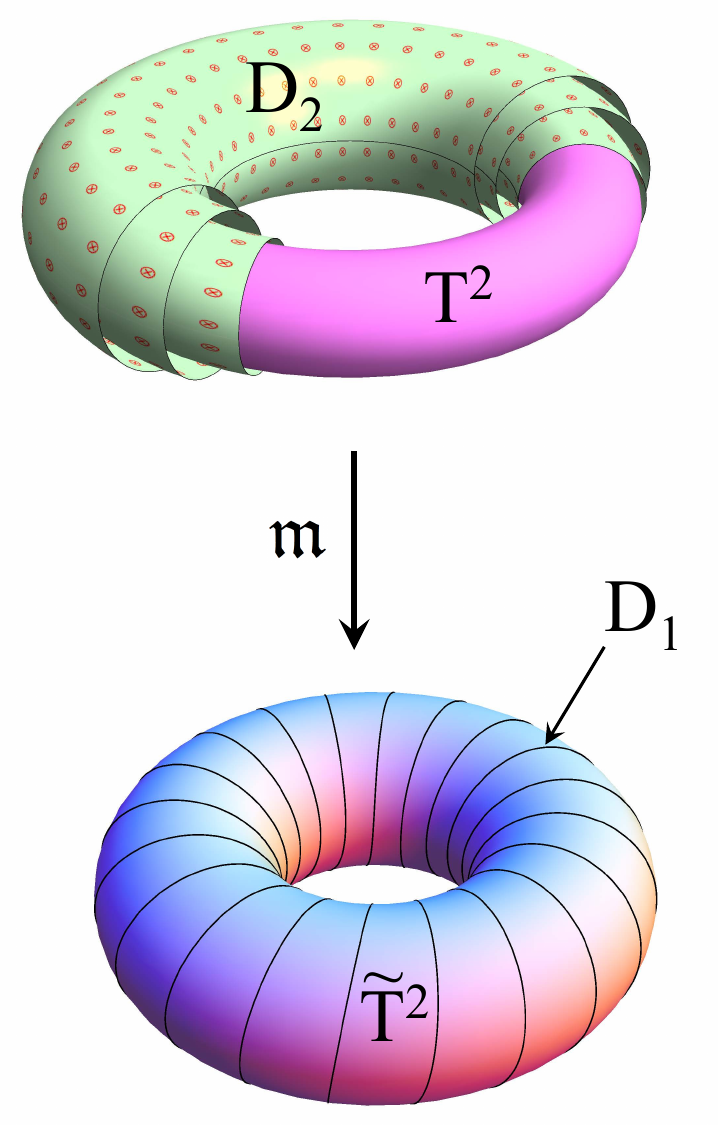}
\caption[Mirror tori]{Cartoon illustrating some of the remarkable properties of Calabi-Yau (CY) 
spaces (compact complex manifolds with vanishing first Chern class) that were discovered in string theory.  
In general a CY manifold is mirrored in a topologically distinct manifold $\widetilde{\rm CY}$, but in one (complex) dimension the only CY space is a torus,  so our
target space $\rm T^2$ and its mirror $\widetilde{\rm T}^2$ are topologically equivalent.
Heuristically, the mirror image of a stack of $D_{2}$-branes wrapped around $\rm T^2$ (top)
(a flat holomorphic vector bundle in the geometric limit) is a $D_1$-brane wrapping $\widetilde{\rm T}^2$ (bottom)
(a gauge invariant Wilson-loop in the geometric limit).
\label{fig:MirrorCartoon}}
\end{figure}

In order to obtain some physical insight that can help us understand the degrees of freedom of this model, 
in particular how their geometrical and topological properties conspire to  explain the robustness of Hall 
quantization, we exploit a quantum equivalence called \emph{mirror symmetry}.
Mirror symmetry maps the original toroidal model to another, equivalent, toroidal model, whose shape 
is  parametrized by the conductivity. In this model the mirror image of a vector bundle is a ``dyonic" 
Wilson loop (a gauge invariant field configuration that is both electrically and magnetically charged), 
whose energy depends on the shape of the mirror torus, i.e., the conductivity. 
The topological charges of these field configurations are simply \emph{winding numbers} on the mirror 
torus. An illustration of this stringy phenomenon is shown in Fig.\;\ref{fig:MirrorCartoon}.

This allows us to argue that for each universality class the location of the attractive fixed point 
(plateau) is determined by the topology of a stable vector bundle on the target torus. 
Each phase of the mirror model is dominated by the dyon that has the lowest energy in 
that phase, and every physically stable dyon defines a phase in this way.  Because the dyonic energy function is 
modular invariant, each stable dyon is uniquely associated with a plateau on the boundary of moduli space,
and the rational value of the Hall conductivity is the ratio of magnetic to electric charge of the dyon.
Reversing the mirror map, each stable dyon maps to a bundle with rational slope, 
whose topology therefore is uniquely associated with a rational Hall plateau. 
Since this bundle is topologically stable, we may conclude that mathematical (topological) stability of 
vector bundles in the original model is equivalent to physical stability of dyons in the mirror model, 
which is easier to understand since it follows directly from charge and energy conservation.

Because this somewhat oblique argument relies only on the topology and geometry of two physically 
equivalent mirror models, it is not burdened by difficult calculations.  The models have already rolled 
non-linear dynamics into the quantum geometry of $D$-branes on a pair of mirror target spaces.
This is encoded in the geometry of the quantum moduli space, in much the same way that instanton corrections 
in one CY manifold (string vacuum) may be obtained from the geometry of complex structures on 
the mirror manifold \cite{ALutkenR:1990, *ALutken:1991a,*ALutken:1991b,*ALutken:1991c,
*KLLutkenW:2000, *LLutkenS:2002, Mirror:2003}.

\subsection{Matter}\label{subsec:matter}

This subsection recalls how matter fields (fermions) may be added to a sigma model whose target space
admits ``spin structures".

It is an experimental fact that the parity of the numerator or denominator of $\sigma^{\smallattractor}_H$ 
(or both) always is constrained. This is a crucial piece of phenomenological information, since it determines
the type of modular symmetry that should be built into the model.  
Fortunately, this constraint is automatically built into a toroidal model with matter fields.

There is in general a topological obstruction to including fermions, and in order to remove this obstruction
the model must be equipped with an additional geometric structure that is called a \emph{spin structure}.
This implies that the fermions must satisfy certain boundary conditions, or, equivalently, that they
couple to certain flat gauge potentials. On a torus there are three possible choices of spin structure.
Mathematically, this is equivalent to ``enhancing the elliptic curve with 2-torsion data" 
(this group torsion has nothing to do with the metric torsion discussed above)
\cite{DiamondS:2005}.  This reduces the modular symmetry to one of the three maximal 
subgroups [$\Gamma_{\rm X}\subset \Gamma(1)$ \mbox{(X = R, S, or T)}] observed in the scaling data. 

In view of the supersymmetry lurking in disordered systems (including the QHE) \cite{Efetov:1997}, 
it is natural to pair each of the two bosons with two fermions, which to leading order this just adds a Dirac 
term to the effective action. Since the central charge of the conformal symmetry of a free boson (fermion)
is $1$ ($1/2$), our naive expectation is that this elliptic model has a total central charge 
$c_\smallqcp = 2\times 1 + 4\times 1/2 = 4$ at critical points associated with non-interacting integer phases,
and modular symmetry carries this over to any fractional transition.

For comparison with experiment we retain only the weaker idea that the central charge at a critical point
should be a natural number ($c_\smallqcp = n\in\mathbb N$). This gives a discrete set of values $\nu_n^\pm$
of the critical exponents. 
Comparing these with values $\nu_{\rm num}^\pm$ obtained from numerical simulations of the delocalization transition using the Chalker-Coddington (CC) model \cite{ChalkerC:1988}, we find that they are in 
remarkable agreement with our naive expectation, $\nu_4^\pm \approx \nu_{\rm num}^\pm$ \cite{LR:2014}.
At  the very least, this shows that viable toroidal models exist.
It is striking that what is arguably the simplest and most natural model of this kind (i.e., containing all the 
geometrical and topological structures dictated by phenomenology), is so close to experimental reality.

\subsection{Summary}\label{subsec:summary}

Our two phenomenological constraints \emph{(i)} and \emph{(ii)} have uncovered a 
conjunction of favorable mathematical circumstances (closely knit topological and geometrical 
structures that appear to be tailor-made for the QHE).  
This invites us to conjecture that \emph{Hall plateaux, and the universality 
classes to which they belong, can be modeled by topologically stable holomorphic vector bundles on a torus
with spin structure.}  In this toroidal sigma model the topological invariant that protects the robust and 
universal plateaux values of the conductivity is the slope $\mu(\mathcal V)$ of stable holomorphic vector bundles 
$\mathcal V$ on the target space, $\sigma_H^{\smallattractor} = \mu(\mathcal V) \in\mathbb Q$.  
In the equivalent mirror model topological protection is provided by more familiar winding numbers.

The non-interacting case is modeled by line bundles $\mathcal L_d$ ($r = 1$), 
from which a familiar looking topological statement about the IQHE is recovered \cite{ThoulessKNdN:1982,
NThouless:1984, NThoulessW:1985, AvronSS:1983, Simon:1983, AvronS:1985, AvronSSS:1989}, 
$\sigma_H^{\smallattractor} = c_1(\mathcal L_d) = d \in\mathbb Z$, albeit in a field theoretic context.  
The connection to the first-quantized theory, which is based on drastically simplified but explicit plateaux 
wave-functions, is not completely obvious. Some remarks on this connection are deferred until the 
discussion in the final section.

In the absence of a theoretical derivation of this EFT from microphysics, the merits of the 
toroidal model can only be decided by comparison with experimental data, 
as we review in Sec.\;\ref{sec:data}
\cite{LR:1992, LR:1993, 
Lutken:1993a, Lutken:1993b,Lutken:2006, 
BLutken:1997, BLutken:1999, 
LR:2007, LR:2014,
LR:2006, LR:2009,LR:2011, LR:2010,  
NLutken:2012a,NLutken:2012b,  
Lutken:2014a, Lutken:2015, OLLutken:2018}.

The goal of the next section is to derive the phase diagram and partition function for the toroidal model 
we have tried to motivate above.  We first review the geometry and topology of generalized sigma models 
with toroidal target spaces, relying heavily on results from string theory.
The partition function is evaluated by exploiting mirror symmetry, 
which in the context of string theory found sufficient traction to spearhead fundamental new 
insights in algebraic geometry.  Scaling properties (the RG flow) of this partition function
is investigated in Sec.\;\ref{sec:RGflow}.

Modular symmetry is a quantum symmetry that only emerges when thermal fluctuations 
are too feeble to overcome quantum ordering, so the extremely cold experiments that 
are now available could easily falsify the model. They have instead provided strong evidence in its favor. 
The geometry of scaling flows extracted from the coldest quantum Hall experiments is in good 
agreement with modular predictions, including the location of many quantum critical 
points \cite{LR:2006,LR:2007,LR:2011,LR:2014, OLLutken:2018}.

We also compare this analytical model with the numerical Chalker-Coddington (CC) model
\cite{ChalkerC:1988}, which has been under intense scrutiny since its conception in 1988, 
because it was designed to emulate the quantum critical delocalization transition in the QHE.
The family of toroidal models discussed here yields a discrete set of candidate delocalization exponents $\nu_n$, 
where $n$ is an integer. One of these is $\nu_4 = \nu_{\rm tor} = 2.6051\dots$, which is in excellent agreement 
with the best available numerical value $\nu_{\rm num} = 2.607\pm\,.004$ \cite{SlevinO:2009, SlevinO:2012}. 
This suggests that our family of toroidal models includes the universality class of the CC model.

In order to investigate if these theoretical (numerical and analytical) constructions faithfully model 
universal properties of the QHE near criticality, a careful examination of real scaling data is required.
The real delocalization exponent may be disentangled from other scaling exponents
in finite size scaling experiments, which has yielded an experimental value $\nu_{\rm exp} = 2.3 \pm .2$ 
that appears to be about 10\% smaller than the theoretical values $\nu_{\rm num} \approx \nu_{\rm tor}$.

However, the modular model suggests that there may be a sensitivity to initial conditions in these 
experiments that has not been fully appreciated, in which case it is possible that no experiment so far has
been in the proper scaling domain for the delocalization exponent. If we use the modular scaling function 
to take this into account, the adjusted experimental value does line up with the theoretical one \cite{LR:2014}. 

Another subtlety that complicates the comparison with data is that there could be more than
one universality class, depending on the type of disorder (essentially long vs. short range),
which may not have the same delocalization exponent.  If so, the original CC model would only
be relevant for (at most) one of them, but it is conceivable that the discrete family of toroidal models
could contain both.  In order to determine if these models really are in (one of) the quantum Hall 
universality class(es), improved finite size scaling experiments are urgently needed.

\section{Toroidal models}\label{sec:toromodels}

In less than four dimensions the classification of spaces equivalent under {homeomorphisms} 
(continuous transformations) and {diffeomorphisms} (smooth transformations) are the same.  
So in dimension one, two and three topological and real geometric structures coincide. 

The only real topological invariant of a two-dimensional Riemann surface is its genus,
i.e., the number of holes.  A finer classification is obtained by considering {complex structures}
(equivalence under holomorphic transformations).

The toroidal sigma model has \emph{two} complex structures $C_\tau$ and $C_\sigma$,
giving the \emph{shape} and \emph{size*} of the target space.
The ``complexified size" \emph{size*}  parametrizes the \emph{complex K\"ahler cone}. 
This is a novel object, not used in mathematics, which is essential for understanding the 
quantum geometry of strings, as well as the quantum field theories that emerge in low energy
(large target space) limits where strings appear point-like. 
It is the key concept on which the following narrative pivots.

\subsection{Target geometry}\label{subsec:targets}

Any two-dimensional Riemannian manifold ${\rm M}(\varphi^1,\varphi^2)$ is equipped with a metric
that may be written
\begin{equation*}
ds^2 = g_{ab} d\varphi^a d\varphi^b = \frac{a}{\Im\tau}\,\vert d\varphi^1 + \tau d\varphi^2\vert^2,
\end{equation*}
where
\begin{equation*}
g_{ab} = \frac{a}{\Im\tau}\left( \begin{array}{cc} 1&\Re\tau\\ \Re\tau&\vert\tau\vert^2 \end{array} \right),
\quad\tau = \frac{g_{12} + a\, i}{g_{11}},
\end{equation*}
and $a  = \volg = \sqrt{\det g}$ is the area of ${\rm M}$.

If ${\rm M}$ is a 2-torus ${\rm T}^2$, then the one-dimensional family $C_\tau$ of distinct complex structures 
(shapes) is given by those values of $\tau$ that are not identified under modular transformations.
It is convenient and natural to pair (``complexify") the area $a  > 0$ with \emph{metric torsion} 
$h_{ab} =   b \,\varepsilon_{ab}$ ($b =  \torg\in\mathbb R$), giving another one-dimensional family 
$C_\sigma$ of \emph{complexified K\"ahler structures} \cite{RobbertVV:1988}, parametrized by the 
size* (size and torsion)
\begin{equation*}
\sigma =  \torg + i\volg = b + a\, i. 
\end{equation*} 
$C_\tau$ and $C_\sigma$ are the building blocks of the moduli space of geometric structures on the target 
space ${\rm T}^2$, on which the renormalization group (RG) acts.

Spinor fields  require the target space ${\rm M}$ to have a spin structure $\# = {\rm PA}$, AP or AA,  
where (A)P means that the fermion is (anti)periodic when transported around a cycle.
Each of the three nontrivial spin structures $\#$ gives a torus ${\rm T}^2_\#$ 
(sometimes called  ``decorated" or ``enhanced") with a reduced modular symmetry that is 
one of the three maximal congruence subgroups $\Gamma_\#$
of the modular group $\Gamma(1) = {\rm PSL}(2,\mathbb Z)$. 
We use one of the generators to uniquely label these subgroups (conventional mathematical notation
is shown in the second column):
\begin{eqnarray*}
\Gamma_{\rm T} &=& \Gamma_0(2) = \langle T, R^2\rangle,\quad
T(z) = z + 1,\\
\Gamma_{\rm R} &=& \Gamma^0(2) = \langle R, {\rm T}^2\rangle,\quad 
R(z) = \frac{z}{1+z},\\
\Gamma_{\rm S} &=& \Gamma_\theta(2) = \langle S, {\rm T}^2\rangle,\quad
S(z) = -\frac{1}{z},
\end{eqnarray*}
where $z = \tau$, $\sigma$ or $\rho = S(\sigma) = -1/\sigma$, depending on context. 
 Quantum Hall phenomenology informs us in no uncertain terms that we must equip our target space with
 a spin structure. Spin polarized scaling data force us to endow the torus with spin structure PA.
The modular symmetry of this ``enhanced"  torus ${\rm T}^2_{\rm T}$ is the subgroup 
$\Gamma_{\rm T}\in\Gamma(1)$ of the modular symmetry $\Gamma(1)$ of the unconstrained torus.
The full geometric structure of the base manifold is therefore locally (i.e., ignoring discrete global identifications)
$(C_\tau\times C_\sigma)_{\rm T}$.

By definition the model ${\rm T}^2_\#$ includes all vector bundles consistent with the spin structure $\#$, 
i.e., all bundles whose slope $\mu = c_1/r$ is given by a suitable subset of all coprime lattice points. 
For the spin polarized quantum Hall model ${\rm T}^2_{\rm T}$ we have $r = 1\mod 2$.
For each spin structure the partition function includes a discrete sum over these topologies. 

The moduli space $\mathcal M$ of this geometry is given by two copies of the upper half plane
divided by the full quantum symmetry group of the model, $\mathcal M(\tau,\sigma) = 
[\mathbb C(\tau)/\Gamma_\# \times\mathbb C(\sigma)/\Gamma_\#]/\mathbb Z_2$,
where $\mathbb Z_2$ is the mirror map \cite{RobbertVV:1988}.  Mirror symmetry is an exact quantum symmetry
that follows from dualizing one of the isometries of the target space, by performing what is essentially 
a Hubbard-Stratonovich transformation on the functional integral in first order form, and integrating out 
the linearized constraint in two different ways \cite{Buscher:1987, *Buscher:1988, Robbert:1994, *Robbert:1998}.

\begin{figure}[tbp] 
\includegraphics[scale = .34]{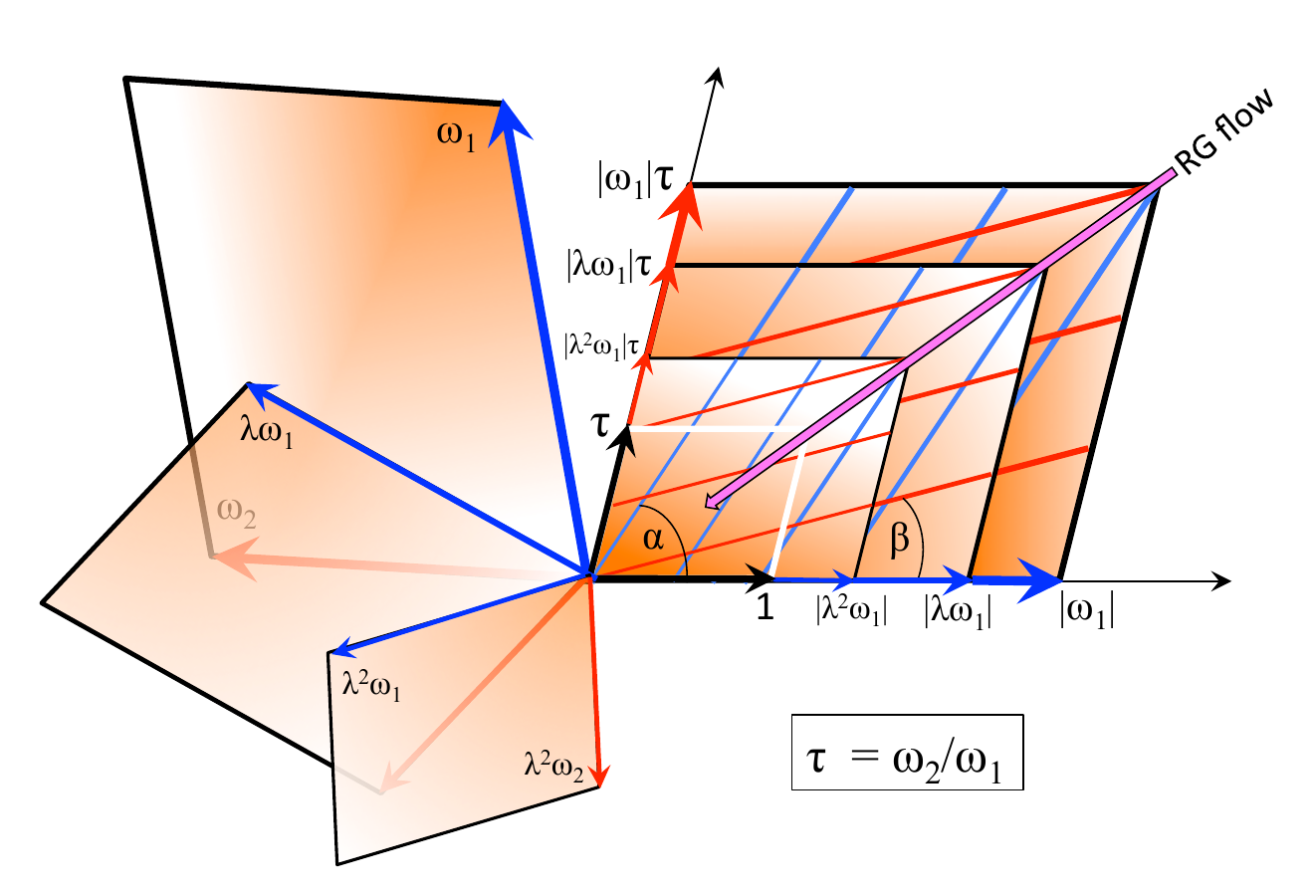}
\caption[Scaling of tori]{ 
An arbitrary flat torus is spanned by basis vectors $\omega_1$ and $\omega_2$.  Multiplying by a complex 
number $\lambda$ (here: $\vert\lambda\vert <1$) rotates and rescales (here: shrinks)  the torus.
A second rescaling by $\lambda$ is also displayed.  Since the orientation of these tori in the complex 
plane is of no physical or mathematical interest, they should be rotated into ``standard position" in the 
first quadrant, i.e. so that the rotated $\omega_1$ is real.  It is then apparent that the rescalings have 
not changed the shape (complex structure $\tau$) of the torus.  
What has changed is the size (area) $a = \Im\sigma = \vert \omega_1\vert^2 \Im\tau$ of the torus. 
\label{fig:ScalingTori}}
\end{figure}

In our model, which is really a family models contained in the extended family of toroidal models 
parametrized by two complex numbers $\tau$ and $\sigma$, the geometry of the target space is given 
by the metric $g_{ab} = \sigma_D \delta_{ab}$ and torsion $h_{ab} = \sigma_H \varepsilon_{ab}$.
The moduli for this sigma model are $\tau = i$ and $\sigma =  \sigma_H + i \sigma_D\in\mathbb C^+$.
Since the complex structure of this model is fixed at $\tau = i$, the target space is a square torus whose 
shape is not renormalized.  Its size* $\sigma$, on the other hand, is not fixed, and it is the RG flow of 
$\sigma$ that will concern us here (cf.\;Fig.\;\ref{fig:ScalingTori}). 

In the strongly coupled domain, which is where critical points appear,
the complexified K\"ahler cone $C_\sigma$ may be transmogrified by topological modes,
and therefore deviate strongly from its classical cousin (the vertical line $\sigma =  a\,i, a > 0$).
We will circumnavigate this problem by using mirror symmetry to find a simpler, and arguably more
intuitive, way to calculate the partition function $\mathcal Z_{\rm T}$.  
This may seem ill-advised in view of the fact that this map is not very
well understood in general, and what is known has been obtained in the context of string theory,
which has far more structure than our model.  However, the string has many geometrical phases,
i.e., phases with an asymptotic limit in string moduli space where the theory reduces to a 
field theoretic sigma model with a CY target space, the simplest being the torus.   
In fact, the elliptic curve considered here is an exception for which the mirror  
map has a solid mathematical foundation \cite{RobbertVV:1988, Robbert:1994, PZaslow:1998}.

Modular symmetries are of course independent of string theory, 
predating strings by a century, and have many applications unrelated 
to strings. That they were being popularized by string theorists 
at the same time as quantum Hall effects were being discovered
was a fortuitous historical accident.

Since the metric is normalized by the effective coupling, the magneto-conductivity 
is playing the role of the inverse coupling constant in a gauge theory, $\sigma_D \sim 1/e^2$.
In the weak coupling limit ($\sigma_D\rightarrow\infty$) the target torus ${\rm T}^2$
degenerates to a sphere, so a spherical sigma model is included in the toroidal model as a
limiting case, but at strong coupling these models are completely different. 
The torus degenerates to a sphere at only one point on the boundary of its moduli space 
(``$\sigma = i\infty$"), and it is not possible to reconstruct the toroidal model from this special case.
In other words, there is no way to infer that a sphere came from a degenerate torus, or for that matter 
from any other topology.  
In the absence of a rigorous derivation of the EFT from microphysics, 
it is only experimental data that can tell us which model is correct at strong coupling. 
Since this is where quantum phase transitions and quantum critical points appear, 
it is the transition region between plateaux that will give us this information.

\subsection{Mirror model}\label{subsec:mirrors}

First discovered in string theory \cite{RobbertVV:1988}, the existence of this particular 
mirror model is now a mathematical fact \cite{PZaslow:1998, Mirror:2003}.
This is a recurrent theme:  frequently  stringy structures have subsequently been found to stay 
up in the low energy field theory limit, without the scaffolding afforded by string theory.
The symmetry between an elliptic sigma model endowed with holomorphic gauge bundles 
($D_{0/2}$-branes), and a mirror model on a dual torus endowed with dyonic Wilson loops 
($D_1$-branes), is one example, which we will exploit here.

\subsubsection{Mirror symmetry on the base space}

If the target manifold ${\rm M}$ has an isometry, then the functional integral has a
remarkable quantum invariance that now is called \emph{mirror symmetry} \cite{Buscher:1987}.
By changing variables in the functional integral, completing squares, and integrating out the isometry, 
a dual sigma model with a different target geometry $\widetilde{\rm M}$ is obtained.

In two dimensions the mirror transformation $\mm$ is 
\begin{eqnarray}
K_{ab} &=&  \frac{\Im\sigma}{\Im\tau}\,
\left( \begin{array}{cc} 1&  \Re\tau\\ \Re\tau & \vert\tau\vert^2 \end{array} \right)
+\left( \begin{array}{cc} 0& \Re\sigma\\ -\Re\sigma & 0 \end{array} \right)\nonumber \\
\mmapdown\kern -7pt&&
\label{eq:mirrormap}\\
\widetilde K_{ab} &=& \frac{\Im\tau}{\Im\sigma}\,
\left( \begin{array}{cc} 1 &  \Re\sigma\\  \Re\sigma& \vert\sigma\vert^2 \end{array} \right)
+\left( \begin{array}{cc} 0&  \Re\tau\\ -\Re\tau & 0 \end{array} \right),\nonumber
\end{eqnarray}
so the mirror map simply swaps the two complex structures $\tau$ and $\sigma$:
$(\tau, \sigma) \stackrel{\mm}{\longrightarrow}  (\tilde\tau, \tilde\sigma) = (\sigma, \tau)$ 
\cite{Buscher:1987, Buscher:1988, RobbertVV:1988, Robbert:1994, Robbert:1998}.
This is the first nontrivial check of mirror symmetry: since there is only one one-dimensional CY manifold, 
the torus must be its own mirror.  ${\rm M} = {\rm T}^2$ and $\widetilde{\rm M} = \widetilde {\rm T}^2$ 
must therefore have the same parameter space, and by virtue of the mirror map $\mm$ they do.

Our quantum Hall sigma model $\mathcal L_{\rm EFT}(\tau,\sigma)$ and its dual 
$\widetilde{\mathcal L}_{\rm EFT}(\tilde\tau,\tilde\sigma)$ are parametrized by
$\tau = i = \tilde\sigma$ and $\sigma = \sigma_H + i \sigma_D = \tilde\tau$, 
so in this case Eq.\;(\ref{eq:mirrormap}) becomes
\begin{equation}
\kern -5pt K_{ab} = \left( 
\begin{array}{cc}\phantom{-}\sigma_D &\sigma_H\\ - \sigma_H &\sigma_D 
\end{array}
\right) 
\stackrel{\mm}{\longrightarrow}
\widetilde K_{ab} =  \frac{1}{\sigma_D} \left(
\begin{array}{cc} 
1 & \sigma_H\\ 
\sigma_H & \vert\sigma\vert^2
\end{array} 
\right) = \tilde g_{ab}.
\label{eq:mirrormetric}
\end{equation}
One of the main virtues of the dual model is that $\widetilde K_{ab}$ is symmetric, 
so that the target space $\widetilde {\rm T}^2$ of the mirror model is 
a conventional torsionless torus,  with periodic coordinates (bosonic fields) 
$(\widetilde\varphi^{\,1}, \widetilde\varphi^{\,2}) \in S^1\times S^1$ and effective action
\begin{equation*}
\widetilde{\mathcal L}_{\rm EFT} 
=   \tilde g_{ab}\, \widetilde\partial_\mu\widetilde\varphi^{\,a}\, \widetilde\partial^\mu \widetilde\varphi^{\,b} + \dots
= \frac{1}{\sigma_D}\, \widetilde\partial_\mu\widetilde \phi \, \widetilde\partial^\mu \overline{\widetilde\phi} + \dots ,
\end{equation*}
where $\widetilde\phi   = \widetilde\varphi^{\,1} + \sigma\widetilde\varphi^{\,2}$ and
$\overline{\widetilde\phi}$ is the complex conjugate field.

Notice that the duality (mirror) transformation has eliminated the unfamiliar torsion,
by exchanging a target space of constant (square) shape ($\tau = i$) and scale dependent size
[$a(t) = \volg = \sigma_D(t)$] for one with constant size [$\tilde a = \voltg  = 1$] 
and scale dependent shape [$\sigma(t) = \sigma_H(t) + i\sigma_D(t)$].
Because $\mm$ is a quantum symmetry, these two models are equivalent, and 
therefore have the same partition function $\mathcal Z_{\rm EFT} = \widetilde{\mathcal Z}_{\rm EFT}$.  

\subsubsection{Mirror symmetry on the fibre}\label{subsubsec:mirroroffibre}
\begin{figure}[tbp] 
\includegraphics[scale = .4]{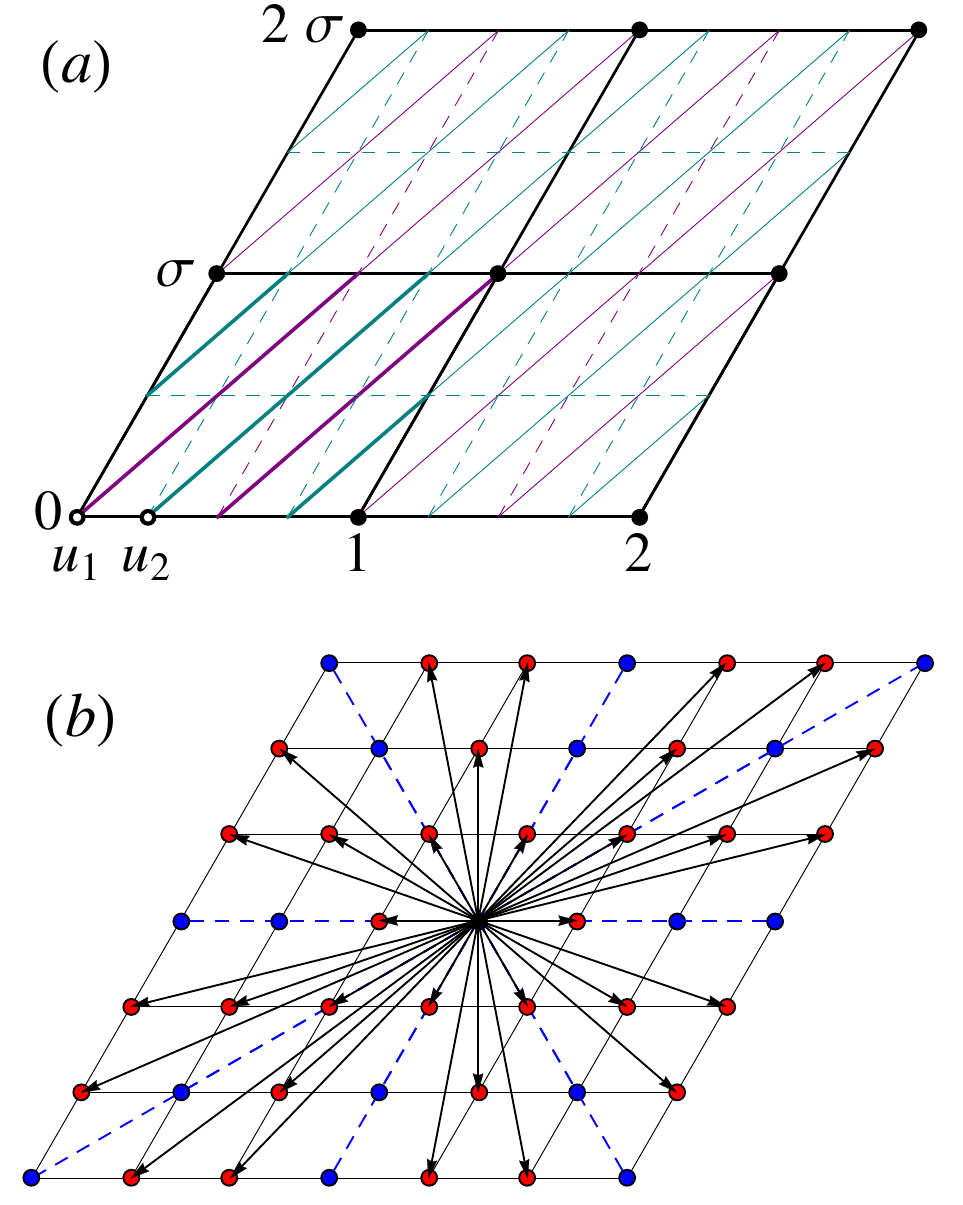}
\caption[Rational branes]{Topological modes in the mirror model $\widetilde {\rm T}^2$ with shape 
(complex structure) $\tilde\tau = \sigma = \exp(\pi i/3)$ and size (K\"ahler structure) $\tilde\sigma = \tau = i$.
(a) The field theory limit of a rational $D_1$-brane on a torus is a straight Wilson loop with three real moduli 
$(\tilde\mu = m/n, u, v)$, which winds $n$ times around one cycle, and $m$ times around the other cycle. 
The centre of mass coordinate $u$ and the value $v$ of the Wilson loop parametrizes a torus that
is isomorphic to the target space, $u + \sigma v \simeq \widetilde\phi\in\widetilde {\rm T}^2$. 
The two parallel closed loops shown here 
(in purple and green) have the same winding numbers ($n_1 = n_2 = 1,\, m_1 = m_2 = 2$) and slope 
($\tilde\mu_1 = \tilde\mu_2 = 2$), but since they have distinct moduli ($u_1 = 0 \neq 1/4 = u_2$) 
they are not the same mode.
(b)  A state has finite energy iff its charge vector $(n,m)$ belongs to this charge lattice, i.e., 
iff the slope $m/n$ of the charge vector is rational.
For a given rational slope, only the point nearest the origin (red), whose charges are coprime, is stable.
The rest of the points (blue), whose charges are not coprime, are unstable.} 
\label{fig:Branes}
\end{figure}

By a toroidal sigma model we mean more than just a toroidal target space:
the base manifold ${\rm T}^2$ comes equipped with a holomorphic structure, 
a spin structure respected by fermions, and gauge fields with nontrivial topology.  
Since ${\rm T}^2$ is not simply connected, even flat bundles (vanishing field tensor) may be 
topologically nontrivial (the Aharonov-Bohm effect). It is the energy of these topological field 
configurations that determines the ground state of the model, and therefore its phase diagram.

The partition function $\mathcal Z_{\rm EFT} = \widetilde{\mathcal Z}_{\rm EFT}$
includes a sum over all topologically distinct field configurations of finite energy.
In order to use mirror symmetry to calculate this function we must therefore investigate the fate of 
flat vector bundles on ${\rm T}^2$ when subjected to the mirror map.

We are fortunate that toroidal mirror symmetry by now is a mathematical fact \cite{PZaslow:1998},
but unfortunately the proof is quite abstract, mapping ``the derived category of coherent sheaves 
on the B-side to the Fukaya category of Lagrangian sub-manifolds on the A-side".  
Our first challenge is to transcribe this into ordinary quantum field theory language.  

In this we are aided by open string theory, from which we know that
mirror symmetry in the field theory limit relates flat vector bundles on a CY manifold to gauge invariant 
objects on certain sub-manifolds of the mirror manifold \cite{Mirror:2003}. 
These sub-manifolds are minimum volume spaces of suitable dimension, 
and when the mirror manifold is a torus they are simply straight lines winding the torus
[cf.\;Fig.\;\ref{fig:Branes}(a)].  
The objects on $\widetilde {\rm T}^2$ that are dual to stable holomorphic bundles on ${\rm T}^2$ 
are gauge-invariant Wilson loops \cite{Mirror:2003}.

The energy density of a $D_1$-brane cannot be smaller than the BPS bound, 
which is determined by the length $\vert Q = q + ip \vert$ of the topological charge vector
\cite{Polchinski:1998},
\begin{eqnarray}
p \propto m \sqrt{\sigma_D},&\quad&
q \propto  (n  + \sigma_H m)/\sqrt{\sigma_D},\nonumber\\
Q_{n,m} &\propto& (n + \sigma m)/\sqrt{\Im\sigma}.\label{eq:dyoncharges}
\end{eqnarray}
In soliton language, the magnetic charge $p$ is given by the winding number $m$.
Similarly, the electric charge $q$ is given by the winding number $n$, but because of the 
Witten effect \cite{Witten:1979} $q$ picks up an additional $m$-dependent contribution from the 
topological piece of the effective action that is parametrized by $\sigma_H$ \cite{footnote:theta-term,
CardyR:1982, *Cardy:1982, *ShapereW:1989}. 

Topological properties of winding modes survive in the low energy (geometric) 
field theory limit, and the total energy $E$ of a Wilson loop of length $L$ should saturate the BPS bound.
In other words, $E \propto L\, \vert Q\vert$, where the constant of proportionality will be determined later.
For the energy of these modes to be finite the loop must be of finite length, and this
happens iff it winds a finite number $n$ times around one cycle of $\widetilde {\rm T}^2$, 
and $m$ times around the other cycle, before returning to the starting point 
(cf.\;Figs.\;\ref{fig:MirrorCartoon}\,-\,\ref{fig:Branes}).
In other words, only a loop $\ell = (\ell^1,\ell^2) = (n,m)$ whose slope is rational [$m/n\in \mathbb Q$, 
cf.\;Fig.\;\ref{fig:Branes}(a)] has finite energy.
The length $L$ of this loop in the mirror metric $\tilde g_{ab}$ [Eq.\;(\ref{eq:mirrormetric})] is given by
\begin{equation}
L_{n,m}^2(\sigma,\bar\sigma) = {\tilde g_{ab}\ell^a \ell^b}  = \frac{ \vert n + \sigma m\vert^2}{{\Im\sigma}}.
\label{eq:length}
\end{equation}
Combining Eqs.\;(\ref{eq:dyoncharges}) and (\ref{eq:length}) we obtain the energy function for 
topologically stable field configurations of finite energy on the mirror torus $\widetilde {\rm T}^2$,
\begin{equation}
E_{n,m} \propto L_{n,m}^2.
\label{eq:BPSmass}
\end{equation}

\subsubsection{Stability}

A vector bundle $\mathcal W$ is called \emph{slope-stable} (or \emph{Mumford-stable}, 
or just \emph{stable} in one dimension) if every sub-bundle $\mathcal V \subset \mathcal W$ has smaller 
slope than $\mathcal W$, $\mu(\mathcal V) < \mu(\mathcal W)$, and it is called \emph{semi-stable} if 
$\mu(\mathcal V) \leq \mu(\mathcal W)$.  A semi-stable bundle $\mathcal V$ for which ${\rm deg}(\mathcal V)$ 
and ${\rm rk}(\mathcal V)$ are coprime is stable. Line bundles [${\rm rk}(\mathcal L) = 1$] are therefore always
stable, as required if they are to model the IQHE (i.e., all integer plateaux are observed, 
and should therefore all correspond to stable bundles).

The mathematical motivation for singling out these bundles is that
slope-stable bundles are indecomposable, in the sense that they cannot be reduced to a direct sum of 
``smaller" bundles, and all holomorphic bundles can be built from these.  
This offers little physical insight, but, since $\mathfrak m$ is an exact 
quantum symmetry, the classification of stable holomorphic bundles on the elliptic curve ${\rm T}^2$ 
should be mirrored in an equivalent classification of dyons on $\widetilde {\rm T}^2$, 
and hopefully one whose physical interpretation is more transparent.
We have already discussed the physical reason (finite energy) why the geometrical slope of an 
$(n,m)$-mode must be rational, as required by the rationality of the topological slope of the dual 
bundle and equivalence under mirror symmetry.  

We must also show that the charges are coprime.  
This follows from the physical fact that a dyon is stable against decay iff its charges are coprime.
In order for two states $\ell_i = (n_i, m_i)$ ($i = 1, 2$) to give a composite state $\ell = (n,m)$, 
it follows from charge conservation that they are related by vector addition on the charge lattice,
so they must satisfy the triangle inequality $\vert\ell\vert \leq \vert\ell_1\vert + \vert\ell_2\vert$. 
From energy conservation it follows that $\ell$ decays into $\ell_1$ and $\ell_2$ iff 
this inequality is saturated, i.e., iff the three states point in the same direction, 
with $\vert n_i\vert \leq \vert n\vert$ and $\vert m_i\vert \leq \vert m\vert$.  
They therefore have the same slope, so  $m_i = n_i m/n\in \mathbb Z$, and since $n$ cannot divide $n_i$,
$n$ must divide $m$.  So a dyon is unstable iff it is not coprime, and it is stable iff it is coprime.

Geometrically, this means that a state is stable iff its charge vector intersects the 
charge-lattice precisely once.  In other words, the lattice points nearest the origin are stable,
and these are the points with coprime coordinates [cf.\;Fig.\;\ref{fig:Branes}(b)].  
All other lattice points are not coprime and not stable, since neither charge nor energy conservation 
forbids them from decaying into shorter (collinear) states that also belong to the lattice.

The energy function $E_{n,m}(\sigma) \propto L_{n,m}^2(\sigma)$ in Eq.\;(\ref{eq:BPSmass})
is invariant under a very large group 
$\Gamma(1) = \textrm{SL}(2,\mathbb Z)$ of discrete symmetries that are called \emph{modular}.  
A modular transformation $\gamma\in\Gamma(1)$ acting on a complex variable $\sigma$ 
is a M\"obius transformation with integer coefficients ($a,b,c,d\in\mathbb Z$) and unit determinant 
($\det\gamma = ad - bc = 1$):
\begin{equation}
\sigma\stackrel{\gamma}{\longrightarrow} \sigma^\prime  = \gamma(\sigma)
=  \frac{a \sigma + b}{c \sigma + d} 
= \Re\sigma^\prime + \frac{\Im\sigma}{\vert c\sigma + d\vert^2} \, i.
\label{eq:modtransform}
\end{equation}

The nice transformation of the ``electron" state $(1,0)$, 
\begin{equation*}
L_{1,0}(\sigma) = \frac{1}{\sqrt{\Im\sigma}} \stackrel{\gamma}{\longrightarrow} 
 \frac{\vert d + \sigma c\vert}{\sqrt{\Im\sigma}} = L_{d,c}(\sigma),
\end{equation*}
signals that the simple form of the energy function is preserved by modular transformations:
\begin{eqnarray*}
L_{n,m}(\sigma)&\stackrel{\gamma}{\longrightarrow}&
L_{n,m}(\sigma^\prime) = L_{n^\prime,m^\prime}(\sigma),\\
(n^\prime, m^\prime)^t &=& \widetilde\gamma\cdot (n, m)^t, \\
\widetilde\gamma = \left(\begin{array}{cc} d&b\\c&a\end{array}\right) 
&=& \left(\begin{array}{cc} \phantom{-}a&-b\\-c&\phantom{-}d\end{array}\right)^{-1},
\end{eqnarray*}
where the modular transformation $\widetilde\gamma$ acts on charge vectors by matrix multiplication.  
We can therefore undo the effect of  $\gamma$ by acting on the charge lattice with $\widetilde\gamma^{\,-1}$, 
leaving the energy unchanged.  Notice that any modular image of the coprime pair $(1,0)$ is also coprime, 
because $(d, c)$ is coprime for any modular transformation.
Stability is therefore preserved by modular transformations.

In order to pick a unique state we must also specify the position $u$ of the Wilson loop, 
together with the value $v$ of the loop integral (monodromy) [cf.\;Fig.\;\ref{fig:Branes}(a)] \cite{Mirror:2003}.
These may be combined into a complex parameter that takes values on a torus 
that is isomorphic to the base space, $u + \sigma v \simeq \widetilde\phi\in\widetilde {\rm T}^2$.
So the classification of topological states on ${\rm T}^2$ and $\widetilde {\rm T}^2$
require the same arithmetic data (a pair of coprime integers and a complex number that 
parametrizes the torus), as required by mirror symmetry \cite{footnote:CYstability}.

In summary, we have learned from string theory the surprising but convenient fact that mathematical 
(topological) stability in our toroidal model is equivalent to physical stability in the mirror model,
thus vindicating our suspicion that the spectrum of mathematically stable holomorphic vector bundles 
on ${\rm T}^2$ determines the physical spectrum of universality classes in this model.
Modular and mirror symmetry guarantees that there is a unique stable fixed point 
compactifying each universality class, which is associated with a unique slope-stable 
holomorphic vector bundle, and the plateau value is given by the topological invariants of that bundle.

\subsubsection{Phase diagram}

The phase structure is determined by a competition between the dyons.
The state of lowest energy depends on the shape of the torus,
and phase transitions occur when the ground state degenerates.

Consider first the family of rectangular tori with complex structure $\sigma = i t\; (0<t\in\mathbb R)$ and
full modular symmetry (no spin structure).  Since the length $L$ of a Wilson loop in this case is given by 
\begin{equation*}
L_{n,m}^2(\sigma = i t) = \frac{n^2}{t} + m^2 t,
\end{equation*}
we see that when $t > 1$ the shortest loop is $(\pm 1, 0)$, when  $t = 1$ the states 
$(\pm 1, 0)$ and $(0,\pm 1)$ are degenerate, and when $t < 1$ the ground state is $(0,\pm 1)$.  
It is known that the only extremal points on the fundamental domain of a modular function 
is a saddle point at $\sigma_{\smallqcp} = i$ and a global extremum at 
$\sigma_{\smallrepulsor} = \exp(2\pi i/3)$ \cite{OPSarnak:1988}. 
Since the partition function is a modular function, we conclude that there is a phase transition
at $t = 1$ (${\qcp} = i$), and at all its modular images.

\begin{figure*}[tbp] 
\hskip -1cm
\includegraphics[scale = 0.33]{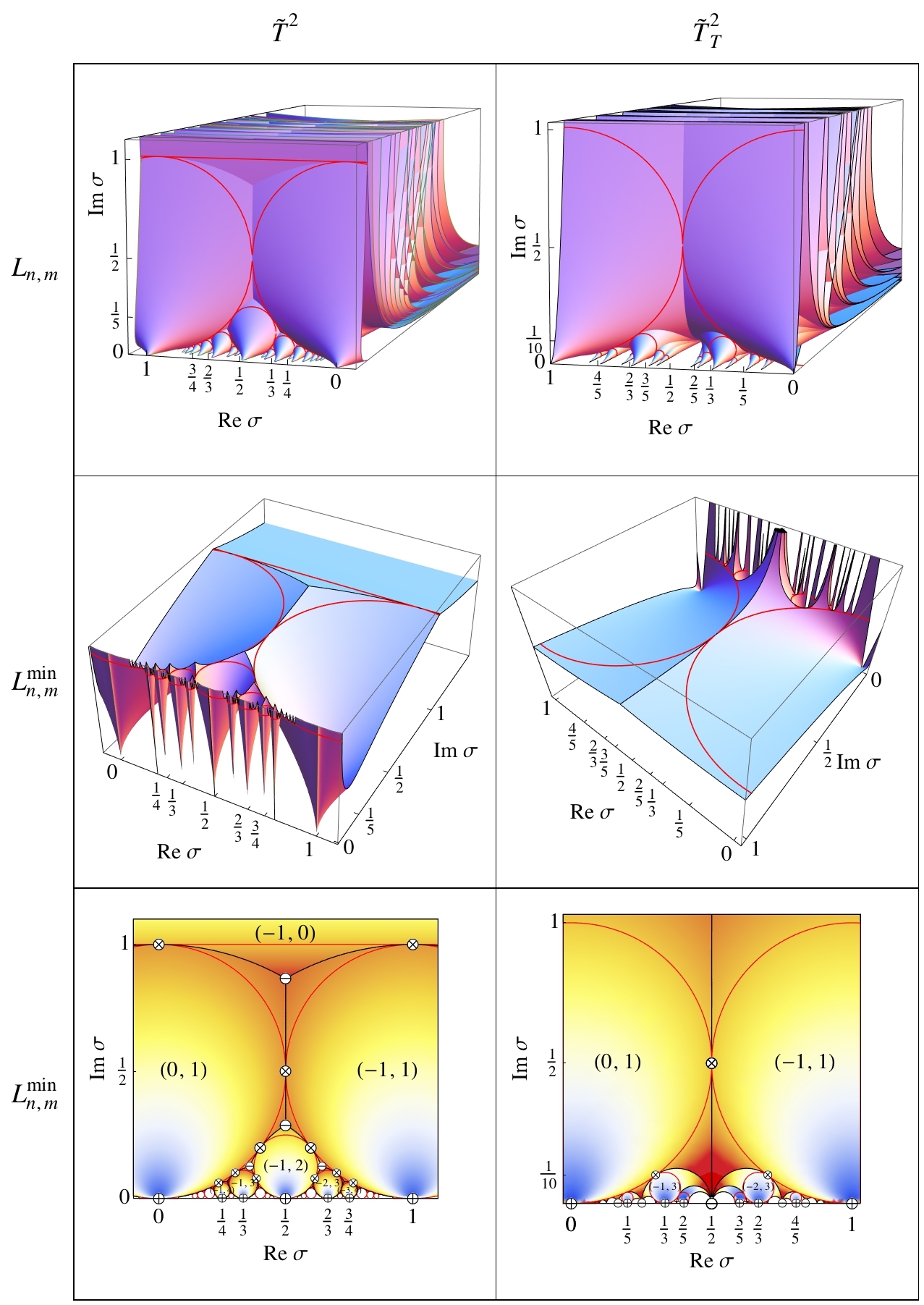}
\caption[modsubmodgrid]{Energy landscapes (top row) and minimum energy landscapes (middle row) 
[$\ln E_{n,m}(\sigma)\propto \ln L_{n,m}(\sigma)$], together with phase diagrams in the moduli space of 
complex structures $\tilde\tau = \sigma$ (bottom row), for the mirror model 
$\widetilde {\rm T}^2$ without spin structure (left), and the mirror model $\widetilde {\rm T}^2_{\rm T}$
with spin structure (right).  Saddle points (${\otimes}$) on the phase boundaries (black) are
candidate quantum critical points.  (The iconography is explained in the main text.)
The red lacework showing level curves of unit height [$L_{n,m} (\sigma) = 1$],
are Apollonian (left) and sub-Apollonian (right) gaskets.}
\label{fig:DyonGrid}
\end{figure*}

Moving away from the imaginary axis, for every stable dyon there is a region of moduli space (a phase)
where that dyon has less energy than all others, cf.\;left-hand column in Fig.\;\ref{fig:DyonGrid}.
The top left panel shows a bottom view of the energy landscape of the mirror model 
$\widetilde {\rm T}^2$.  Each flake of this ``dyonic millefeuille" gives the energy
of a specific dyon as a function of the complex structure $\tilde\tau = \sigma$. 
This landscape, derived from the unconstrained integer 
charge lattice $\Lambda$, has maximal modular symmetry $\Gamma(1)$.

The middle row shows the minimum energy landscape as a function of the complex structure 
$\tilde\tau = \sigma$ on $\widetilde {\rm T}^2$ (left) and $\widetilde {\rm T}^2_{\rm T}$ (right).
Quantum phase transitions take place along the black ridges where neighboring states are 
degenerate in energy.

The bottom left panel shows the phase diagram in the moduli space of 
complex structures $\tilde\tau = \sigma$ on $\widetilde {\rm T}^2$.
Color-temperature is proportional to energy.  In each phase, labelled by its dyon charge $(n, m)$, 
the energy has a unique global minimum (${\attractor}$) at the rational point $\sigma_{\smallattractor} = - n/m$ 
($n,m\in\mathbb Z$) on the boundary of moduli space, where $E_{n,m}({\attractor}) = 0$. 
The quantum critical points (${\qcp}$) are points of minimum energy  
on the self-similar tree of phase boundaries [$L_{n,m} ({\qcp})=1$].
Bifurcation points (${\repulsor}$), which appear when three phases are degenerate, 
have maximum energy [$L_{n,m}^2({\repulsor}) = 2/\sqrt{3} \,$].

For the subgroup $\Gamma_{\rm T}\in\Gamma(1)$ of relevance to the spin polarized QHE, 
only dyons with odd magnetic charge have finite energy, and this has a dramatic
effect on the phase diagram, cf.\;right-hand column in Fig.\;\ref{fig:DyonGrid}.
The top right panel shows a bottom view of the dyonic energy landscape of the 
mirror model $\widetilde {\rm T}^2_{\rm T}$, in which the charge vector is restricted to a sub-lattice 
$\Lambda_{\rm T} \subset \Lambda$ that includes only odd magnetic charges.
Compared to the unrestricted case (left), this removes half the phases and all bifurcations (triskelions),
and the modular symmetry is reduced to a subgroup $\Gamma_{\rm T}\subset \Gamma(1)$.
Each phase has a unique global minimum (${\attractor}$) at the rational point $\sigma_{\smallattractor} = - n/m$ 
($n\in\mathbb Z,\, m\in2\mathbb Z + 1$) on the boundary of moduli space, where $E_{n,m}({\attractor}) = 0$.
These plateaux values are protected by topology because $(m,n)$ are winding numbers that cannot be 
continuously deformed. They can only jump by integer amounts, by provoking a quantum phase
transition to a different ground state. 
Since these charges are mirror reflections of $(r, c_1)$ in the original sigma model, 
we also have $\sigma_{\smallattractor} = \sigma_H^{\smallattractor} = c_1/r\in\mathbb Q$,
which reduces to the IQHE ($\sigma_H^{\smallattractor} = c_1\in\mathbb Z$) for line bundles ($r = 1$).

The bottom right panel shows the phase diagram in the moduli space of complex structures 
$\tilde\tau = \sigma$ on $\widetilde {\rm T}^2_{\rm T}$. In this restricted case all maxima ${\repulsor}$ 
have moved to even denominator rationals and acquired infinite energy 
[$E_{n,m}({\repulsor}_{\rm T})\rightarrow \infty$]. This fractal phase diagram, including the location of 
critical points, is in accurate agreement with the spin polarized QHE \cite{OLLutken:2018}.

So far we have only used the geometry of $\widetilde {\rm T}^2$ and $\widetilde {\rm T}^2_{\rm T}$, 
which is built into the dyonic energy function.  We have seen that this is sufficient to extract the
phase and fixed point structure, and we have verified that the model is manifestly invariant under 
the modular symmetry that respects the chosen spin structure on the target space.
In order to study scaling (RG flows) and calculate critical exponents we now turn to the 
partition function for the model $\widetilde {\rm T}^2_{\rm T}$, which has been ``enhanced" with a 
spin structure that is appropriate for the spin polarized QHE.

\subsection{Partition function}\label{subsec:partitions}

We want to calculate the partition function of the mirror model, and in order to simplify notation we
temporarily set $x = \sigma_H$, $y = \sigma_D$ and $Z_\# = \widetilde{\mathcal Z}_{\rm EFT}^\#$, 
where $\# = {\rm T, R, S}$ labels the choice of spin structure (PA, AP, AA) on the target torus 
$\widetilde {\rm T}^2_\#$.
We focus here on $Z_{\rm T}$; similar discussions for the R and S spin structures may be found in the Appendix. 

Physical and mathematical considerations, including the requirements that the bosonic part $Z_b$ 
of the partition function should be real, nonsingular, and invariant under both modular and mirror transformations, gives a function that is independent of spin structure, and therefore the same for all spin structures:
\begin{equation*}
Z_b \propto \frac{1}{\sqrt{\Im\sigma} \left\vert\eta(\sigma)\right\vert^2}\times
 \frac{1}{\sqrt{\Im\tau} \left\vert\eta(\tau)\right\vert^2}.
\end{equation*}
This function is invariant under all modular transformations, because the \emph{modular weights} 
\cite{footnote:weights} $(w,\overline w) = (1/2,1/2)$ of the squared modulus of the Dedekind function 
\begin{equation}
\eta(z) = e^{\pi i z/12}\prod_{n=1}^\infty\left(1 - e^{2\pi i n z}\right),\quad(z = \sigma\;{\rm or}\;\tau)
\label{eq:Dedekind}
\end{equation}
are cancelled by the weights $(w,\overline w) = (-1/2,-1/2)$ of $\sqrt{\Im z}$  
[cf.\;Eq.\;(\ref{eq:modtransform})]. Since $\eta(i) = \Gamma(1/4)/(2\pi^{3/4}) = 0.768225\dots$ is finite, 
in our case ($\tau = i$) this simplifies to
\begin{equation}
Z_b (\sigma, \bar\sigma)\propto \frac{1}{\sqrt{y}\left\vert\eta(\sigma)\right\vert^2}.
\label{eq:Zboson}
\end{equation}
$Z_b$ is invariant under modular transformations of the target space complex structure, 
but it is neither holomorphic nor holomorphically factorized.

The dyonic contribution $Z_d$ to the $\Gamma_{\rm T}$-invariant partition function
$Z_{\rm T} = Z_b Z_d$ is obtained by tracing the Boltzmann weight over the sub-lattice 
\begin{equation*}
\Lambda_{\rm T} =  \mathbb Z {\oplus} (2\mathbb Z+1)\sigma \subset \Lambda 
= \mathbb Z {\oplus} \mathbb Z \sigma.
\end{equation*}
The spectrum found in Sec.\;\ref{subsubsec:mirroroffibre} [Eq.\;(\ref{eq:BPSmass})] gives
\begin{eqnarray*}
Z_d(\sigma,\bar\sigma) &=& {\rm Tr}_{\Lambda_{\rm T}} {z}_{n, m}(\sigma, \bar\sigma),\\
z_{n, m}  &=&  e^{-\beta E_{n,m}} = e^{-\tilde\beta L_{n,m}^2}
= e^{-(\pi\alpha^2/2)\vert n + \sigma m\vert^2/y},
\end{eqnarray*}
where $\tilde\beta = \pi\alpha^2/2$ is an unknown factor parametrized by $\alpha$ for future 
convenience \cite{footnote:alphanorm}.
The remaining trace is over field configurations of finite energy that respect the spin structure 
$\# = {\rm T}$ ($n\in \mathbb Z, m\in 2\mathbb Z +1$).

In order to check the symmetry of the partition function, consider first the action of the generators 
of the modular group on the effective action:
\begin{eqnarray*}
\vert n + \sigma m\vert^2/y&\stackrel{T}{\longrightarrow}& \vert n + m + \sigma m\vert^2/y, \\
\vert n + \sigma m\vert^2/y &\stackrel{S}{\longrightarrow}& \vert m - \sigma n\vert^2/y  ,
\end{eqnarray*}
where we have used that $S(y) = y / \vert \sigma\vert^2$.  
Notice that the duality transformation $S$ switches the role of electric and magnetic charges.
This gives
\begin{eqnarray*}
 {\rm Tr}_{\Lambda_{\rm T}} z_{n, m}(T\sigma) &=& {\rm Tr}_{\Lambda_{\rm T}}z_{n + m,m}(\sigma)
=  {\rm Tr}_{\Lambda_{\rm T}} z_{n, m}(\sigma),\\
 {\rm Tr}_{\Lambda_{\rm T}} z_{n, m}(S\sigma) &=&  {\rm Tr}_{\Lambda_{\rm T}} {z}_{m, n}(\sigma) 
=  {\rm Tr}_{\Lambda_{\rm R}}{z}_{n, m}(\sigma),
\end{eqnarray*}
where $R = STS$ and $\Lambda_{\rm R} =  (2\mathbb Z+1) {\oplus} \mathbb Z \,\sigma$.
The $\Lambda_{\rm T}$-trace is therefore unchanged by $T$ but not by $S$,
as expected since  $T$ is in $\Gamma_{\rm T}$ and $S$ is not.  
Suppressing traces, the duality generator $D = R^2 = ST^2S$ in $\Gamma_{\rm T}$ gives: 
\begin{equation}
D: \;\; {z}_{n, m} \stackrel{S}{\longrightarrow} {z}_{m, n} \stackrel{T^2}{\longrightarrow} {z}_{m, n} 
\stackrel{S}{\longrightarrow} {z}_{n, m} ,
\label{eq:duality}
\end{equation}
which shows that the effective action is invariant under the duality transformation $D$ on the lattice 
$\Lambda_{\rm T}$. Since both $Z_b$ and $Z_d$ are $\Gamma_{\rm T}$-invariant, 
so is $Z_{\rm T} = Z_b Z_d$. 

After a Poisson resummation 
\begin{equation*}
\sum_{n\in\mathbb Z} e^{ 2 i A n - B n^2} = \sqrt{\frac{\pi}{B}}\, \sum_{n\in\mathbb Z} e^{-(A + \pi n)^2/B},
\end{equation*}
$Z_d$ takes a more familiar form,
\begin{eqnarray*}
Z_d &=&  \frac{\sqrt{2y}}{\alpha} \,  {\rm Tr}_{\Lambda_{\rm T}}\,
e^{2\pi i \left\{xmn + y \left[\left(n/\alpha\right)^2 + \left(\alpha m/4\right)^2\right] i \right\}}\\
&=&  \frac{\sqrt{2y}}{\alpha}  \,  {\rm Tr}_{\Lambda_{\rm T}} \,q^{h_+^2(\alpha)}\bar q^{\,h_-^2(\alpha)},
\end{eqnarray*} 
where  $q = \exp(2\pi i\sigma)$, and $h_\pm (\alpha) = (n/\alpha \pm \alpha m/2)/\sqrt{2}$.  
Notice that the nonholomorphic pre-factor $\sqrt{y}$ in $Z_d$ is cancelled by the denominator of 
$Z_b$ [Eq.\;(\ref{eq:Zboson})]:
\begin{equation*}
Z_{\rm T} = Z_b Z_d \propto \zeta_{\rm T} = \frac{1}{2\left\vert\eta\right\vert^2} \,
{\rm Tr}_{\Lambda_{\rm T}} \,q^{h_+^2(\alpha)}\bar q^{\,h_-^2(\alpha)}.
\end{equation*}

At critical points conformal symmetry imposes severe restrictions on a quantum field theory.
This will be discussed in the next section, but in order to avoid suspending our current calculation
we anticipate one result: the partition function should be
holomorphically factorized in $\sigma$  [cf.\;Eq.\;(\ref{eq:triality})].
With $\alpha^2 =1$ we obtain after some work the factorized form
\begin{equation}
 \zeta_{\rm T} = \frac{1}{4}\bigg\vert\frac{\theta_2}{\eta}\bigg\vert^2 = 
\bigg\vert\frac{\eta(2\sigma)}{\eta(\sigma)}\bigg\vert^4
= \vert \eta_+ \vert^4,
\label{eq:ZT}
\end{equation}
where 
\begin{equation*}
\theta_2(q) = \sum_{n\in\mathbb Z} q^{(n + 1/2)^2/2} = 2q^{1/8} \prod_{n=1}^\infty (1-q^n)(1+q^n)^2
\end{equation*}
is a Jacobi theta-function, and we have defined
\begin{equation}
\eta_{\pm}(q) = q^{1/24}\prod_{n=1}^\infty (1 \pm q^n).
\label{eq:etapm}
\end{equation}
The function $\eta_-$ is the usual Dedekind $\eta$-function defined in 
Eq.\;(\ref{eq:Dedekind}), but $\eta_+(\sigma) = \eta(2\sigma)/\eta(\sigma)$ is less 
conventional \cite{footnote:eta_remarks}.
We therefore choose $\alpha = \pm 1$, which gives the ``free energy"
\begin{equation*}
f_{\rm T} = - \ln\zeta_{\rm T} = -4\ln\vert\eta_+\vert
 = - 2\left( \varphi_{_{\rm T}}+ \overline \varphi_{_{\rm T}}\right)
\end{equation*}
in terms of the $\Gamma_{\rm T}$-invariant holomorphic potential
\begin{equation}
\varphi_{_{\rm T}}(\sigma) = \ln\eta_+(\sigma)= \ln\eta(2\sigma) - \ln\eta(\sigma) .
\label{eq:Tpot}
\end{equation}

We shall see in the next section that the RG potential ($C$-function) for this model also derives
from the \mbox{``thermodynamic} potential" $\varphi_{_{\rm T}} =  \ln\eta_+$. The reason for this
surprising and useful simplification is that modular vector fields in these 
models derive directly from modular scalar fields, so 0-forms and 2-forms are paired,
and therefore equally unique.

This means that the deceptively simple looking function $\eta_+$ encodes most universal properties 
of the model, including the location of critical points and the flow geometry between these. 
Provided that it is properly (physically) normalized, it also gives the flow rates near critical points, 
i.e., critical (delocalization) exponents.  

In summary, the high degree of modular symmetry in this model means that the potential $\varphi_{_{\rm T}}$ 
is extremely rigid (essentially unique).  It is therefore perhaps surprising, and encouraging,  
that thirty years of experiments not only have failed to falsify the model, 
the agreement appears to be improving with time as the technology
needed to reach the scaling domain evolves (cf.\;Sec.\;\ref{sec:data}).

\section{RG flow}\label{sec:RGflow}

In order to investigate scaling properties of these models we exploit some geometric properties
of RG flows. The physics of this flow is determined by a vector field 
$\beta^a  = \dot\lambda^a = d\lambda^a/dt$ tangent to the parameter space ${\mathcal M}(\lambda)$ 
spanned by the relevant effective couplings $\lambda^a$ of the quantum field theory under consideration.

\subsection{RG potential}\label{subsec:RGpot}

Explicit perturbative calculations of scalar $\varphi^4$ theories in $4-\varepsilon$ 
dimensions up to three loop order showed that the RG flow is a gradient flow 
\begin{equation}
\beta^a(\lambda) = g^{ab}_{_Z}  \partial_b C 
= g^{ab}_{_Z} (\lambda) \,\frac{\partial C(\lambda)}{\partial\lambda^b},
\label{eq:betafn}
\end{equation}
where $C$ is a scalar potential (the \emph{RG potential}) and $g_{_Z}$ is an invertible metric on 
${\mathcal M}$ \cite{WallaceZ:1974, *WallaceZ:1975}. Since this eliminates pathological limit cycles,
it was believed to be a general property of any effective field theory in its domain of validity.
Further evidence was obtained from the study of non-linear sigma models in two dimensions 
\cite{Friedan:1980, *Friedan:1985}, and finally established as the so-called ``$C$-theorem" in \cite{Zamo:1986}.

Here we consider only two-parameter flows \mbox{(${\rm dim}\,\mathcal M = 2$)}, 
in which case we can work with complexified coordinates $\lambda = \lambda^1 + i \lambda^2$ and 
$\bar\lambda = \lambda^1 - i \lambda^2$ without loss of generality. This gives the complexified beta function
$\beta^\lambda = d\lambda/dt = \dot\lambda^1 + i \dot\lambda^2 = \beta^1 + i \beta^2$,
and its complex conjugate $\beta^{\bar\lambda} = d\bar\lambda/dt = \beta^1 - i \beta^2$.
Our moduli space $\mathcal M$ is the space of conductivities ($\lambda = \sigma$)
[or, equivalently, the space of resistivities ($\lambda = \rho$)].

Renormalization must respect the symmetries of a model, and in our case this includes the infinite discrete 
group of modular transformations. Consequently, the physical beta function, which is a contravariant vector field 
$\beta^\sigma = \dot\sigma$, must transform like a modular form of weight $w = -2$.  Since no modular form 
of this weight exists, $\beta^\sigma$ it is not a modular form: it is a nonholomorphic vector field that 
transforms like a modular form of negative weight.

A covariant vector field $\beta_\sigma$, on the other hand, transforms like a modular form of weight $w = 2$, 
and such forms usually do exist [but not for the full modular group $\Gamma(1)$]. 
They first appear for congruence subgroups at level two. 
For each of the maximal symmetries $\Gamma_{\rm X}$ (X = T, R, S) considered here, 
the $w = 2$ form $\mathcal E_2^{\rm X}$ is a unique Eisenstein function, up to an overall normalization. 

Eq.\;(\ref{eq:betafn}) is transposed to complex coordinates by using
$\partial = \partial_\sigma = \partial/\partial\sigma$ and 
$\bar\partial = \partial_{\bar\sigma}= \partial/\partial{\bar\sigma}$. 
The complexified beta function $\beta^\sigma$ has the properties required by the $C$-theorem
if the modular scalar $C = \Phi + \bar\Phi$ is given by a holomorphic potential 
$\Phi(\sigma)$:
\begin{equation*}
\beta^\sigma = \frac{d\sigma}{dt} = g^{\sigma\bar\sigma}_{_Z}\, \bar\partial C
= g^{\sigma\bar\sigma}_{_Z}\, \bar\partial\bar\Phi.
\end{equation*}
Since $\bar\partial C = \bar\partial\bar\Phi$ is a modular form of weight $(w,\overline w) = (0,2)$,
in the most symmetric cases ($\Gamma = \Gamma_{\rm X}$, for which there
is only one  2-form) we must have 
\begin{equation*}
\beta^\sigma_{\rm X} = g^{\sigma\bar\sigma}_{_Z}\, \bar\partial\bar\Phi_{\rm X}
\propto g^{\sigma\bar\sigma}_{_Z}\, \bar{\mathcal E}_2^{\rm X}.
\end{equation*}
Fortunately, potentials $\Phi_{\rm X}$ whose derivatives are the Eisenstein functions 
$\mathcal E_2^{\rm X}$ do exist.

In order to check the transformation properties of this form, notice that an inverse
modular metric transforms like a mixed tensor of weight $(-2, -2)$. The antiholomorphic weight
is cancelled by the weight of $\mathcal{\bar E}_2$, and the physical beta function 
$\beta^\sigma$ is therefore a nonholomorphic vector field of weight $(-2,0)$, as required. 

It is remarkable that the requirement that RG transformations commute with modular 
transformations is so strong that (for the cases we consider) the beta function is all but unique,
and the potential $\Phi_{\rm X}$ must be equally unique. For example, 
if the modular symmetry is $\Gamma_{\rm T}$, then we have no choice but to take 
$\Phi_{\rm T} \propto \varphi_{_{\rm T}} = \ln\eta_+$ [cf.\;Eq.\;(\ref{eq:Tpot})], and it follows that the 
RG potential $C_{\rm T}$ must be proportional to the free energy found in the previous section, 
$C_{\rm T}\propto f_{\rm T}$.

It is not surprising that the partition function and RG potential
are intimately related near a critical point, since they both in some sense count degrees of freedom
at these ``geostationary" points in parameter space.  There is, however, usually no universality 
associated with the RG flow beyond leading order in an expansion around a critical point, 
as the higher order expansion coefficients normally are scheme dependent. 
A modular symmetry is so rigid that it removes this arbitrariness, forcing the RG flow to take a unique 
geometric form, up to an overall (physical, scheme independent) normalization that parametrizes
our residual ignorance about these toroidal models.

We would like to fix the value of this parameter, since it determines the values of the critical exponents.
We will argue that the remaining ambiguity is integer valued, which, if true, is enough to provide a very 
strong test of the model. Furthermore, if the supersymmetry of disordered materials \cite{Efetov:1997}
forces this integer to be 4, then the critical exponent has a precise value that is in remarkable agreement with
numerical simulations (at the per mille level, cf.\;Fig.\;\ref{fig:exponents}).

\subsection{Conformal field theory}\label{subsec:CFT}

At quantum critical points ($\qcp$) the finite space-time (Poincar\'e) symmetry expands to an 
infinite conformal symmetry, and this imposes severe restrictions on the quantum field theory.
These constraints may be extracted by wrapping the ``world-sheet" $\Sigma^2$ on a torus 
${\mathcal T}^2(\rho)$ with complex structure $\rho$ (not to be confused with resistivity), 
\begin{equation*}
\Sigma^2 \stackrel{\smallqcp}{\longrightarrow}{\mathcal T}^2(\rho)
\stackrel{\varphi}{\longrightarrow} {\rm T}^2(\tau,\sigma)
\stackrel{\mm}{\longrightarrow}\widetilde {\rm T}^2(\sigma,\tau),
\end{equation*}
in which case the critical partition function ${\mathcal Z}_\smallqcp(\rho,\tau,\sigma)$ 
has three moduli (three complex parameters) that take values in three copies of the upper half plane
[$\mathbb C^+(\rho)\times\mathbb C^+(\tau)\times\mathbb C^+(\sigma)$].

The toroidal partition function ${\mathcal Z}(\tau,\sigma)$ is unchanged when 
complex and K\"ahler structures are exchanged, because the mirror transformation 
is a quantum equivalence. When the model is critical this \emph{duality} invariance 
(a $\mathbb Z_2$ symmetry) is enhanced to a $\mathbb Z_3$ symmetry (\emph{triality})
that allows us to permute the moduli \cite{RobbertVV:1988}:
\begin{equation}
{\mathcal Z}_\smallqcp(\rho,\tau,\sigma) = {\mathcal Z}_\smallqcp(\rho,\sigma,\tau) 
= {\mathcal Z}_\smallqcp(\sigma,\rho,\tau).
\label{eq:triality}
\end{equation}

Since a conformal partition function is holomorphically factorized in $p = \exp(2\pi i\rho)$ 
\cite{DiFMS:1997}, by triality it is also factorized in $q = \exp(2\pi i\sigma)$,
which we used above (in Sec.\;\ref{subsec:partitions}) to evaluate the partition function.

In order to obtain the leading order expansion coefficients $y^\pm = 1/\nu^\pm$ at a critical point $\qcp$, 
we need to approximate the RG potential $C$ near $\qcp$. We simplify notation by omitting
references to the symmetry $\Gamma_{\rm T}$ considered here ($Z = Z_{\rm T}$, 
$\varphi = \varphi_{_{\rm T}}$, etc.).
   
Since $Z\propto\zeta$ and $C\propto \ln Z$ we have two undetermined constants $A$ and $B$, 
with $C = B \ln(A\zeta)$. Near a saddle point ($\sigma\rightarrow\sigma_\smallqcp$) we can 
trade $B$ for the central charge $c_\smallqcp$,
\begin{equation}
C\stackrel{\smallqcp}{\longrightarrow} 
c_\smallqcp \,\frac{\ln Z(q)} {\ln Z_\smallqcp(p)}
= c_\smallqcp \,\frac{\ln\zeta(q) + \ln A}{\ln\zeta_\smallqcp(p) + \ln A} 
 \stackrel{\smallqcp}{\longrightarrow} c_\smallqcp,
\label{eq:ccharge}
\end{equation}
valid for $q\approx q_\smallqcp = \exp(2\pi i \sigma_\smallqcp)$ and any $p = \exp(2\pi i\rho)$.
Because of triality at the critical point we expect 
$Z_\smallqcp(p) = Z(p,q_\smallqcp) = A \zeta_\smallqcp(p)$
to have the same functional form as $Z(q) = A \zeta(q)$.

The asymptotic form ($p\rightarrow 0$) of the conformal partition function is
\begin{eqnarray*}
\ln Z(q) &\stackrel{q\rightarrow\raisebox{0.4ex}{\smallqcp}} {\longrightarrow} & \ln Z_\smallqcp(p) 
\stackrel{p\rightarrow 0}{\longrightarrow}  \ln A + \alpha c_\smallqcp \ln\vert p\vert,
\end{eqnarray*}
for some constant $\alpha$.  
This double limit gives an approximation of the central charge,
 \begin{equation*}
c_\smallqcp \approx \frac{\ln Z(q) - \ln A}{\alpha\ln\vert p\vert} 
\approx c_\smallqcp \frac{\ln\zeta(q)}{\ln\zeta_\smallqcp(p)},
\end{equation*}
that may be compared with Eq.\;(\ref{eq:ccharge}), and we conclude that the appropriate 
normalization is $Z = \zeta$ ($A = 1$).

Returning to the original geometry, these considerations suggest that the $C$-function 
near a quantum critical point is approximated by
\begin{equation*}
C \stackrel{{\smallqcp}}{\longrightarrow} 
c_{\smallqcp} \;\frac{\ln\zeta(\sigma)} {\ln\zeta(\qcp)}
=  c_{\smallqcp}\;\frac{\Re\varphi(\sigma)}{\Re\varphi({\qcp})},
\end{equation*}
where $\varphi(\sigma) = \varphi_{_{\rm T}}(\sigma) = \ln\eta_+(\sigma)$.
Our ``only" remaining challenge is to determine the value of $c_\smallqcp$.

The elliptic sigma model contains two free bosons, which contribute  $c_b = 2$ to the
total central charge $c_\smallqcp = c_b + c_f = c_b + c_d$, where $c_f$ ($c_d$) is the 
contribution from fermions (dyons) in the ${\rm T}^2$ ($\widetilde{\rm T}^2$) model.
Both numerical and real experiments (cf.\;Sec.\;\ref{subsec:flowrates}) are consistent
with $c_\smallqcp = 4$, in which case $c_f = c_d = 2$. 
Each fermion contributes $c = 1/2$, so there are two fermions
for every boson.  This is the signature of supersymmetry, and there is indeed a global
supersymmetry in disordered models of localization in the QHE \cite{Efetov:1997}.

If this observation could be formalized, then the most conspicuous (supersymmetric) elliptic model
has  $c_\smallqcp = 4$, and we would obtain a sharp prediction for the values $\nu^+ = -\nu^-$ 
of the critical exponents. This would-be prediction is in excellent agreement with numerical 
experiments, and perhaps also real experiments, as explained below (in Sec.\;\ref{subsec:flowrates}).

\subsection{Beta function}\label{subsec:beta}

In order to extract critical exponents from the beta function we need to expand the 
potential $\varphi =  \varphi_{_{\rm T}}$ at $\qcp = \sigma_{\smallqcp} = (1 + i)/2$ 
as a power series in $\delta\sigma = \sigma - \sigma_{\smallqcp}$:
\begin{equation}
\varphi(\sigma) = \sum_{n=0}^\infty \frac{a_n}{n!} \,\delta\sigma^n =
a_0 +  a_1 \delta\sigma + \frac{a_2}{2}\delta\sigma^2 + \dots
\label{eq:PotExpansion}
\end{equation}
The coefficients $a_n = a_n(\qcp)$ ($n\geq 0$) are evaluated in the Appendix
by exploiting transformation properties of modular and quasimodular forms. 
This verifies by explicit computation that $a_1(\qcp) = 0$  [cf.\;Eq.\;(\ref{eq:a1})], 
so that the beta function indeed has a simple zero at every quantum critical point.
We also find that $a_2(\qcp) = \pi^2 G^4/3 = \pi_\infty^4/(3\pi^2)$ [cf.\;Eq.\;(\ref{eq:a2})], 
where $G$ is Gauss' constant and $\pi_\infty = \pi G = 2.6220\dots$ is called the
 \emph{lemniscate constant} (a geometrical interpretation of $\pi_\infty$ may be found in the Appendix).
In conjunction with the $C$-theorem this allows us to evaluate the critical exponents
\cite{LR:2007, LR:2014}.

The remaining ingredient in Zamolodchikov's $C$-theorem is the parameter space metric 
$g_{_Z}$ \cite{Zamo:1986}.  
Since the EFT should be well-defined near quantum critical points, 
this metric is nonsingular and invertible near ${\qcp}$, and because the torus is a one-dimensional 
CY manifold we can use a result obtained in string theory.   At least to leading order in sigma model 
perturbation theory, which is all we need,  Zamolodchikov's metric and the Weil-Petersen metric 
coincide for CY spaces \cite{CandelasHS:1990}.  In our case we should therefore use 
$g_{_Z} = g_{_H}$, where $g_{_H} = 1/y^2$ is the hyperbolic metric on the upper half plane that 
covers $\mathcal M(\sigma)$.

Collecting everything, the $C$-theorem gives
\begin{equation*}
\beta^\sigma = g_{_H}^{\sigma\bar\sigma} \beta_{\bar\sigma} 
= - \frac{\sigma_D^2}{12} \;\bar\partial C \stackrel{{\smallqcp}}{\longrightarrow} 
-\frac{c_\smallqcp}{24}\;\frac{(\Im\sigma_\smallqcp)^2}{\Re\varphi(\qcp)}\,\bar\partial \bar\varphi .
\end{equation*}
Expanding $\varphi$ near the critical point $\sigma_{\smallqcp}= (1 + i)/2$,
as explained above, we find the beta function to leading nonvanishing order,
\begin{equation*}
\beta^\sigma  = \frac{\delta\bar\sigma}{\nu} +\dots,\quad \nu = \frac{4}{c_{\smallqcp}}\nu_{\rm tor},
\end{equation*}
where the \emph{toroidal exponent} is defined to be
\begin{equation}
\nu_{\rm tor}  =  18 \ln 2 /(\pi^2 G^4) =  18\pi^2\ln 2/\pi_\infty^4  = 2.6051\dots
\label{eq:torexpo}
\end{equation}

We argued above that we expect $c_\smallqcp = 4$ to give the simplest and most natural 
elliptic sigma model, in which case $\nu = \nu_{\rm tor}$. However, in the absence of any compelling 
argument why this is the preferred EFT of the QHE, for the purpose of comparing with experiments 
we prefer for now to retain only the weaker idea that the central charge should be a natural number 
($c_\smallqcp = n\in\mathbb N$), and ask first if the value $\nu_{\rm num}$ of the delocalization 
exponent, obtained from numerical simulations using the CC model \cite{ChalkerC:1988}, 
is close to any of the exponents $\nu_n = 4\nu_{\rm tor}/n$.

We shall find in the next section that $\nu_{\rm num}$ is \emph{very} close to $\nu_4 = \nu_{\rm tor}$
($\nu_{\rm tor}/\nu_{\rm num} = 0.999\dots$), which we interpret as strong evidence that the $n=4$ 
elliptic sigma model is in the same universality class as the CC model. Comparison with experimental 
values $\nu_{\rm exp}$ is more ambiguous (cf.\;Sec.\;\ref{subsec:flowrates}).

In summary, the toroidal model gives a very strong constraint on the 
experimental value of the delocalization exponent: we should find that 
$\nu_{\rm exp} \approx 4 \nu_{\rm tor}/n$ for some positive integer $n$.  
Since $\nu_{\rm tor}$ is very close the numerical value $\nu_{\rm num}$ calculated
in the CC model \cite{ChalkerC:1988}, which is supposed to be in the quantum Hall universality class, 
the pressing experimental question is if
\begin{equation*}
\nu_{\rm exp} \stackrel{?}{\approx} \nu_{\rm num} 
\stackrel{\surd}{\approx} \nu_{\rm tor}  \approx \pi_\infty\approx 21/8 . 
\end{equation*} 
The best available data are reviewed in the next section.

\begin{figure}[tbp] 
\includegraphics[scale = .38]{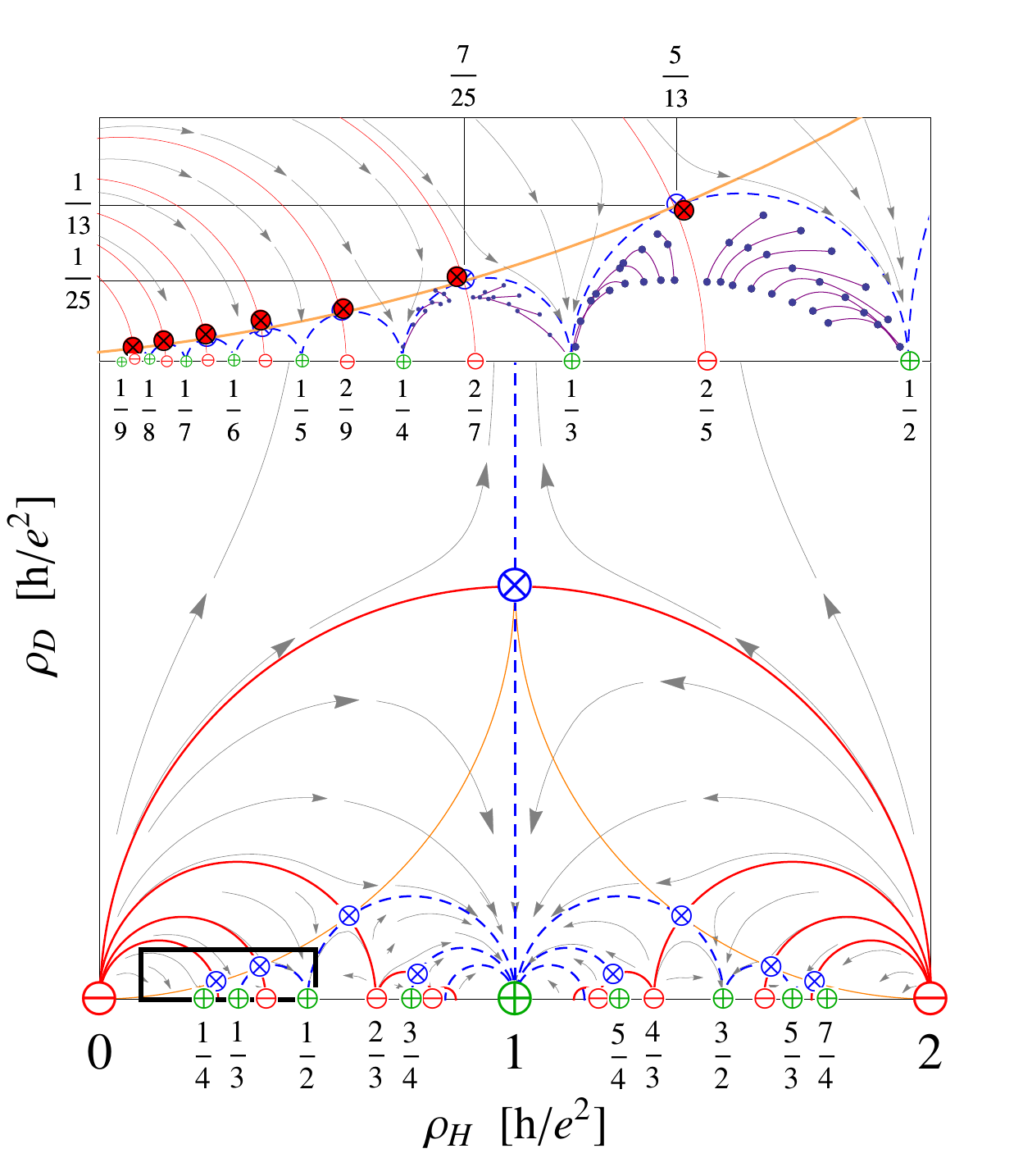} 
\caption{ The top inset is a magnification of the small black rectangle in the main 
modular diagram, which has $\Gamma_{\rm T}$-symmetry, shown here in order to facilitate comparison 
with other spin polarized experiments that probe different parts of the phase diagram, 
and other symmetries (cf.\;Fig.\;\ref{fig:TRSgrid}).
The inset contains a comparison of the experimental location of seven critical points (solid red icons),
reconstructed from Ref.\;\cite{LVXPTsuiPW:2009}, with the modular prediction (blue ${\qcp}$). 
(The iconography is explained in the main text.) 
The size of the red plot-marker is a rough estimate of the experimental uncertainty 
in the data (no error-analysis is provided in \cite{LVXPTsuiPW:2009}).  Purple curves are 
temperature driven flow lines, obtained by quadratic interpolation of points (black bullets) sampled 
at different temperatures for a dozen arbitrary fixed values of the magnetic field.
\label{fig:7Sisters}}
\end{figure}

\begin{figure*}[tbp] 
\vskip 0cm
\begin{center}
\includegraphics[scale = .35]{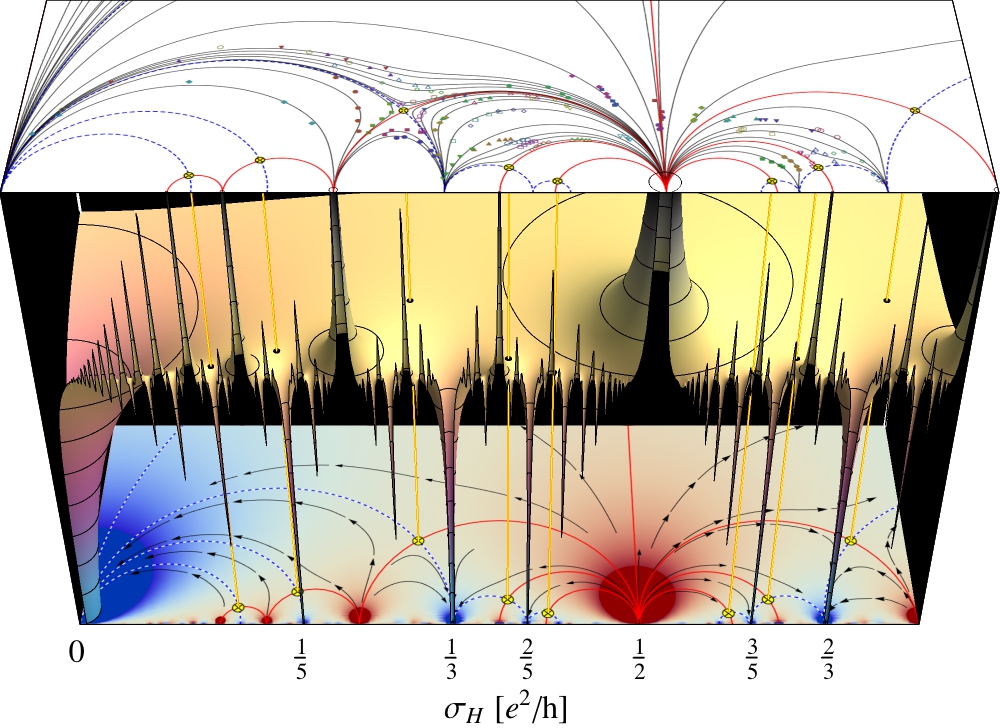}
\end{center}
\vskip -2mm
\caption{Comparison of the toroidal model ${\rm T}^2_{\rm T}$ with experimental conductivity 
data reconstructed from Ref.\;\cite{MurzinDMJ:2005}  (colored plotmarkers on top face).  
RG flow lines (thin black curves) and phase boundaries (red semi-circles) on the top and bottom 
of the box are generated by the gradient of the potential surface $\vert\varphi(\sigma)\vert$ inside the box.
Dashed blue semi-circles are separatrices for the flow.
Vertical orange lines are visual guides connecting critical points in the three diagrams  
(blue and yellow icons $\otimes$ on the top and bottom face; inflection points on the surface). 
Most of the data are probing the quantum phase transition between the fractional plateau 
$\sigma_{\smallattractor} = 1/3$ and the insulator phase ($\sigma_{\smallinsulator} = 0$) 
(cf.\;Fig.\;\ref{fig:TRSgrid}), revealing a quantum critical point very close to the modular 
prediction $\sigma_{\smallqcp} = (3 + i)/10$.}
\label{fig:Landscape}
\end{figure*}

\section{Comparison with data}\label{sec:data}

We have found that the toroidal model makes a robust and precise statement
about universal properties of the QHE: \emph{the geometry of the scaling flow should be
modular, and the flow rates near a critical point should be given by 
$y^+_n = - y^-_n =  1/\nu_n = (n /4)\, y_{\rm tor}$ for some positive integer $n$, 
where $y_{\rm tor} = 1/\nu_{\rm tor} = 0.3838\dots\approx 8/21$
is the natural flow unit in this model.} If $n=4$, as seems to be the case in the toroidal mirror model
discussed here, then $\vert y^\pm_4\vert /y_{\rm tor} = \nu_{\rm tor}/\nu_4 = 1$, and the values of 
the relevant ($\nu^+>0$) and irrelevant ($\nu^-<0$) delocalization exponents are 
$\nu^+ = - \nu^- = \nu_4 = \nu_{\rm tor} = 2.6051\dots\approx 21/8$.

We now confront this statement with ``experimental data", by which we mean results from both 
numerical simulations and real experiments. The experimental information required for this 
comparison is obtained by answering the following questions about the scaling flow;

\noindent
\emph{(i) Where are the boundary fixed points (sources ${\repulsor}$ and sinks ${\attractor}$) of the flow?} 
(i.e., which plateaux are observed?)

\noindent
\emph{(ii) Where are the semi-stable fixed points (saddle points ${\qcp}$) of the flow?}
(i.e., where are the critical points?)

\noindent
\emph{(iii) What is the geometry of the flow between fixed points?} 
(i.e., what is the shape of the flow lines?)

\noindent
\emph{(iv) How fast is the flow near a critical point?} 
(i.e., how big are the critical exponents?)

In a system with modular symmetry this is sufficient to identify the symmetry, 
and to map out the topology and geometry of its parameter (moduli) space.
We address these questions by giving a summary comparison of modular predictions
with some of the experimental data collected over the past three decades.
A more detailed comparison may be found in Refs.\;\cite{LR:1992, LR:1993, LR:2006, LR:2007, 
LR:2009, LR:2010, LR:2011, LR:2014, Lutken:1993a, Lutken:1993b, Lutken:2015, BLutken:1997, 
BLutken:1999, NLutken:2012a, OLLutken:2018}.

\begin{figure}[tbp] 
\begin{center}
\includegraphics[scale = .15]{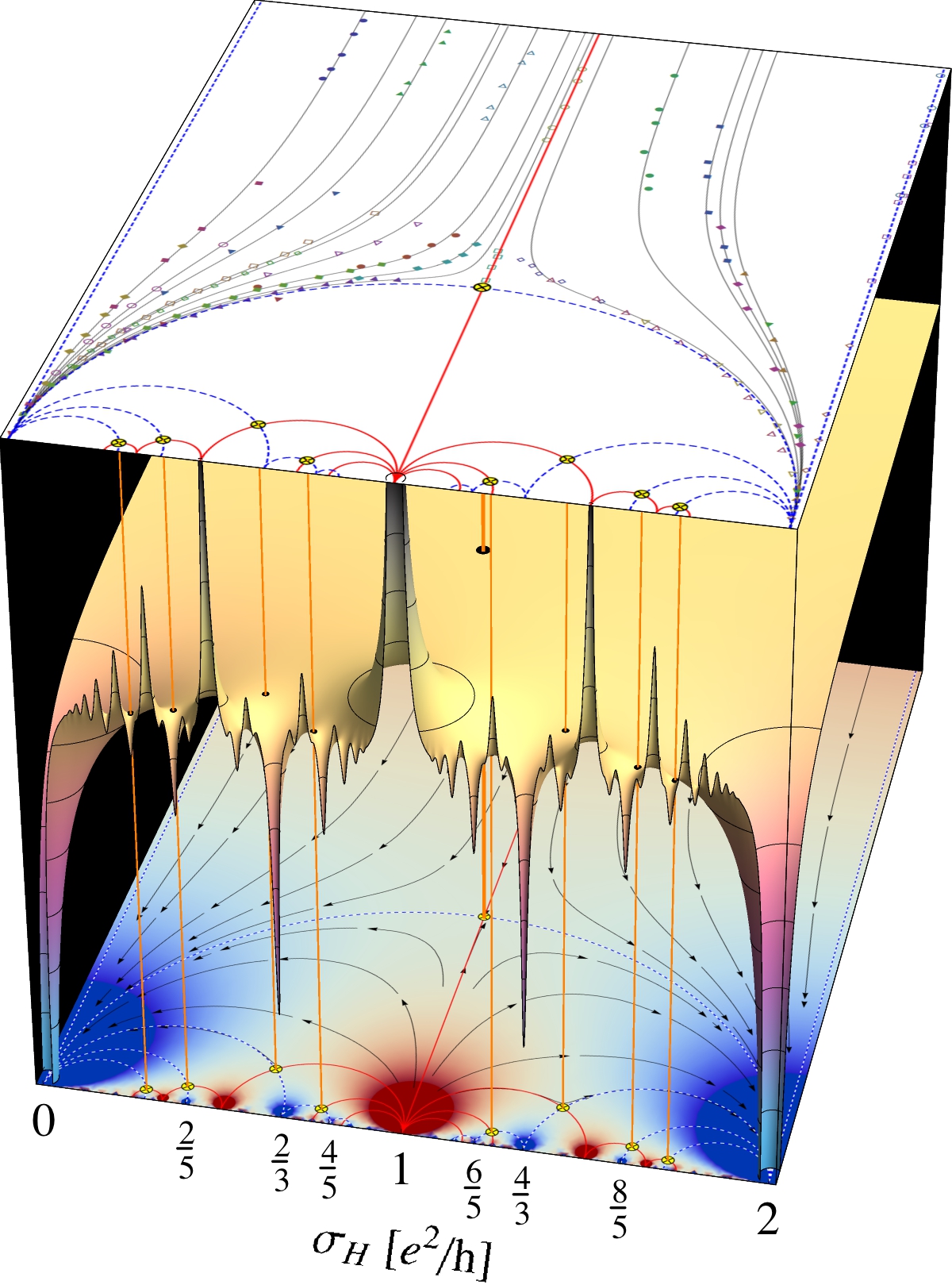}
\end{center}
\caption{Comparison of the toroidal model ${\rm T}^2_{\rm R}$ with experimental conductivity 
data reconstructed from  Ref.\;\cite{MurzinWJE:2002}  (colored plotmarkers on top face),
obtained for temperatures ranging over two decades from 4.2 K down to 40 mK.  
The surface inside the box is the potential $\vert\varphi_{-}(\sigma)\vert$ discussed in the Appendix,
whose gradient generates RG flow lines (black curves) and phase boundaries (red semi-circles)
shown on the top and bottom of the box. (The iconography is explained in Fig.\;\ref{fig:Landscape}.)
In this case it is the quantum phase transition between the unpolarized plateau $\sigma_{\smallattractor} = 2$
and the insulator phase ($\sigma_{\smallinsulator} = 0$) that is being probed (cf.\;Fig.\;\ref{fig:TRSgrid}),
revealing a quantum critical point very close to the modular prediction ${\qcp_-} = 1 + i$.}
\label{fig:Saddle}
\end{figure}

\subsection{Critical points}\label{subsec:qcps}

The easiest way to identify a modular symmetry in experimental data is to find the location of a few 
quantum critical points.  Each modular subgroup has a unique signature of fixed points, 
as explained in the Appendix.  Since it is easy to get these confused, an ``atlas" of both 
conductivity and resistivity flow diagrams is appended (cf.\;Fig.\;\ref{fig:TRSgrid}). 
This should make it easier to compare future experimental results with modular symmetries.

Figure \ref{fig:7Sisters} shows some theoretical (modular) critical points, together
with experimental critical points obtained in a spin polarized experiment
\cite{LVXPTsuiPW:2009,LR:2014}. The toroidal model predicts that
transitions between neighboring plateaux (green ${\attractor}$) at 
$\rho_{\smallattractor} = 1/(1+n)$ and $1/n$
 ($n = 0,1,2,\dots$) are controlled by saddle points of the 
torus potential $\varphi$, which are located at rational points 
$\rho_{\smallqcp} = (1+2 n+i)/(1+2 n+2 n^2)$  (blue ${\qcp}$ on orange circles), and 
repulsors (red ${\repulsor}$) at $\rho_{\smallrepulsor} = 2/(1+2 n)$.  In this phase diagram 
red semi-circles are phase boundaries, thin black curves (arrows) are modular flow lines
sampled from the gradient of $\varphi$, and dashed blue semi-circles are separatrices for the flow.
The modular and experimental critical points coincide, within the uncertainty in the data.

Figure \ref{fig:Landscape} extends this comparison of toroidal and experimental critical points to the FQHE.  
The data on the top face of the box-diagram are reconstructions of spin polarized conductivity
data reported in Ref.\;\cite{MurzinDMJ:2005}, most of which were obtained for
the insulator-plateaux transition $0={\attractor}\leftarrow\qcp\rightarrow {\attractor}' = 1/3$ (this notation is 
explained in  the Appendix).  Modular symmetry predicts a critical point at 
$\sigma_{\smallqcp} = (3 + i)/10$ for this transition (cf.\;Fig.\;\ref{fig:TRSgrid}), 
which is in good agreement with this experiment.

An example of a different symmetry is provided by Fig.\;\ref{fig:Saddle},
where unpolarized quantum Hall data is used to identify a critical point \cite{MurzinWJE:2002}.
This simply involves a rescaling of the polarized case by a factor of two, and the relevant 
symmetry is now $\Gamma_{\rm R}$. It has a  critical point at ${\qcp_-} = 1 + i$
(cf.\;Fig.\;\ref{fig:TRSgrid}), in agreement with this experiment.

Some recent experiments on graphene have probed the fixed point structure
in sufficient detail to enable a comparison with modular predictions.  Because of the peculiar
Fermi surface in this material the symmetry $\Gamma_{\rm S}$ comes into play (cf.\;Fig.\;\ref{fig:TRSgrid})
\cite{BurgessD:2007a}, and sometimes a rescaling of the observed conductivity by a factor of two is required, 
but otherwise it is again a simple transposition of the spin polarized case. 
A preliminary analysis found good agreement with modular symmetry also in this case \cite{Lutken:2015}.

\begin{figure}[tbp] 
\begin{center}
\includegraphics[scale = .17]{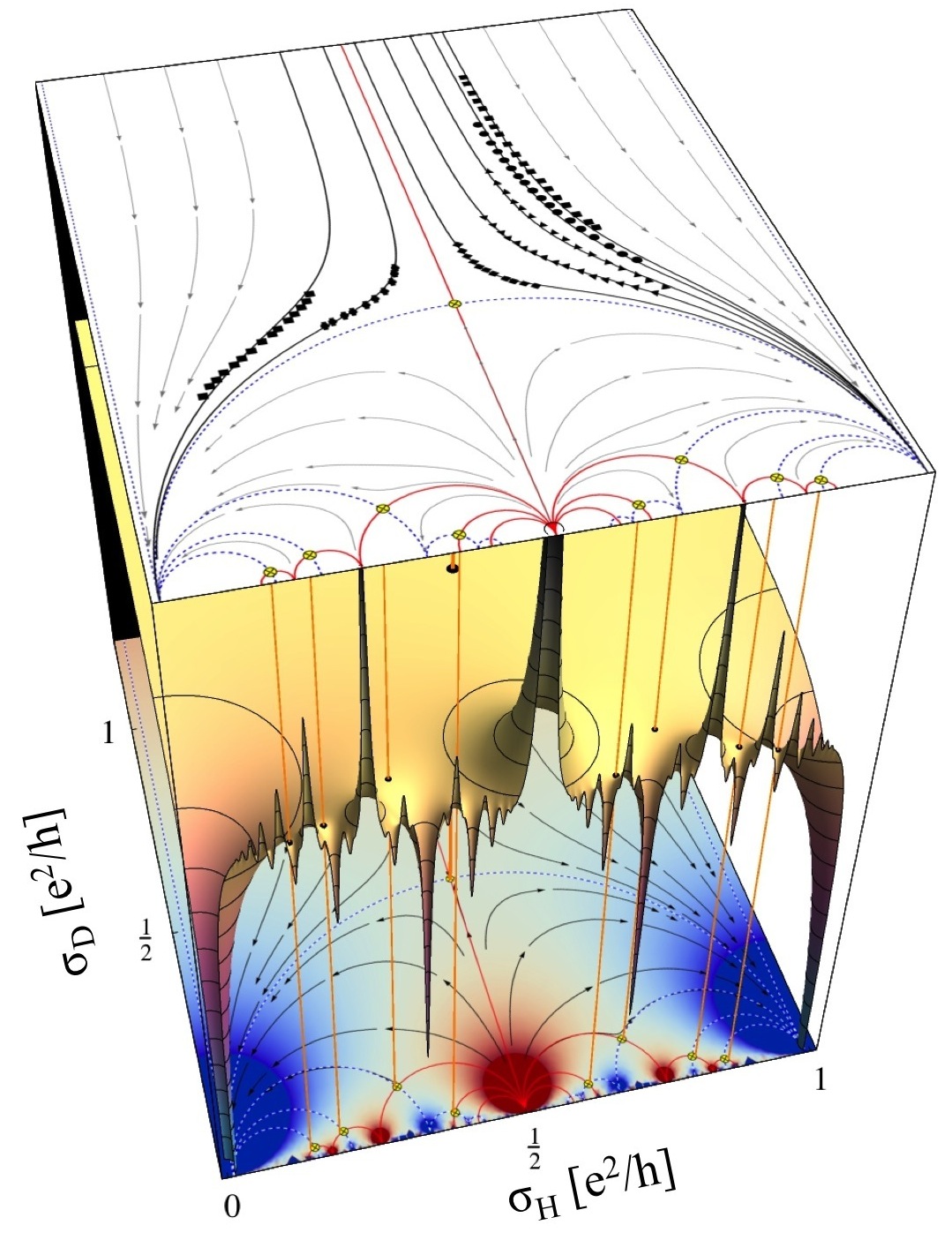}
\end{center}
\vskip -2mm
\caption{Comparison of experimental conductivity data reconstructed from 
Ref.\;\cite{Yoshimi:2014}  (black plotmarkers on top face) with the toroidal model ${\rm T}^2_{\rm T}$. 
RG flow lines (black curves) and phase boundaries (red semi-circles) on the top and bottom of the box 
are derived from the potential surface $\vert\varphi(\sigma)\vert$ inside the box. 
 (The iconography is explained in Fig.\;\ref{fig:Landscape}.)
The experimental flow reveals a quantum critical point consistent with 
the modular fixed point $\sigma_\smallqcp = (1 + i)/2$ for the quantum phase transition 
$0 = \insulator\connector{{\attractor}} = 1$ between the plateau $\sigma_{\smallattractor} = 1$ and 
the insulator phase $\sigma_{\smallinsulator} = 0$ (cf.\;Fig.\;\ref{fig:TRSgrid}).  
The geometry of the experimental flow is in good agreement with the modular prediction
(black curves on the top face). }
\label{fig:NewLandscape}
\end{figure}

\subsection{Flow lines}\label{subsec:flowgeometry}

The paucity of scaling experiments designed to probe the geometry of flow lines makes
it difficult to test this part of the model in detail, but 
Figs.\;\ref{fig:Saddle} and \ref{fig:NewLandscape} show two examples where
it is possible to carry out a comparison.

Thin black arrows on the bottom of these box-diagrams are theoretical flow lines 
sampled from the gradient of the potential shown inside the box, chosen
in order to give a clear and uncluttered picture of the phase and fixed-point structure.

Thin black curves on the top face are also theoretical flow lines, this time chosen 
for comparison with the experimental data
plotted on the same side of the box  (discrete icons).
All plotmarkers of the same shape and color belong to a data-series obtained by changing 
only the scale parameter (here temperature), holding all other variables fixed.
The starting point of each series is given by choosing a sample and a value of the magnetic field, 
which is then held fixed as the temperature is lowered.
While the starting point of each theoretical flow line may be chosen at will, the shape of the
chosen line is completely fixed by the symmetry.  It is this geometry that should be 
compared with experiment, by choosing flow lines that pass as close as possible to all data points
(not just the first one in each series).  While many details are not resolved by these experiments, 
we observe a remarkable overall agreement.  

Furthermore, by using standard temperature-driven scaling theory it is possible to compare the steepness 
of the modular potential with how quickly the data points spread out as the temperature is lowered, 
i.e., to compare flow rates.  Within experimental uncertainty, which unfortunately is quite large, 
there does not appear to be any inconsistency (cf.\;Ref.\;\cite{LR:2011} for further details).

It is sometimes possible to extract a flow diagram from families of 
published temperature-dependent transport data, as illustrated in Fig.\;\ref{fig:7Sisters}.
Experimental flow lines (purple curves in the inset near the critical points 
$\frac{1}{4}{\qcp}\frac{1}{3}{\qcp}\frac{1}{2}$) were obtained by interpolation 
of data points (bullets) reconstructed from Ref.\;\cite{LVXPTsuiPW:2009}.

Most scaling experiments focus on critical regions, with the goal of extracting flow rates
close to a critical point. We shall argue that it is a pressing problem to
determine when ``close" is close enough for a reliable estimation of critical exponents.

\subsection{Critical exponents}\label{subsec:flowrates}

We focus now on the \emph{delocalization transition} between Hall plateaux,
whose critical exponents should characterize the quantum Hall universality class.
The localization length $\xi$ of electron states in the centre of an impurity-broadened Landau
level depends on the value of the magnetic field $B$, and is believed to scale 
with $\delta B = \vert B - B_{\smallqcp}\vert$ when $B$ is near one of the critical values $B_{\smallqcp}$
where the localization length $\xi$ diverges \cite{Pruisken:1990}.
The critical \emph{correlation (delocalization) length exponent $\nu$} is defined by the scaling 
relation $\xi\sim \delta B^{ -\nu} + \dots$, where the dots represent corrections to scaling.  
Since $\xi$ is not directly observable, we need to swap it for another scaling variable that is more accessible.

\subsubsection{Numerical scaling}

\begin{figure}[tbp]
\hskip -1mm
\includegraphics[scale = .16]{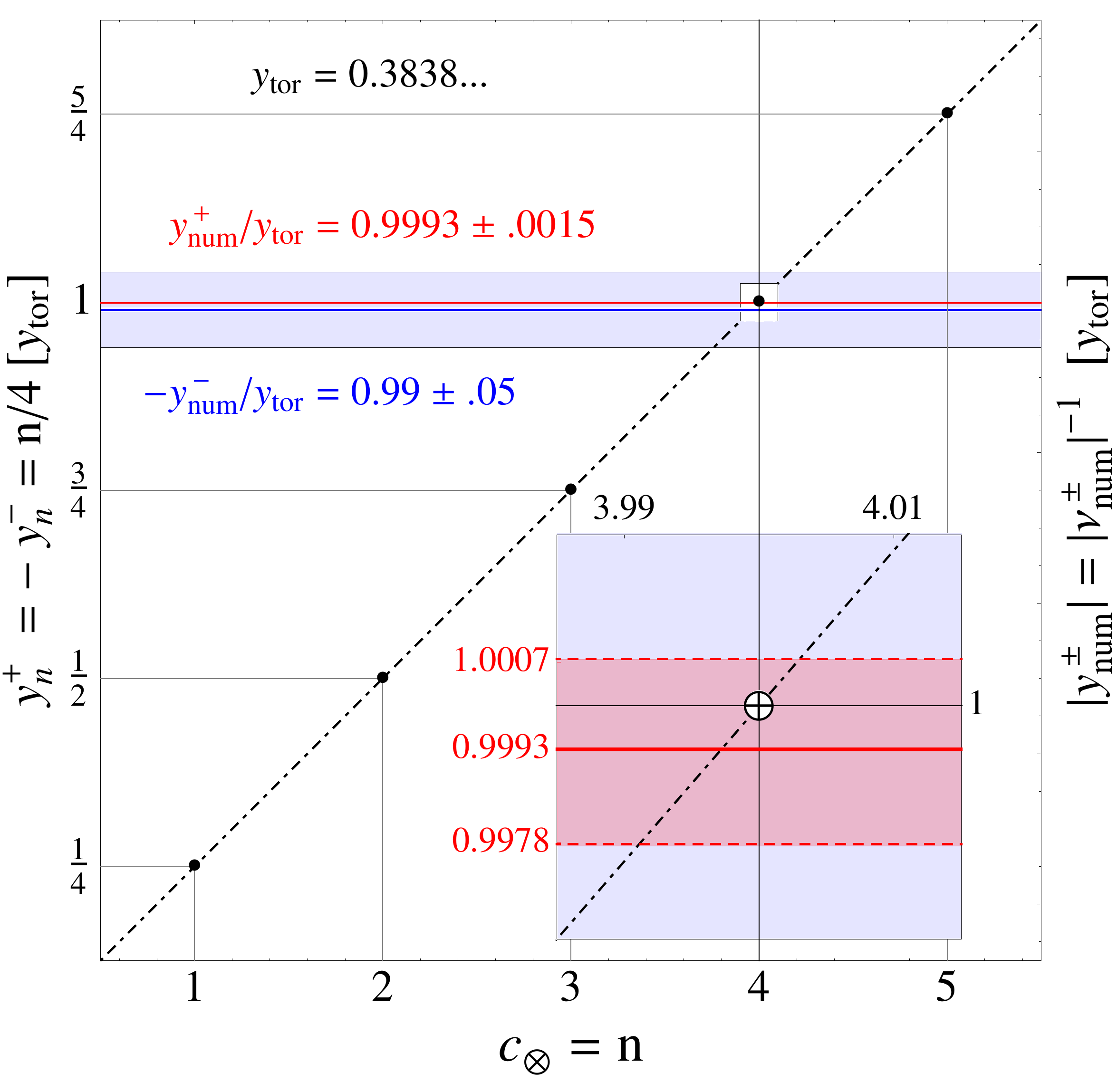} 
\caption[exponents]{Toroidal flow rates $\vert y^\pm_n\vert$ evaluated at integer values
$c_\smallqcp = n\in\mathbb N$ of the central charge of the conformal symmetry at critical points 
of the elliptic model (black dots), compared to the best available numerical values $y^\pm_{\rm num}$ 
obtained from the CC model \cite{SlevinO:2012} (red and blue lines). The inset shows a small part of the 
white box surrounding a bullet in the main diagram, hugely magnified so that the value of the toroidal exponent 
$y^+_4 = 1/\nu^+_4$ (white $\oplus$) can be distinguished from the numerical result 
$y^+_{\rm num} = 1/\nu^+_{\rm num}$ (red line bisecting the pink $\pm 1\sigma$ band, which is thinner than the 
red line in the main diagram).}
\label{fig:exponents}
\end{figure}

Figure \ref{fig:exponents} shows the toroidal flow rates 
$y^+_n = - y^-_n = (n/4) \, y_{\rm tor}$ derived in the previous section, 
evaluated at integer values $c_\smallqcp = n$ of the central charge (black dots),
and measured in the natural toroidal flow unit $y_{\rm tor} = 1/\nu_{\rm tor} = 0.3838\dots$.  
This should be compared with the best available numerical values, 
$y_{\rm num}^+/y_{\rm tor} = 0.9993\pm .0015$ ($\nu^{\,+}_{\rm num} = 2.607\pm .004$) (horizontal red line),
and $y_{\rm num}^-/y_{\rm tor} = - 0.99\pm .05$ ($\nu^{\,-}_{\rm num} = - 2.63\pm .14$) 
(horizontal blue line)  \cite{SlevinO:2012}. 
The uncertainty in $y^{\,+}_{\rm num}$ is smaller than the thickness of the red line in the main diagram. 
If one of the toroidal models is in the same universality class as the CC model, then the ``experimental" line 
should pass near one of the theoretical  black dots, and it does in fact appear to intersect one of them. 

The inset shows a small part of the intersection region covered by a white square in the main diagram, 
greatly magnified in order to appreciate the accuracy of the numerical work. The red band covers one 
standard deviation from the mean of $y^{\,+}_{\rm num}$.  The relevant exponents $y^+_4 = y_{\rm tor}$ 
and $y^{\rm +}_{\rm num}$ agree at the per mille level.

Similarly, the blue band is the uncertainty in $y^{\,-}_{\rm num}$.  This shows that the value of the
irrelevant flow rate $y^{\,-}_{\rm num}$ is consistent  with the antiholomorphic scaling relation 
$y^- = - y^+$ (within experimental uncertainty), predicted by any holomorphic modular 
symmetry \cite{LR:2006}.

\subsubsection{Temperature scaling}

The easiest and most common approach is to study the temperature ($T$) dependence of 
$y(T) = y(T, w, B, \dots)$, where $y$ is either the inverse half-width $\Delta B^{-1}$ of the peak in $\rho_D$ 
or the maximum slope $\rho_H^\prime = \max\left(\partial\rho_H/\partial B\right)$ of $\rho_H$
in the transition region between two plateaux, while holding all other variables fixed (size $w$ 
of the Hall bar, magnetic field $B$, etc.)  
Provided the sample is not too hot or too cold ($0 < T_s < T < T_q < \infty$),
both are observed to scale with temperature, $\Delta B^{-1} \sim T^{-\kappa_D}$ and 
$\rho_H^\prime \sim T^{-\kappa_H}$, with the same exponent $\kappa_D\approx\kappa_H$, 
but the value of this exponent does not appear to be universal (cf.\;Fig.\;\ref{fig:ExponentGrid}).

Another disadvantage of this approach is that a third exponent is needed in order to relate $\kappa$ to $\nu$.
This is most easily seen from the scaling relation  $\kappa\nu z = 1$, where the dynamical frequency 
exponent $z$ seems to be as inaccessible as $\nu$.   We can use the scaling relation $p z = 2$ to swap
$z$ for the temperature exponent $p$ of the inelastic scattering length, but this is not much help.
Even if $z$ and $p$ are no more universal than $\kappa$, it is still possible that $\nu$ is universal.

\subsubsection{Finite size scaling}\label{subsubsec:fss}

In order to investigate this, Koch \emph{et al.} \cite{KHKlitzingP:1991, KHKlitzingP:1992} devised a 
method to measure $\nu$ directly in a finite size scaling experiment.   
The idea is that we expect the width $w$ of the physical Hall bar to be the most relevant 
scale at sufficiently low temperature, so that we can use $w$ as the scaling variable  in this regime.  
As the temperature is lowered into the quantum regime ($T<T_q$), 
temperature scaling has been observed over two decades in some
quantum Hall devices, but this must eventually stop at some finite \emph{saturation
temperature} $T_s > 0$, because the inelastic scattering (phase coherence) length 
$L_{\rm in}$ depends on temperature.  When the material is cold enough this length
exceeds the size $w$ of the Hall bar, so temperature scaling should ÒsaturateÓ
and the response functions should become independent of $T< T_s$.

\begin{figure}[tbp]
\vskip 0mm
\includegraphics[scale = .27]{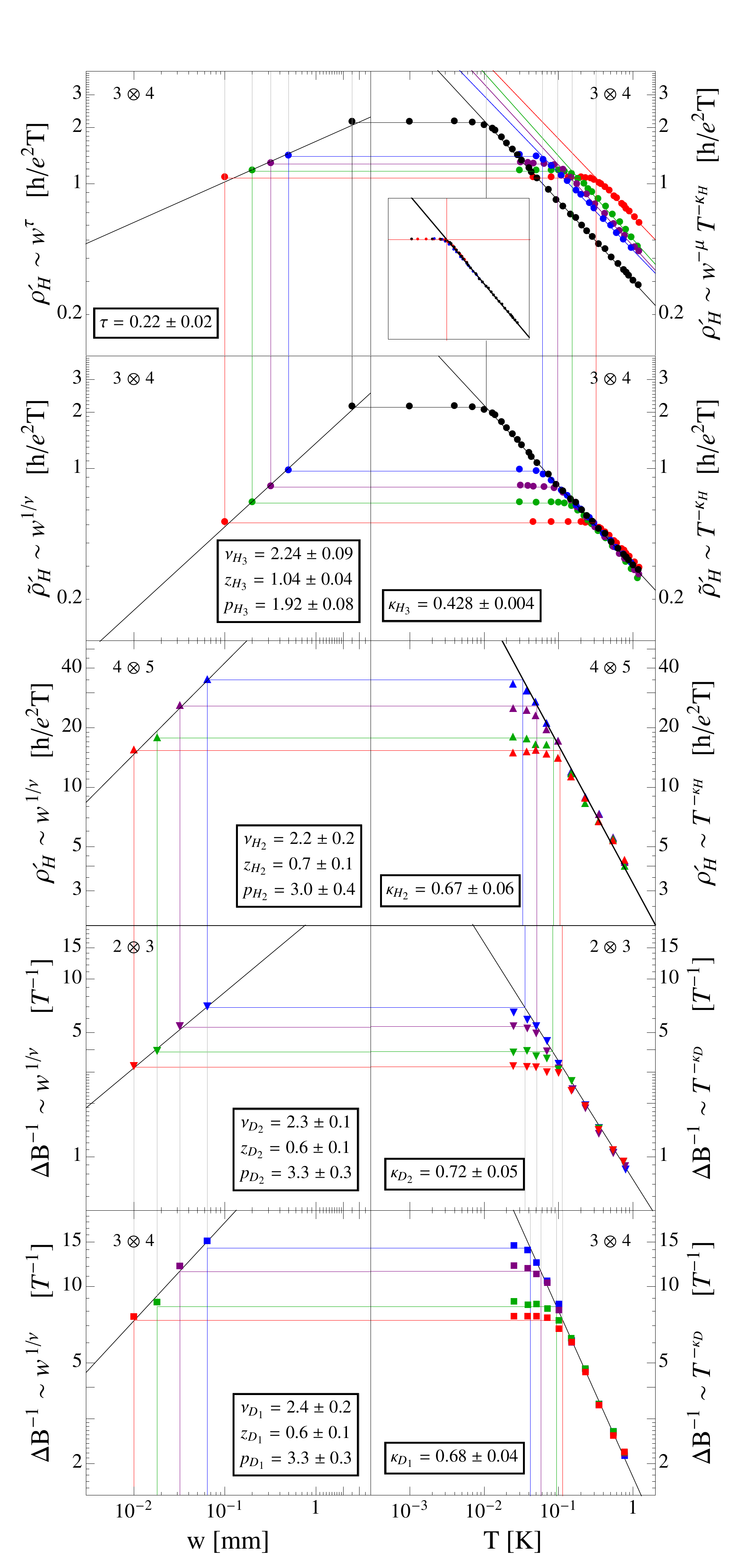}
\caption{ Compilation of finite size scaling data, reconstructed from 
Refs.\;\cite{KHKlitzingP:1991} (bottom), \cite{KHKlitzingP:1992}  (rows 3 and 4),
and \cite{LVXPTsuiPW:2009, Li:2007} (top).  The meaning of the various scaling exponents 
is explained in the text.  Each panel is labelled by the 
delocalization transition $n {\qcp}_\sigma n+1\; (n = 2,3,4)$ for which the data were obtained. 
The slope of high-temperature ($T_s < T < T_q$) data on the right gives $\kappa$,
while scaling in $w$ of the saturation data ($T \lesssim T_s$) on the left allows $\nu$ 
to be determined independently.  
The inset in the top right panel shows how the published data \cite{LVXPTsuiPW:2009} collapse
after using the double scaling law (revealed by the top left panel, where $\tau = 1/\nu - \mu$) 
to rescale the data \cite{LR:2014}. Observe that the delocalization exponents $\nu$ appears 
to be more or less universal, but that the temperature scaling exponent $\mu$ is not (see the text
for more details).}
\label{fig:ExponentGrid}
\end{figure}

This is indeed what they observed in a GaAs-AlGaAs heterostructure etched with several Hall bars 
of different size $w$ (constant aspect ratio), cf.\;the bottom three panels on the right in 
Fig.\;\ref{fig:ExponentGrid}, where some of the data from Refs.\;\cite{KHKlitzingP:1991, KHKlitzingP:1992}
has been reconstructed and replotted so as to facilitate comparison with other experiments.

They also found that in this material all the unsaturated data ($T_s < T < T_q$) collapse to a single line in 
a log-log plot, suggesting some kind of universality, but the slope $\kappa$ of this line is not universal.
However, provided the high-temperature data collapse, as they do in this experiment,
the scaling argument in Ref.\;\cite{KHKlitzingP:1991} predicts that the saturation value $y_s(w) = y(T_s,w)$
($y = \Delta B^{-1}$ or $\rho_{H}^\prime$) should be directly related to $\nu$ by a scaling relation 
$y_s(w) \sim w^{1/\nu}$, allowing the delocalization exponent to be measured independently of $\kappa$. 
Their data verifies this, cf.\;the bottom three panels on the left in Fig.\;\ref{fig:ExponentGrid}, 
and for this exponent they observed a remarkable degree of universality, 
with $\nu = 2.3 \pm 0.1$ for a number of different samples and Hall transitions.

Subsequent work has shown that the type of disorder determines if the temperature scaling exponent 
$\kappa$ is universal \cite{STsuiSBC:1995, LVXPTsuiPW:2009, Li:2007}.
The top right panel in Fig.\;\ref{fig:ExponentGrid} shows a reconstruction of more recent finite size 
scaling data, obtained for an AlGaAs-AlGaAs heterostructure suitably doped so that the disorder 
potential is dominated by short-range fluctuations \cite{LVXPTsuiPW:2009, Li:2007}.
For each width $w$ of the Hall bar, high-temperature scaling of $\rho_H^\prime$ 
was again observed, with what appears to be a reasonably universal 
value $\kappa = 0.42\pm 0.01$ of the scaling exponent.  But now a new problem has surfaced: 
the high-temperature data for different widths $w$ no longer collapse to a single line (in a log-log plot), 
so a direct and independent determination of $\nu$ no longer seems possible.

A way around this was proposed in Ref.\;\cite{LR:2014}, by observing that the high-temperature data  
($T>T_s$) satisfy a \emph{double scaling law} $\rho_H^\prime \sim w^{-\mu} T^{-\kappa}$,
where $\mu$ is a new scaling exponent (cf.\;top two rows of Fig.\;\ref{fig:ExponentGrid}).
The best fit of this scaling form to the 71 unsaturated data points in Refs.\;\cite{LVXPTsuiPW:2009, Li:2007}
is obtained with $\kappa = 0.428\pm0.004$ and $\mu =  0.230\pm 0.004$.
We have $\mu \approx 0$ in the older experiments on materials with long-range 
disorder \cite{KHKlitzingP:1991, KHKlitzingP:1992},  because the data above saturation collapse.

The exponent $\mu$ is perhaps related to a scaling exponent $\mu_{\rm mf}$ 
that is expected to appear in systems with \emph{multi-fractal wave-functions}, 
which includes the numerical CC model \cite{ChalkerC:1988} and systems with 
short-range disorder \cite{HuckesteinK:1990, Mieck:1990, Huckestein:1992}.
A range of values $\mu_{\rm mf} \in \{0.22 - 0.26\}$ have been reported for the 
CC model \cite{CainRR:2003, ObuseSFGL:2010, DahlhausETB:2011, 
AmadoMSD-A:2011, ObuseGE:2012, SlevinO:2009, SlevinO:2012}.
Long-range interactions break the self-similar scale invariance giving rise to multi-fractals,
so presumably $\mu_{\rm mf} = 0$ in materials with long-range disorder.
If $\mu = \mu_{\rm mf}$ this would explain why the value of $\mu$ is different in materials 
with long-range \cite{KHKlitzingP:1991} and short-range \cite{LVXPTsuiPW:2009} disorder.

Since all the rescaled high-temperature data  
$\tilde\rho_H^\prime = w^{\mu} \rho_H^\prime \sim T^{-\kappa}$ collapse to a single line, 
we can now apply the original scaling argument to the saturation value
$\tilde y_s(w) \sim w^{1/\nu}$ of $\tilde\rho_H^\prime$.
This is equivalent to the scaling relation $y_s(w) \sim w^{\tau}$ for the raw
saturation value of $\rho_H^\prime$, where  $\tau = 1/\nu - \mu = 0.22\pm 0.02$ 
can be read off directly from the data, cf.\;top row of Fig.\;\ref{fig:ExponentGrid}.
Combined with the best-fit value of $\mu$ this gives $\nu_{\rm exp} = 2.24 \pm 0.09$, consistent
with the approximately universal value previously found in materials with long-range 
disorder \cite{KHKlitzingP:1991, KHKlitzingP:1992}.
 
Since these experiments show that both $y(T)$ and $y_s(w)$ ($y = \rho_H^\prime$ or $\Delta B^{-1}$)
are log-log-linear, it follows that the saturation temperature also scales with sample size,
$T_s \sim w^{-z}$, where $z$ satisfies the scaling relation $\kappa \nu z = 1$.
This is consistent with standard scaling theory if $z$ is the dynamical frequency scaling exponent.

We therefore expect the inelastic scattering length to scale as $L_{\rm in}\sim w \sim T_s^{-1/z}\sim T_s^{-p/2}$, 
where $p = 2/z$ is the temperature scaling exponent of $L_{\rm in}$. 
Combining this with $\kappa \nu z = 1$, we obtain a scaling relation $\kappa = p/2\nu$ that is 
consistent with standard scaling theory. 

The values of $z$ and $p$ given in Fig.\;\ref{fig:ExponentGrid} are derived from the independently 
measured values of $\kappa$ and $\nu$ by using the scaling relations $\kappa \nu z = 1$ and $p z = 2$.
Notice that $z\approx 1$ and $p\approx 2$ only when the disorder is short-range (top row).
We observe that the delocalization exponent $\nu$ appears to be fairly universal 
(independent of material and experimental details),  but neither $\kappa$, $z$ nor $p$ are found to be 
universal in these finite size scaling experiments \cite{footnote:z_expo}.

\subsubsection{Concordance?}

The excellent agreement between the exact toroidal value
$\nu_{\rm tor}  =  18\pi^2\ln 2/\pi_\infty^4  = 2.6051265833\dots$, and the numerical value 
$\nu^{\,+}_{\rm num} = 2.607\pm .004$  obtained from the CC model 
\cite{ChalkerC:1988,SlevinO:2009, SlevinO:2012}, presumably means that these two theoretical 
(mathematical) models are in the same universality class 
($\nu_{\rm tor}/\nu^{\,+}_{\rm num} = 0.999\dots$).

It was argued in Ref.\;\cite{LR:2014} that this theoretical value close to $2.61$
not necessarily is in conflict with 
the experimental value $\nu_{\rm exp} = 2.24 \pm 0.09$ of the delocalization exponent (extracted 
from finite size scaling data), because the two flow rates are obtained at different points in moduli space.
These points appear to the naked eye to be quite close together, but they are not close enough.
When the modular flow rate is evaluated at the same point in parameter space where the experiments were 
performed, the new value $2.23$ is consistent with the experimental flow rate
(cf.\;Ref.\;\cite{LR:2014} for a more detailed quantitative discussion). This suggests that the theoretical 
(toroidal and numerical) models are in the quantum Hall universality class.

In other words, it is possible that the size of the scaling domain has been 
overestimated in the past, in which case real experiments have yet to reach the proper 
scaling domain required for a reliable estimate of the critical exponent.  
If this is true, improved scaling data should see $\nu_{\rm exp}$ trailing $\nu^{\,+}_{\rm num}$,
which in recent years has crept upwards until it now agrees
with the toroidal value $\nu_{\rm tor} \approx \nu^{\,+}_{\rm num} \approx 2.61$, within a 
very small uncertainty in the numerical value \cite{SlevinO:2012} (cf.\;Fig.\;\ref{fig:exponents}).

\section{Discussion}\label{sec:discussion}

In summary, in the toroidal sigma model discussed here, Hall quantization is protected by slope-stable
holomorphic vector bundles over a target torus with spin structure
($\sigma_H^{\smallattractor} = \mu \in\mathbb Q$).  In the equivalent mirror model 
topological protection is provided by more familiar winding numbers.
The non-interacting case is modeled by line bundles $\mathcal L$ ($r = 1$), from which a familiar looking 
topological statement $\sigma_H^{\smallattractor} = c_1(\mathcal L) \in\mathbb Z$ about the IQHE is recovered 
\cite{ThoulessKNdN:1982, *NThouless:1984, *NThoulessW:1985,* AvronSS:1983, *Simon:1983, 
*AvronS:1985, *AvronSSS:1989}, albeit in the context of effective field theories. 

The connection to the first-quantized theory, which is based on hugely simplified but explicit 
plateaux wave-functions, is not completely obvious, but if the holomorphic vector bundles discussed here 
can be identified with ``vacuum bundles" on the so-called ``flux torus" (a section of this bundle is a ground 
state), then $\mu$ is presumably the same as the topological invariant first discussed 
by Thouless et al.\;in 1985 \cite{NThoulessW:1985}. 
Although it took another decade before a connection to stable bundles was recognized 
 \cite{Varnhagen:1995, footnote:thankstoKSO}, with hindsight we now see a remarkable mathematical 
similarity with some aspects of the stringy ideas that were used in the construction of the toroidal model.
This does not include modular and mirror symmetry, and it would be interesting to find a 
first-quantized interpretation of the topology (winding numbers) of the mirror model.

If a relationship of this kind could be firmly established, it would provide an appealing convergence of the 
``bottom-up" approach based on wave-functions, and the ``top-down" approach based on effective 
(emergent) quantum fields that has been discussed here. This would mean that they provide complementary
descriptions of the plateaux, as they should, because the idea that powers both the first- and 
second-quantized approach is that universality can be exploited to give a simplified description
that retains only the most relevant information.

Universality means that the macroscopic response is \emph{robust}, i.e., insensitive to microscopic details,
which is why a simplified wave-function can stand in for the  messy real wave-function.
Although they are quite different microscopically, provided that they are in the same universality class
the sanitized avatar will emulate the long-distance properties of the real system.  Usually it is sufficient
to respect the particle type and symmetry of the real system, but in the QHE there seems to be a stronger
version of universality at work, enforced by modular symmetry. 

In the model discussed here quantum phase transitions are associated with topology change, 
rather than spontaneous symmetry breaking.  Furthermore, the experimentally observed emergent 
modular symmetry polices  a ``super-universality" that forces all critical points to have the same exponents.  
In essence, for each spin structure (polarized vs. unpolarized spins, etc.) there is only one quantum Hall 
universality class.

The first- and second-quantized approaches are not, however, in general equivalent, 
since the EFT is much more ambitious.
It aspires to describe the \emph{global} structure of the compactified moduli space 
$\overline{\mathcal M} = \mathcal M\cup\partial\mathcal M$,  including quantum critical points 
deep inside ${\mathcal M}$, not only the local structure at or near plateaux on the
boundary $\partial\mathcal M$ of $\overline{\mathcal M}$.

The sigma model discussed here is not an ad hoc construction, although this
would have been quite acceptable as long as the model is falsifiable.  We have been guided in 
its construction by two powerful empirical observations about universal properties of the QHE,
by discovering that the mathematics capable of modeling these observations more or less
naturally leads us to consider holomorphic structures on elliptic curves. The simplest viable sigma model 
therefore appears to be toroidal, which is superficially similar to, but much richer than, 
the more familiar spherical sigma model.

This ``leap of faith" is arguably no worse than what is done in first-quantized theory to arrive at the conventional 
topological interpretation of Hall quantization.  Both physical and mathematical assumptions are needed, 
regarding the degeneracy, boundary conditions and universality of the sanitized many-particle 
wave-functions used in the computation of transport coefficients, 
and the validity of these assumptions is not easy to check.  
The bird's-eye view provided by ``macroscopic phenomenology" reveals new features 
that are not apparent in the first-quantized approach.  Crucially, experimental data force us to consider 
the emergence of an infinite global discrete symmetry that appears to be wed to a rigid 
holomorphic structure.   

As mentioned in the introduction, there have been other attempts to construct an EFT for the QHE.
Although they look quite different, because there is an intricate web of ``dualities" between 
low-dimensional theories that we do not fully understand, it is not obvious that they are (all) inequivalent.
Of course, if they model the same physics they must in some sense (or some domain) be equivalent. 

At first sight the tensor sigma model \cite{EfetovLK:1980, Khmelnitskii:1983, LevineLP:1983, 
*LevineLP:1984a, *LevineLP:1984b, *LevineLP:1984c, *Pruisken:1984, *LevineL:1985}
looks like a relative of the toroidal model, but sharp predictions in the scaling region 
between plateaux that can be falsified have yet to be obtained from this model. 
Nor does it seem to be rich enough to model the FQHE, so there is little to compare with experimental data, 
or the modular model. The toroidal model degenerates to a similar model in the weak coupling limit, 
in a sense that can be made mathematically precise, but the latter is irrelevant for the strongly coupled 
physics that is responsible for quantum phase transitions \cite{footnote:wcl}.

A Chern-Simons Landau-Ginzberg (CSLG) theory was developed 
in \cite{ZHanssonK:1989, *LeeZ:1991, *KLZ:1992a, *KLZ:1992b}, 
independently and around the same time as the modular model \cite{LR:1992}.
By identifying a so-called ``law of corresponding states" in this model, heuristic phase diagrams
that look (very) roughly modular may be constructed. The ``law" is superficially similar to a modular 
transformation acting on the filling factor (a real number).  It is in fact quite different, 
not least because the modular symmetry acts directly on response functions, and only makes 
contact with the filling factor at trivial fixed points (plateaux). 
A careful analysis of the experimental discovery of duality in the QHE \cite{STsuiSSS:1996} shows that 
this ``law" does not sit well with the duality data \cite{LR:2010}.  Modular duality, on the other hand, 
is in complete agreement with the data (within experimental uncertainty).

An important difference between these models is that although the CSLG-model provides an appealing 
physical picture of anyonic quasiparticles, it appears to be an uncontrollable mean-field
description that does not address scaling (renormalization) since it only describes plateaux.
One consequence of this is that the CSLG-model does not respect any complex structure, 
while in the modular model this is, as we have seen, one (actually two) of the main ingredients.

We have here relied heavily on some results from string theory, both in the conception and analysis 
of the toroidal model. This is perhaps not ideal, but when we need a new paradigm in physics,
as we surely do in the QHE, this often requires that we expand our mathematical vocabulary.  
Strings have gifted us a cornucopia  of new mathematical tools for analyzing non-perturbative 
properties of quantum field theories, and many of these we barely know how to use.
Central to several important developments in string theory are two novel and exotic types of 
symmetry that are inextricably linked to the geometry and topology of Calabi-Yau manifolds.
They have so far not been much used in condensed matter physics,
but in our narrative here modular and mirror symmetry take center stage.

We believe that the QHE is the first real system to exhibit an infinite discrete (modular) symmetry,
and we have argued that when this is combined with the mirror symmetry that is hardwired into 
toroidal models, then the geometry and topology of the EFT is so severely constrained that 
a simple viable model presents itself. 

The rigidity of this model is arguably its greatest virtue.  Experimental predictions extracted from 
the model are sharp and non-negotiable, rendering it extremely vulnerable to ``the untimely intrusion of reality".  
As technology has improved and error-bars have shrunk, it could easily have been falsified by 
experimental results obtained over the past three decades, but this has not happened.  
We have seen that the geometry of scaling flows extracted from quantum Hall experiments is in 
good agreement with modular predictions, including the location of several quantum critical points 
and the shape of some RG flow lines.

The mirror symmetry of toroidal models is not observable, because it is a quantum equivalence that
essentially amounts to a change of field-variables in the functional integral.  It equates two mathematical
models that appear to be quite different, but which in fact are completely equivalent, so they model
the same physics and no experiment can tell them apart. However, we have seen that mirror symmetry 
takes us to an intriguing ``alternate reality" on the other side of the looking glass, i.e., a different way of 
looking at the role of topology in the QHE.  
Mirror symmetry is also very useful for calculations, allowing us to sidestep difficult non-perturbative 
computations and conclude that the toroidal model appears to include the universality class of 
the numerical CC model \cite{ChalkerC:1988,SlevinO:2009, SlevinO:2012}.

The alleged experimental value of the delocalization exponent may be seen as another vindication 
of the toroidal model, because if the modular flow rate is evaluated at the same point in parameter space
where experiments have been done, then it is consistent with the experimental value.
This experimental number may therefore not be the true value of the critical exponent, 
if real experiments have yet to reach the proper scaling domain near a critical point.

This tentative agreement with available experimental data is encouraging, 
but in the absence of a derivation that makes direct contact with microphysics
there is only one way to determine if the theoretical (analytical and numerical) models really are 
in the quantum Hall universality class(es): 
\emph{improved finite size scaling experiments  are urgently needed.}

\begin{acknowledgments}
The author gratefully acknowledges a long-standing collaboration with 
Graham G. Ross that made this work possible, and the hospitality 
of the Theory Division at CERN where some of this work was done.
\end{acknowledgments}

\appendix* 
\section{Modular mathematics} 

\begin{figure*}[t]
\vskip 3mm
\hskip -3mm
\includegraphics[scale = .29]{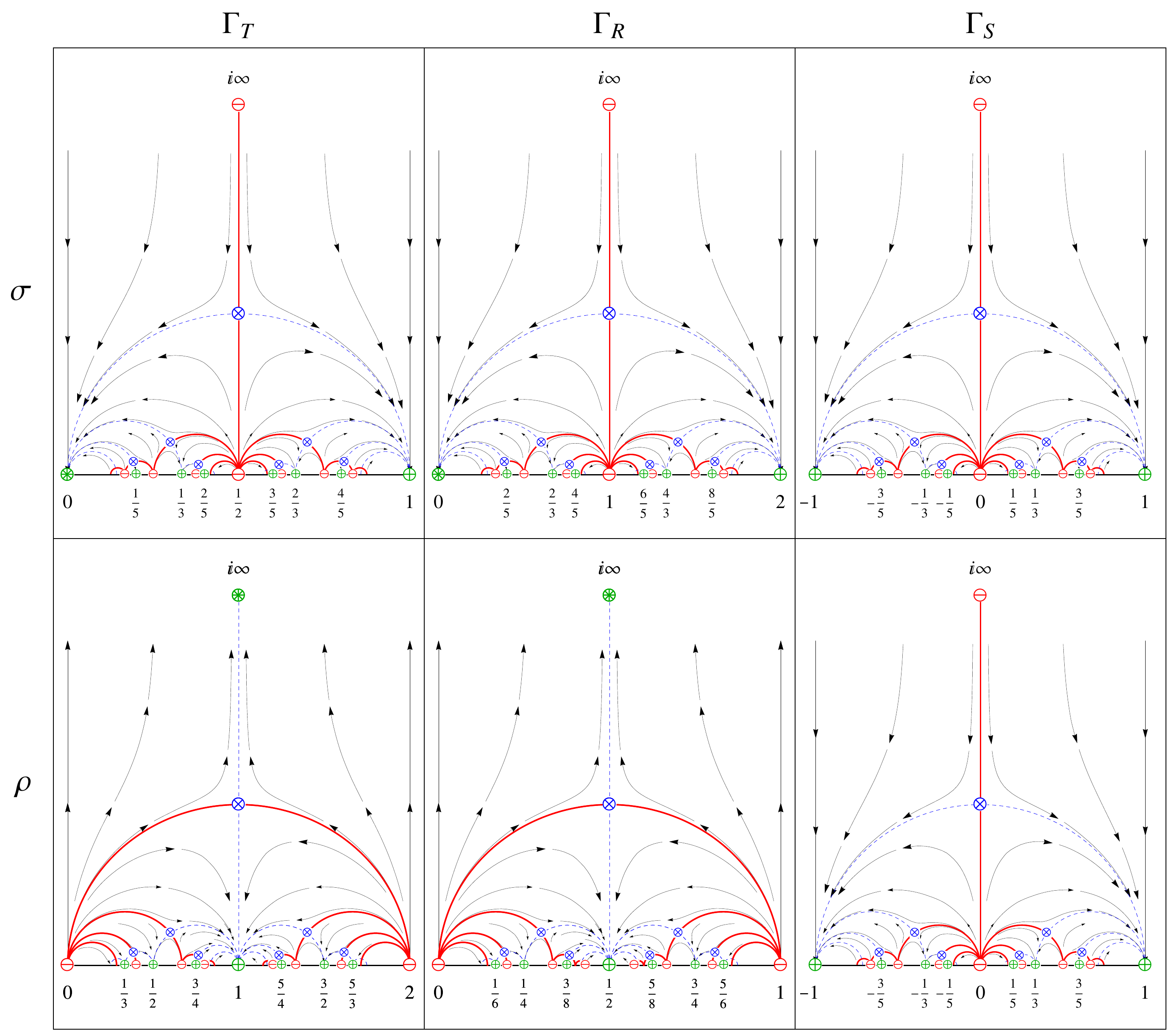}
\caption[TRSgrid]{Phase diagrams with modular symmetry in the space of complexified conductivities 
$\mathbb C^+(\sigma)$ (top row), and equivalent diagrams in the space of resistivities 
$\mathbb C^+(\rho)$ (bottom row), for the congruence subgroups $\Gamma_{\rm T}$,  
$\Gamma_{\rm R}$ and  $\Gamma_{\rm S}$ of the modular group $\Gamma(1)$.
Notice that the $\Gamma_{\rm R}$-diagram is the $\Gamma_{\rm T}$-diagram rescaled by a factor of 2, 
and that the $\Gamma_{\rm S}$-diagram does not have an insulator phase ($\insulator$).
Black arrows are RG flow lines, which cannot cross the red phase boundaries (the iconography is
explained in the text).
\label{fig:TRSgrid}}
\end{figure*}

\subsubsection*{Modular phase diagrams}

In order to facilitate the comparison of experimental data with modular symmetries, 
an ``atlas" of modular conductivity and resistivity diagrams is provided in Fig.\;\ref{fig:TRSgrid}.  
In general, if modular symmetries are discovered outside the QHE, 
``resistivity" is any complex parameter $z$ on which modular symmetries act by 
holomorphic fractional linear (M\"obius) transformations.

We always use the top row of the transport matrix to define the corresponding complex parameter, 
so $\rho = \rho_{12} + i\rho_{11}  = -\rho_H + i\rho_D$, since it is $\rho_H = \rho_{21} = - \rho_{12}$ 
that is measured when the current vector $(I,0)^t$ is aligned with the horizontal axis 
(Ohm's law in matrix form: $(V_D, V_H)^t = (V_1, V_2)^t = (R_{11}, R_{21})^t I = (R_D, R_H)^t I$).
Defining the Hall  and magneto-conductivities as  $\sigma_H = \rho_H/\vert \rho\vert^2$
and $\sigma_D = \rho_D/\vert \rho\vert^2$, matrix inversion gives
$\sigma = \sigma_{12} + i\sigma_{11} = \sigma_H + i\sigma_D$.

Observe that in addition to the obvious arithmetic simplification obtained by converting matrices into 
numbers, these complex variables are tailor-made for modular mathematics, because $\rho$ and $\sigma$ 
are related by a modular duality transformation, $\sigma = S(\rho) = -1/\rho$ and $\rho = S(\sigma) = -1/\sigma$. 
$S$ is its own inverse, so $S^2 = 1$, which is one of the two relations used to give an abstract
definition of the modular group [the other is $(ST)^3 = 1$]. This greatly facilitates the transposition of 
modular conductivity diagrams into resistivity diagrams, and vice versa.
For the full modular group $\Gamma(1) = {\rm SL}(2, \mathbb Z) = \langle T,S\rangle$, 
and any subgroup  containing $S$, the dual diagrams are identical
(cf.\;left column in Fig.\;\ref{fig:DyonGrid}, and third column in Fig.\;\ref{fig:TRSgrid}), 
but if $S$ is not a symmetry they are different (cf.\;first two columns in Fig.\;\ref{fig:TRSgrid}).

$\Gamma(1)$ is too strong for the QHE, but its largest subgroups  are not.
They are obtained by relaxing the translation symmetry ($T\rightarrow T^n$), or the duality 
symmetry [$S\rightarrow R^n = (TST)^n$], or both.
Three of these so-called ``congruence subgroups at level two" preserve parities, 
which means that each of them groups the fractions into two equivalence classes.
Because $p$ and $q$ in ${\attractor} = p/q$ are relatively prime,
there are only three types of fractions with well defined parities.
With ``$o$" representing odd integers and ``$e$" representing even integers, we have
$p/q \in o/o$, $o/e$ or $e/o$, and it is easy to verify that the equivalence classes are:
\begin{eqnarray*}
\Gamma_{\rm T}&=& \langle T, R^2\rangle: 
\quad\kern -1pt \left\{\frac{e}{o},\frac{o}{o}\right\}_{\attractor}\;\cup\kern 1pt\;\quad\left\{\frac{o}{e}\right\}_{{\repulsor}},\\
\Gamma_{\rm R} &=& \langle R, T^2 \rangle: 
\quad\quad \left\{\frac{e}{o}\right\}_{\attractor}\;\cup\;\left\{\frac{o}{o}, \frac{o}{e}\right\}_{{\repulsor}}  ,\label{eq:subgroups}\\
\Gamma_{\rm S} &=& \langle S, T^2\rangle:  
\quad\quad \left\{\frac{o}{o}\right\}_{\attractor}\;\kern 1pt\cup\;\left\{\frac{o}{e},\frac{e}{o}\right\}_{{\repulsor}}.
\end{eqnarray*}
A class is indexed by ${\attractor}$ if the fractions are
sinks (attractive fixed points) for the scaling flow in the $\sigma$-plane, 
and by ${\repulsor}$ if they are sources (repulsive fixed points).
This assignment follows from the requirement that the direction of the flow is downward at the top of the 
conductivity plane. The fixed point at vanishing 
coupling must therefore be repulsive, $i\infty ={\repulsor}$.  Since $\infty = 1/0\in o/e$, and all fixed points 
in a given class are mapped into each other by the symmetry, 
all fractions in the class containing $o/e$ must be repulsive.

Because the duality transformation $S$ swaps $e/o$ and $o/e$, leaving $o/o$ unchanged,
the direction of the flow in the $\rho = S(\sigma)$-plane is reversed if the symmetry acting on 
$\sigma$ is $\Gamma_{\rm T}$ or $\Gamma_{\rm R}$, but not if the symmetry is $\Gamma_{\rm S}$
(which contains $S$).

The fixed point at the origin of the $\sigma$-plane (at $i\infty$ in the $\rho$-plane)
has a special significance. If it is attractive this means that the system has an insulating phase,
which we call the quantum Hall insulator (QHI) and assign the special symbol $\insulator$.  
Since $0 = 0/1\in e/o$, we conclude that a model with $\Gamma_{\rm T}$ or $\Gamma_{\rm R}$ 
symmetry does have this phase, but that a $\Gamma_{\rm S}$-symmetric model does not.

Any symmetry $\Gamma \subseteq \Gamma(1)$ partitions the parameter space into universality classes, 
with each phase ``attached" to a unique attractive fixed point ${\attractor}$ (plateau) on the (extended) real line.
A phase is by definition the set of all points in $\overline{\mathbb H}$ that flow to a given plateau
${\attractor}$ (IR fixed point), and is uniquely labelled by this limit point on the real axis.  
A direct phase transition between two plateaux ${\attractor} = f = p/q$ and ${\attractor}' = f' = p'/q'$ is permitted 
by the symmetry iff there is a saddle point ${\qcp}$ 
(quantum critical point) located on the semi-circle (separatrix) in $\overline{\mathbb H}$ connecting 
${\attractor}$ and ${\attractor}'$, which we write as ${\attractor}\leftarrow{\qcp}\rightarrow{\attractor}'$.
If one of the attractors is at $i\infty = 1/0$ this ``semi-circle" has infinite radius, i.e., it is a vertical line.
If we let ${\rm k_ X}$ be the minimum translation in $\Gamma_{\rm X}$ (so ${\rm k_T} = 1$ and  
${\rm k_R} =  {\rm k_S} = 2$) and $0 < f < f'$, then two phases share a critical point iff 
\begin{equation}
\left\vert\begin{array}{cc}
 p' & p\\
 q' & q  
\end{array}\right\vert
= p' q - q' p  = {\rm  k_X}  \;\implies\;{\qcp} = \frac{p + ip'}{q + iq'}.
\label{eq:qcploc}
\end{equation}
If $f$ or $f'$ is negative the signs are tricky and it is easier to exploit the reflection symmetry of the phase diagram.

Consider first the symmetry $\Gamma_{\rm T}$ and the transition in the conductivity plane 
between the insulator $\sigma_{\smallinsulator} = 0$ and the first integer plateau $\sigma_{\smallattractor} = 1$.
Since $\qcp = (1 + i)/2$  is fixed by $TR^{-2}\in\Gamma_{\rm T}$, and also belongs
to the semi-circle connecting  $\sigma_{\smallinsulator}$ and  $\sigma_\smallattractor$
(cf.\;top left panel in Fig.\;\ref{fig:TRSgrid}), we expect spin polarized experiments to show a 
direct transition between these two phases.
Since there are no fixed points of $\Gamma_{\rm T}$ on the semi-circles connecting 
$\sigma_{\smallinsulator}$ to the other integer plateaux, we do not expect to find direct transitions 
between these phases.  These predictions are consistent with a recent comprehensive analysis of scaling 
data \cite{OLLutken:2018}, some of which is shown in Figs.\;\ref{fig:7Sisters} and \ref{fig:NewLandscape}.

Having found one allowed transition we have them all, because $\Gamma_{\rm T}$ maps
$0\leftarrow (1 + i)/2 \rightarrow 1$ to all other allowed transitions, and vice versa.
For example, since both $T^n(\sigma) = \sigma + n$ and $R^{2n}(\sigma) = \sigma/(1 + 2n\sigma)$ 
belong to $\Gamma_{\rm T}$ for all $n\in \mathbb Z$, the ``image" transitions
\begin{eqnarray*}
&&\quad\;n\;\longleftarrow\;\frac{1 + 2n + i}{2}\;\longrightarrow\;1 + n,\\
&&0\;\longleftarrow\;\frac{1 + 2n + i}{2(1 + 2n + 2n^2)}\;\longrightarrow\;\frac{1}{1 + 2n}
\end{eqnarray*}
are also allowed (cf.\;top left panel in Fig.\;\ref{fig:TRSgrid}).
This includes the $n = 1$ transition $0\leftarrow (3 + i)/10 \rightarrow 1/3$,
in agreement with the experimental flow data shown in Fig.\;\ref{fig:Landscape}.

When this modular symmetry acts on the resistivity
the attractive QHI fixed point is $\rho_{\smallinsulator}= i\infty$.  
The transition $1 \leftarrow 1 + i  \rightarrow i\infty$ to the integer plateau $\rho_{\smallattractor} = 1$ is now 
mediated by the fixed point at $\rho_{\smallqcp} = 1 + i$, and the transitions become 
\begin{eqnarray*}
&&\frac{1}{1+n}\;\longleftarrow\;\frac{1 + 2n + i}{1 + 2n + 2n^2}\;\longrightarrow\;\frac{1}{n},\\
&&\;\;1 + 2n\;\longleftarrow\;1 + 2n + i\;\longrightarrow\; i\infty
\end{eqnarray*}
(cf.\;bottom left panel in Fig.\;\ref{fig:TRSgrid}).

The analysis of the unpolarized $\Gamma_{\rm R}$-symmetric 
case is very similar to the preceding spin polarized case.  The simplest way to obtain the 
phase diagrams shown in the middle column in Fig.\;\ref{fig:TRSgrid} is to apply the substitutions 
$\sigma\rightarrow 2\sigma$ and $\rho\rightarrow\rho/2$ to the diagrams in the left column.
The principal critical point of the conductivity flow is now $\qcp_- = 1 + i$.

Since $\Gamma_{\rm S}$ has a saddle point at $\qcp_+ = i$, it admits a direct transition
$-1\leftarrow i \rightarrow 1$ between the plateaux $\sigma_\attractor = -1$ and $\sigma_\attractor^\prime= 1$
(cf.\;top right panel in Fig.\;\ref{fig:TRSgrid}).
Since both $T^{2n}$ and $R^{2n}$ belong to $\Gamma_{\rm S}$ for all $n\in \mathbb Z$, the transitions
\begin{eqnarray*}
&&\;\;\,2n-1 \;\longleftarrow\; 2n + i \;\longrightarrow\; 2n+1,\\
&&\frac{1}{2n-1} \;\longleftarrow\; \frac{2n + i}{1 + 4n^2} \;\longrightarrow\; \frac{1}{2n+1}
\end{eqnarray*}
are also allowed.  The simplest way to obtain the $\Gamma_{\rm S}$-symmetric phase diagram shown in the
top right panel in Fig.\;\ref{fig:TRSgrid} is to apply the substitution $\sigma\rightarrow\sigma - 1$ 
to the top middle panel. The resistivity diagram is the same in this case, because $S\in\Gamma_{\rm S}$.

\begin{table*}[tbp]  
\vskip 1cm
\begin{center}
\begin{tabular}{|c||c|c|c|c|c|c|}  
\hline
  & $\Gamma(1)$ & $\emptyset$  &  $\Gamma_{\rm T}$ & $\Gamma(1)$  &  $\Gamma(1)$  &  $\Gamma(1)$ \\ 
\hline
$\sigma$  &  $\;\;J(\sigma)\;\;$  &  $\;\;E_2(\sigma)/G^{2}\;\;$  &  $\;\;\;E_2^+(\sigma)/G^{2}\;\;\;$  &  
$\;\;\;E_4(\sigma)/G^{4}\;\;\;$  &  $\;\;\;E_6(\sigma)/G^{6}\;\;\;$  &  $\;\;\;\Delta(\sigma)/G^{12}\;\;\;$ \\
\hline
\hline
$i$  &  $1$ &  $3a$  &  $3/2$  & $3$  &  $0$   &  $2^{-6}$ \\ 
\hline
$\;\;(1 + i)/2\;\;$  &  $1$   &  $6a$  &  $0$  &  $-12$  &  $0$ &  $-1$ \\ 
\hline
$i/2$  &   $\;\;\;1331/8\;\;\;$  &  $\;6 a - 3\;$  &  $3$  &  $33$  &  $-189$  &  $2^{-3}$\\
\hline
$2 i$  &  $1331/8$    &  $\;\;3 a/2+ 3/4\;\;$  &    &  $33/16$  &  $189/64$ &  $2^{-15}$ \\ 
\hline
\end{tabular}
\end{center}
\caption[tablecaption]{Some useful values of the modular invariant function $J$ 
($J$ is not a form because it is not entire), 
the quasimodular form $E_2 = E_2^-$, the (sub)modular forms $E_2^+$, $E_4$, $E_6$, and 
the weight $12$ cusp form $\Delta$. The top row shows the symmetry of the function below it. 
The rings of modular and quasimodular forms on $\Gamma(1) = {\rm SL}(2,\mathbb Z)$ 
are generated by $(E_4, E_6)$ and $(E_2, E_4, E_6)$, respectively.
The ring of modular forms on $\Gamma_{\rm T} = \Gamma_0(2)$  is generated by $(E_2^+, E_4, E_6)$.
$G$ is Gauss' constant,  and $a = 1/\pi G^2 = \pi/\pi_\infty^2$ parametrizes the anomalous 
contribution to quasimodular forms. Notice that the ``dimension" of a form of weight $w$ is $G^w$.}
\label{tab:Table1}
\end{table*}

\subsubsection*{Modular potentials}

We collect here more details about the spin polarized potential 
$\varphi(\sigma) = \varphi_{_{\rm T}}(\sigma) = \ln\eta(2\sigma) - \ln\eta(\sigma)$ analyzed in the 
main part of this paper, which is symmetric under the subgroup $\Gamma_{\rm T} = \Gamma_0(2)$ 
of the modular group $\Gamma(1) = {\rm SL}(2,\mathbb Z)$.   
By the \emph{Hall group} $\Gamma_{\rm H}$ we mean the full symmetry of the 
spin polarized QHE: $\Gamma_{\rm H} = {\rm Aut}\, \Gamma_{\rm T} = {\rm Aut}\,\Gamma_0(2) = \{J, T, R^2\}$, 
where $J$ is the nonholomorphic generator of outer automorphisms (``particle-hole symmetry"), 
which is reflection in the vertical axis, $J(z) = -\bar z$.

In order to extract critical exponents from the beta function we need to expand the 
potential $\varphi =  \varphi_{_{\rm T}}$ at $ {\qcp} = \sigma_{\smallqcp} = (1 + i)/2$ 
as a power series in $\delta\sigma = \sigma - \sigma_{\smallqcp}$ [cf.\;Eq.\;(\ref{eq:PotExpansion})].
The coefficients $a_n = \partial^n\varphi({\qcp})$ ($n\geq 0$) can be evaluated by exploiting 
transformation properties of modular and quasimodular forms. It is convenient to introduce
\begin{equation*}
E_w^\pm(\sigma) = 1 \pm \frac{2w}{B_w} 
\sum_{n=1}^\infty\frac{n^{w-1} q^n}{1 \pm q^n}, \quad w\geq 2,
\end{equation*}
where $B_2 = 1/6$, $B_4 =-1/30$, $B_6 =  1/42,\;\dots$ are Bernoulli numbers, 
and observe that [cf.\;Eq.\;(\ref{eq:etapm})]
\begin{eqnarray}
E_2^\pm(\sigma) &=& \frac{12}{\pi i}\partial \ln \eta_\pm(\sigma),
\label{eq:instantonforms}\\
E_w^+(\sigma) &=& 2 E_w^-(2\sigma) - E_w^-(\sigma).\nonumber
\end{eqnarray}
Notice that $\eta_-(\sigma) = \eta(\sigma)$ and $\eta_+(\sigma) \,\eta_-(\sigma)  = \eta(2\sigma)$,
where $\eta$ is Dedekind's $\eta$-function.
$E_w = E_w^-$ are the conventional Eisenstein series, which are modular forms, 
\begin{eqnarray*}
E_w(T\sigma &=& \sigma + 1) = E_w(\sigma),\\
E_w(S\sigma &=& -1/\sigma) = \sigma^w E_w(\sigma),
\end{eqnarray*}
iff the weight $w\geq4$.
The $\eta$-function and its log-derivative $E_2 = E_2^-$ are not modular forms, because 
they have ``anomalous" transformation properties:
\begin{eqnarray*}
\eta(T\sigma) &=& e^{\pi i/12} \;\eta(\sigma) ,\;\;
E_2(T\sigma) = E_2(\sigma) \label{eq:notforms}, \\
\eta(S\sigma) &=& \sqrt{-i\sigma} \;\eta(\sigma) ,\;\;
E_2(S\sigma) = \sigma^2 E_2(\sigma) + \frac{6\sigma}{\pi i} . \nonumber
\end{eqnarray*}
These transformations allow us to evaluate functions at $\sigma = i$ instead of at ${\qcp} = S T^{-1}i$.

Notice that $E_2$ transforms like a connection on moduli space, rather than a modular form
(tensor), and is sometimes called a \emph{quasimodular form}.  
The reason for this is that a proper modular form of weight two derives from a weightless form, 
and for the full modular group there are no such invariant holomorphic forms (potentials).  

This is intimately related to the existence of \emph{cusp forms}, i.e., forms that vanish at infinity.
For the full modular group the ring of all modular forms is generated by the two Eisensteine 
$E_4$ and $E_6$, which are normalized so that their leading (constant) term in the 
$q$-expansion is one. 
The first cusp form is therefore $\Delta\propto E_4^3 - E_6^2$, and this is the only one.
$E_2$ is the log-derivative of  $\Delta = \eta^{24}$, but this form is not weightless [$w(\Delta) = 12$].
By contrast, for sub-modular symmetries there are at least two independent cusp-forms, whose
ratio provides a weightless potential.  For example, we saw above by explicit calculation that
$E_2^+\propto \partial\ln\eta_+(\sigma) =  \partial\ln[\eta(2\sigma)/ \eta(\sigma)] 
\propto\partial\ln[\Delta(2\sigma)/\Delta(\sigma)]$.   Since $\Delta(2\sigma)$
is a new (independent) cusp form on the modular subgroup $\Gamma_{\rm T}$,
$E_2^+$ is a modular form on $\Gamma_{\rm T}$ of weight two (a vector-field).

The most spectacular consequence of this fact is that Fermat's last theorem is true, since any would-be 
counter-example implies that this form does not exist. Fortunately it does, since it is our only viable 
candidate beta function (up to normalization) for the QHE.

Observe that $\varphi(\sigma) = \left[\ln\Delta(2\sigma) - \ln\Delta(\sigma)\right]/24$,
where the discriminant (of the elliptic curve in Weierstrass form)
$\Delta = \eta^{24} = (E_4^3 - E_6^2)/12^3$
is the cusp form (vanishing at infinity) of weight 12 that plays a very special role in the theory of 
modular forms.  From the transformation laws $\Delta(T\sigma) = \Delta(\sigma)$ and  
$\Delta(S\sigma) = \sigma^{12} \Delta(\sigma)$ it follows that
\begin{equation*}
\Delta(2{\qcp}) = \Delta(i),\quad
\Delta({\qcp}) =  \Delta(S(i-1)) = -2^6 \Delta(i) .
\end{equation*}
This gives $a_0$ and the normalization of the partition function $\zeta_{\rm T} = \exp(-f_{\rm T})$:
\begin{equation*}
\Re a_0 =\Re\varphi(\qcp)  = - \frac{\ln2}{4},\quad
f_{\rm T}(\qcp) = - \ln\zeta_{\rm T}(\qcp) = \ln 2.
 \end{equation*}

These observations immediately give the holomorphic derivative
\begin{equation}
\partial\varphi(\sigma) =\partial \ln\eta_+(\sigma) = \partial\ln\left[\frac{\eta(2\sigma)}{\eta(\sigma)} \right]
 = \frac{\pi i}{12} \,E_2^+(\sigma).
\label{eq:grad}
\end{equation}
While $E_2$ is only quasimodular on $\Gamma\subseteq\Gamma(1)$,
because $\varphi$ is a weightless form on $\Gamma_{\rm T}$, 
$E_2^+ \propto \partial\varphi$ is a modular form of weight 2 on  $\Gamma_{\rm T}$.
In order to evaluate this form at the critical point we use:
\begin{eqnarray*}
E_2({\qcp}) &=& \frac{12{\qcp}}{\pi} - 2i E_2(i),\\
E_2(2{\qcp}) &=& E_2(Ti) = E_2(i).
\end{eqnarray*}
Since $i$ is fixed by $S$, the transformation property of $E_2$ under $S$ gives
\begin{eqnarray}
E_2(i) &=& E_2(Si) = \frac{6}{\pi} - E_2(i) = \frac{3}{\pi},\nonumber\\
E_2({\qcp}) &=& 2E_2(2{\qcp})  = \frac{6}{\pi},\nonumber\\
a_1 &=& \partial\varphi({\qcp}) = \frac{\pi i}{12} \, E_2^+({\qcp}) = 0,\label{eq:a1}
\end{eqnarray}
which shows that  ${\qcp}$ and all its $\Gamma_{\rm T}$-images are critical points of $\varphi$.
Since we know from the $C$-theorem that $\partial\varphi$ is essentially the beta function, 
we have also located the quantum critical points of our model.  A simple and constructive way of 
calculating the exact location of these points is given by Eq.\;(\ref{eq:qcploc}).

\begin{figure}[tbp]
\includegraphics[scale = .1]{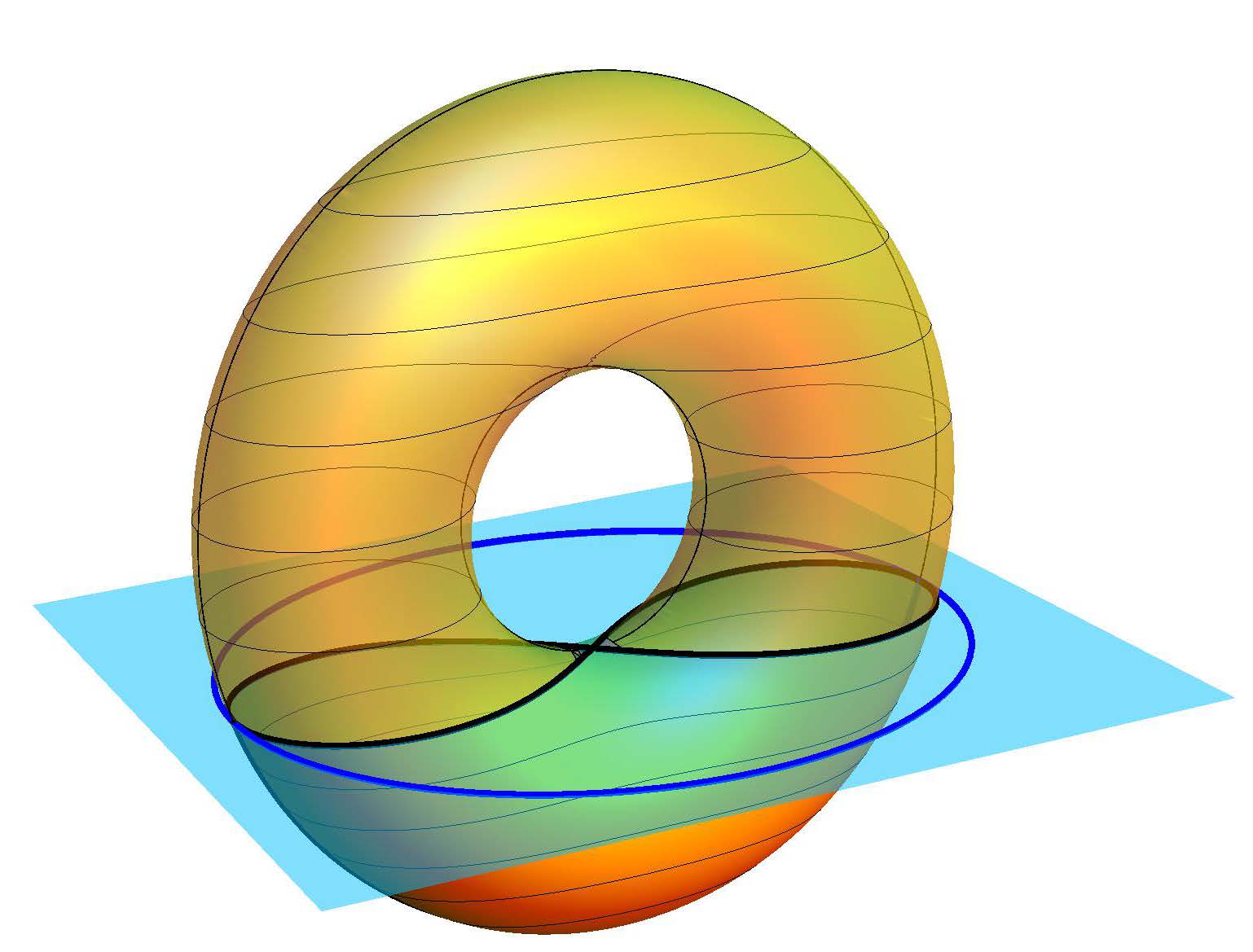}
\caption[Lemniscate]
{Geometric meaning of the fundamental mathematical constant $\pi_\infty$.
The polar curve $r^2 = \cos(2\theta)$ is a \emph{lemniscate}, i.e., a ``bow-tied" closed 
curve that looks like $\infty$.  It is shown here as the intersection of a square torus with a plane 
tangent to the hole of the torus, circumscribed by a unit circle of length $2\pi$. The length $2\pi_\infty$ 
of this lemniscate, where $\pi_\infty = 2.6220575542\dots$, is shorter than the circle 
by a factor $G = \pi_\infty/\pi  = 0.8346268416\dots\approx 5/6$ that is called Gauss' constant.  
\label{fig:Lemniscate}}
\end{figure}

The second derivative of $\varphi$ is obtained by using the first Ramanujan identity
\begin{equation*}
\partial E_2(\sigma) = \frac{\pi i}{6} \left[E_2^2(\sigma) - E_4(\sigma)\right] ,
\end{equation*}
which gives a polynomial in quasimodular forms
\begin{equation*}
\partial^2\varphi(\sigma) = \frac{\pi^2}{72} \left[E_2^2(\sigma) - E_4(\sigma) 
- 4 E_2^2(2\sigma) + 4 E_4(2\sigma)  \right] .
\end{equation*}
The simplest way to evaluate $E_4$  is to exploit
\begin{equation*}
E_4^3(\sigma) = 12^3J(\sigma)\, \Delta(\sigma) ,
\end{equation*}
where $J(\sigma)$ is Klein's modular invariant function. 
Using the well known value $J(i) = 1$, and Gauss' constant
\begin{equation*}
G =   \sqrt{2}\, \eta(i)^2 =  \frac{\Gamma(1/4)^2}{(2\pi)^{3/2}} = 0.8346\dots ,
\end{equation*}
it follows that
\begin{eqnarray*}
E_4(2{\qcp}) &=& E_4(i) = 3 \,G^4 ,\\
E_4({\qcp}) &=& -4 E_4(i) = -12  \,G^4 .
\end{eqnarray*}
Combined with the values of $E_2$ obtained above, we get the second expansion coefficient
\begin{equation}
a_2 = \partial^2\varphi({\qcp}) =  \frac{E_4(i)}{E_2^2(i)} = \frac{\pi^2G^4}{3} = \frac{\pi_\infty^4}{3\pi^2}.
\label{eq:a2}
\end{equation}  
The geometrical meaning of the \emph{lemniscate constant} $\pi_\infty = \pi G$ is explained in 
Fig.\;\ref{fig:Lemniscate}.  It has a rational approximation $\pi_\infty\approx 21/8$ that 
is surprisingly similar to Archimedes' result $\pi \approx 22/7$ for the circle; 
both accurate to about {1\textperthousand}.
Notice also the useful mnemonic: $\nu_{\rm tor}\approx\pi_\infty\approx 21/8$ (they differ by less than 1\%).

With hindsight it is perhaps not too surprising that the lemniscate constant makes an appearance.
It is a basic mathematical constant,
as fundamental to the torus as $\pi$ is to the circle.  They are geometric siblings:  the total arc length of the
unit lemniscate is  $2 \pi_\infty = \Gamma(1/4)^2/\sqrt{2\pi} =  2\times 2.62\dots$,
which is analogous to the circumference $2\pi =  2\;\Gamma(1/2)^2 =  2\times 3.14\dots$  of the unit circle.
The doubly periodic lemniscate functions have periods $2 \pi_\infty$ and $2 i\pi_\infty$,
similar to the trigonometric functions that have period $2\pi$. 
It is an historical fact that the problem of computing the arc length of the lemniscate ($\infty$), 
dating back to the seventeenth century, catalyzed the theory of elliptic functions. 
The lemniscate is therefore one of the corner stones of modern algebraic geometry 
(and the only mathematical figure to be found in Abel's work).  
Elliptic functions are genuinely new, since they cannot be reduced to elementary functions,
and $\pi_\infty$ cannot be reduced to elementary constants.

This suffices for our objective here, which is to compute the critical exponents
of the family of $\widetilde {\rm T}^2_{\rm T}$-models at ${\qcp}$, but for completeness we
show how to evaluate the expansion coefficient at any order for any of the main congruence subgroups
of the modular group. 

The third derivative of  $\varphi$ is obtained by using the second Ramanujan identity:
\begin{eqnarray*}
\partial E_4(\sigma) &=& \frac{2\pi i}{3} \left[ E_2(\sigma) E_4(\sigma) - E_6(\sigma)\right]\implies\\
\partial^3\varphi(\sigma) 
&=& \frac{\pi^3 i}{216} [ E_2^3(\sigma) -  3 E_2(\sigma) E_4(\sigma) + 2 E_6(\sigma) \\
&-& 8 E_2^3(2\sigma) + 24 E_2(2\sigma) E_4(2\sigma) - 16 E_6(2\sigma)].\nonumber
\end{eqnarray*}
Evaluating $\Delta = (E_4^3 -E_6^2)/1728$ at $\sigma = i$ we find
\begin{equation*}
E_6(2{\qcp}) = E_6({\qcp}) = E_6(i) = 0 .
\end{equation*}
Together with the values of $E_2$ and $E_4$ obtained in above this gives the third expansion coefficient
\begin{equation*}
a_3 = \partial^3\varphi({\qcp})=\frac{2\pi^3 i}{9} E_2(i) E_4(i) = 2i\alpha ,
\end{equation*}
where $\alpha =1/ a^2 = \pi^2 G^4 = \pi_\infty^4/\pi^2$.

The fourth derivative of  $\varphi$ is obtained by using the third Ramanujan identity
\begin{equation*}
\partial E_6(\sigma) = \pi i \left[E_2(\sigma) E_6(\sigma) - E_4^2(\sigma)\right] .
\end{equation*}
Note that no new forms are generated this time;  a consequence of
 the fact that the ring of quasimodular forms is generated
by $E_2$, $E_4$ and $E_6$.   We find
\begin{eqnarray*}
\partial^4\varphi(\sigma) 
= - \frac{\pi^4}{432}&\big[&\kern -3pt E_2^4(\sigma) -  6 E_2^2(\sigma) E_4(\sigma) + 8 E_2(\sigma)E_6(\sigma)\\
 &-& 3E_4^2(\sigma)\kern-1pt -\kern-1pt 16 E_2^4(2\sigma)\kern-1pt +\kern-1pt  96 E_2^2(2\sigma) E_4(2\sigma)\nonumber\\
&-&\kern3pt 128 E_2(2\sigma)E_6(2\sigma) + 48 E_4^2(2\sigma)\big] .
\end{eqnarray*}
Inserting the values of $E_2$, $E_4$ and $E_6$ obtained above,
we get the fourth expansion coefficient
\begin{equation*}
a_4 = \partial^4\varphi({\qcp}) =  -\frac{4\pi^4}{9} E_2^2(i) E_4(i) =  - 12 \alpha .
\end{equation*}

In summary, we have carried out the expansion of the RG potential $\varphi$ to fourth order in 
$\delta\sigma$.   It is clear that this procedure can be iterated to any desired order,
using only the values of the forms that we already have, 
collected in Table\;\ref{tab:Table1} for easy reference.
The expansion coefficients are always polynomials of odd degree in $\alpha$,
and for weights less than $12$ (where the cusp form $\Delta$ appears) 
they are linear in $\alpha$.  The coefficients given by quasimodular forms of weight $w \leq 24$ are:
\begin{eqnarray*}
a_0 &+& \bar a_0 = - \ln2/2\nonumber,\\
a_1 &=& 0,\quad\quad a_{12} = -8 363 520 \alpha (105 + 294 \alpha ^2 + \alpha ^4),\nonumber\\
a_2 &=& \alpha/3  , \quad\quad a_{11} =126 720 i \alpha (525 + 882 \alpha^2+ \alpha ^4),\nonumber\\
a_3 &=& 2 i \alpha,\quad\quad\quad a_{10} = 1152 \alpha (4725 + 4410 \alpha ^2 + \alpha ^4),\nonumber\\  
a_4 &=& -12 \alpha,\quad\quad\quad\quad  a_9 = -32 256 i \alpha (15 + 7 \alpha ^2),\nonumber\\
a_5 &=& -80 i \alpha, \quad\quad\quad\quad\quad a_8 = -9408 \alpha (5 + \alpha ^2),\nonumber\\ 
a_6 &=& 8 \alpha(75 + \alpha ^2), \quad\quad\quad\quad a_7 = 336 i \alpha (15 + \alpha ^2).
\end{eqnarray*}

For  $\Gamma_{\rm R}$ and $\Gamma_{\rm S}$ the partition functions 
$Z_- = Z_{\rm R}$ and $Z_+ = Z_{\rm S}$ are $Z_\pm =  \left\vert\zeta_\pm \right\vert^4$, with potentials
\begin{equation*}
\varphi_\pm(\sigma) = \ln \zeta_\pm(\sigma) = \ln\eta_\pm(\sigma/2) -\ln \eta_\pm(\sigma)
\end{equation*}
given by the instanton forms $\eta_\pm$ defined in Sec.\;\ref{subsec:beta} [cf.\;Eq.\;(\ref{eq:instantonforms})].
The expansion of the $\Gamma_{\rm R}$-symmetric potential $\varphi_-(\sigma) = - \varphi(\sigma/2)$
near the saddle point $\qcp_- =  1 + i$ is very similar to the expansion of $\varphi$,
since no additional values of modular forms are needed.

The $\Gamma_{\rm S}$-symmetric potential  $\varphi_+$ satisfies
$\varphi + \varphi_{-} + \varphi_{+} = 0$, 
and in order to expand it near the saddle point $\qcp_+ = i$ 
we need the two bottom rows in Table \ref{tab:Table1}. 
These may be obtained by using the doubling formulas
\begin{equation*}
\frac{\Delta(2\sigma)}{\Delta(\sigma)} = \frac{1}{2^8} \frac{\lambda^2}{1 - \lambda},\quad
J(2\sigma) = \frac{1}{108} \frac{(\lambda^2 - 16 \lambda+16)^3}{\lambda^4(1 - \lambda)},
\end{equation*}
where the \emph{elliptic modular lambda function} $\lambda  = (\theta_2/\theta_3)^4$ 
(sometimes called the ``hauptmodul") is a weightless form
on $\Gamma(2) = \langle T^2, ST^2S\rangle$ \cite{Rankin:1977}.
Using $\lambda(i) = 1/2$ gives $\Delta(i) = 2^9\Delta(2i)$ and 
$J(2i) = (11/2)^3$. $S(2i) = i/2$ gives $\Delta(i/2) = 2^3\Delta(i)$,
while $J$ is invariant,  $J(2i) = J(i/2)$. 
Combining this with previous expressions gives the missing values of $E_4$ and $E_6$.

In order to calculate the required values of $E_2$ we derive a doubling formula 
by differentiating the doubling formula for $\Delta$,
\begin{equation*}
E_2(2\sigma) = \frac{1}{2} E_2(\sigma) + \frac{1}{4} \theta_4^4(\sigma) 
\frac{2 - \lambda(\sigma)}{1 - \lambda(\sigma)},
\end{equation*}
where we have used the well known relation $i\pi\,\theta_4^4 =  \lambda^\prime/\lambda$.
Using $\theta_4^4(i) = G^2$ we find the rest of the table.
This gives expansion coefficients $\partial^n\varphi_\pm(\qcp_\pm) = - a_n/2^n\;(n \geq 0)$, 
and we find in both cases the same critical exponents as obtained  above for the 
$\Gamma_{\rm T}$-symmetric case.

\subsubsection*{Modular and quasimodular beta functions}

The observed modular target space symmetry is so 
constraining that we in simple cases can find the exact beta function (up to an overall normalization), 
mainly because it must transform as a modular (co-)vector field, i.e., a modular form of weight $w = \pm 2$. 
This includes non-perturbative corrections, and it is instructive to try to anticipate how these 
will affect the beta function.  

Expanding a gauge theory with coupling constant $e$ [and fine structure constant $\alpha = e^2/(4\pi)$] 
around the weak coupling limit ($\alpha\rightarrow 0$), the amplitude of the leading instanton contribution is 
exponentially damped, with a phase that is determined by the vacuum angle $\theta$.  
In an EFT of the QHE it is the conductivities $\sigma_D$ and $\sigma_H$ that are playing the roles of 
$e$ and $\theta$ \cite{Khmelnitskii:1983}, and the effective action should be a Fourier expansion in
\begin{equation*}
q = e^{i\theta}\,e^{-2\pi/\alpha} = e^{2\pi i\sigma},\quad 
\sigma = \sigma_H + i\sigma_D = \frac{\theta}{2\pi} + \frac{i}{\alpha},
\end{equation*}
and $\bar q = \exp(-2\pi i\bar\sigma)$.

Our first guess might be that the physical beta function 
\begin{equation*}
\beta^\sigma = \dot\sigma = \frac{d\sigma}{dt} =  \frac{d\sigma_H}{dt} + i \frac{d\sigma_D}{dt} = \beta^H + i \beta^D
\end{equation*}
is a modular vector field whose non-perturbative part 
has a holomorphic Fourier expansion in $q$. There are at least two reasons why this is a non-starter:
a holomorphic beta function does not have saddle points, which are needed to model quantum critical points,
and the contravariant beta function $\beta^\sigma$ would transform with weight  $w = -2$. 
No such modular form exists.  

It is therefore only the covariant beta function $\beta_\sigma$ that can be a modular form ($w = 2$), 
since such forms (ususally) do exist. According to the $C$-theorem we can obtain the contravariant beta 
function $\beta^\sigma = g_{_Z}^{\sigma\bar\sigma}\beta_{\bar\sigma}$ by using Zamolodchikov's moduli space 
metric $g_{_Z}$, which is nonsingular and nonholomorphic \cite{Zamo:1986}. 
In our (toroidal) case this is the hyperbolic metric $g_{_H}$ on the upper half of the complex conductivity plane, 
so $g_{_Z}^{\sigma\bar\sigma} = g_{_H}^{\sigma\bar\sigma} = \sigma_D^2$.
We therefore expect the physical quantum Hall beta function $\beta^\sigma$ to have an \emph{antiholomorphic} 
Fourier expansion in $\bar q$, rather than a holomorphic expansion in $q$ \cite{BLutken:1997, footnote:qexpansion}.  

Because renormalization (the RG flow and its beta functions) must respect the symmetries of the model
under consideration, modular symmetry implies that the Fourier expansion of the beta function 
sums to a modular 2-form $\bar{\mathcal E}_2(\bar\sigma)$.

For example, if the modular symmetry is $\Gamma_{\rm T}$ (appropriate for the spin polarized QHE), 
then ${\mathcal E}_2^{\rm T}(\sigma) = {\mathcal N} E_2^+(\sigma)$ is, up to an overall normalization 
$\mathcal N = i N$ ($N>0$) \cite{footnote:normalization},  a \emph{unique} 2-form $E_2^+$ 
[cf.\;Eq.\;(\ref{eq:instantonforms})], and the beta function is
\begin{eqnarray}
\beta^\sigma &=& g_{_Z}\bar{\mathcal N} {\bar E}_2^+ 
= g_{_Z} \bar{\mathcal N}\,\left(1 + 24 \sum_{n=1}^\infty\frac{n\bar q^{\,n}}{1+\bar q^{\,n}} \right)\label{eq:wcl}\\
&=& - i g_{_Z} N\,\left[1 + 24 (\bar q + \bar q^{\, 2} + 4 \bar q^{\, 3} +  \dots)\right].\nonumber
\end{eqnarray}
In more conventional component notation this gives
\begin{eqnarray}
\beta^H &\stackrel{\rm wcl}{\longrightarrow}& 
- 24 N \sigma_D^2 \,\sin(2\pi\sigma_H) \,e^{-2\pi\sigma_D} + \dots,\label{eq:holoexpandH}\\
\beta^D &\stackrel{\rm wcl}{\longrightarrow}&
-N \sigma_D^2\,\left[1 + 24 \cos(2\pi\sigma_H)\, e^{-2\pi\sigma_D} + \dots\right].
\quad\label{eq:holoexpandD}
\end{eqnarray}

Notice that the \emph{weak coupling limit} (wcl) $\sigma_D \rightarrow \infty$ 
($\sigma\rightarrow i\infty$, $q\rightarrow 0$) exhibited in Eqs.\;(\ref{eq:holoexpandH}) and (\ref{eq:holoexpandD}),
which is where perturbative calculations might be useful, leaves no trace of the modular symmetry
of the exact, non-perturbative functions. 
The leading [$\mathcal O(\bar q)$] instanton contribution, also shown, 
is periodic in $\sigma_H$, i.e., invariant under \emph{translations}
$T\in\Gamma_{\rm T}:\sigma_H\rightarrow\sigma_H + 1$ (as is observed in the spin polarized QHE).
But it is not invariant under the other generator $D$ of $\Gamma_{\rm T}$, which is a 
\emph{duality transformation} [cf.\;Eq.\;(\ref{eq:duality})].
More generally, duality symmetry, and therefore modular symmetry, is lost if the instanton expansion 
(Fourier expansion in $q$ or $\bar q$) is truncated at any finite order. 

Notice also that Eq.\;(\ref{eq:wcl}) does not do justice to the perturbative beta function, 
because perturbation theory is an expansion around a singular geometry, 
i.e., a degenerate torus that is ``nodal" or ``pinched". 
This generates nonholomorphic contact terms that are polynomial in $\sigma_D$, which
in string theory are called \emph{holomorphic anomalies} \cite{BCOVafa:1993}. 
This is captured by \emph{quasiholomorphic} forms (or, equivalently, \emph{quasimodular} forms)
\cite{Shimura:1986,*Shimura:1987, Robbert:1994, Lutken:2006}, first studied by Ramanujan
(who called them ``mock theta functions") \cite{Rama:2000}, and by Hecke\;\cite{Hecke:1927}.

A quasiholomorphic form transforms like a modular form, but because it is not holomorphic it is 
not a modular form. An example is provided by the quasiholomorphic 2-form under $\Gamma_{\rm T}$,
which gives the beta function \cite{Lutken:2006}
\begin{eqnarray}
\beta^\sigma_\infty &=&
 - i g_\infty N \,\left(\frac{1}{\pi\sigma_D} +16 \sum_{n=1}^\infty\frac{n\bar q^{\,n}}{1-\bar q^{\,2n}} \right)
\nonumber\\
&=& - 16 i g_\infty N \,\left(\frac{1}{16\pi\sigma_D} + \bar q + 2 \bar q^{\, 2} + 4 \bar q^{\, 3} +  \dots\right),
\quad\label{eq:quasibeta}
\end{eqnarray}
\begin{eqnarray}
\beta^H_\infty &\stackrel{\rm wcl}{\longrightarrow}& 
-16 g_\infty N \,\sin(2\pi\sigma_H) \,e^{-2\pi\sigma_D} + \dots,\nonumber\\
\beta^D_\infty &\stackrel{\rm wcl}{\longrightarrow}&
- 16 g_\infty N\,\left[\frac{1}{16\pi\sigma_D} + \cos(2\pi\sigma_H)\, e^{-2\pi\sigma_D} + \dots\right].\nonumber
\end{eqnarray}
The new notation ($g_{_Z}\rightarrow g_\infty$) signals that it is not clear which metric should be used in this
singular  asymptotic expansion. 

An explicit evaluation of the leading order instanton correction to beta functions in 
the tensor model of the IQHE was carried out in \cite{LevineLP:1983}.  Although their dilute instanton gas
approximation seems to be fraught with difficulties, it is perhaps of interest to observe that with a 
suitable normalization [corresponding to $g_\infty N = 1/(2\pi)$] their beta function (transcribed to our notation)
\begin{equation}
\beta^\sigma_{\papillon}\propto - i \left(\frac{1}{16\pi\sigma_D} + \sigma_D \bar q +\dots\right)
\label{eq:tensorbeta}
\end{equation}
only differs from the $\Gamma_{\rm T}$-symmetric quasiholomorphic beta function $\beta^\sigma_\infty$ 
in Eq.\;(\ref{eq:quasibeta}) by the factor $\sigma_D$ in the leading instanton term. 
It is not clear if the beta functions $\beta^\sigma_\infty$ and $\beta^\sigma_{\papillon}$ 
in Eqs.\;(\ref{eq:quasibeta}) and (\ref{eq:tensorbeta}) can be fully reconciled.

Since the perturbative part of the beta function is a polynomial in $\sigma_D$ 
with no modular signature, we conclude that neither perturbation theory nor dilute instanton gas 
approximations can detect a modular symmetry.

The $C$-theorem presents a potentially lethal threat to modular beta functions,
since it requires the RG flow to be a gradient flow [cf.\;Eq.\;(\ref{eq:Ricci})]. 
In the example with $\Gamma_{\rm T}$-symmetry discussed above, 
given that the covariant beta function is uniquely given by the 2-form $E_2^+$ (up to normalization), 
it follows that $E_2^+$ must be the derivative of some scalar modular function  $\varphi_{_{\rm T}}$ 
that is invariant under $\Gamma_{\rm T}\subset\Gamma(1)$. No such function exists for $\Gamma(1)$, 
but it does exist for the congruence subgroups we consider, and lo and behold, 
$E_2^+\propto \partial \varphi_{_{\rm T}}$ [cf.\;Eq.\;(\ref{eq:grad})], 
so $C_{\rm T} \propto  \varphi_{_{\rm T}} + c.c.$   
satisfies the $C$-theorem.

If the beta functions are harnessed by a modular symmetry, then the critical (delocalization) exponents 
are also severely constrained. They are obtained by expanding the beta function near a critical point 
$\sigma_\smallqcp$ (where it by definition has a simple zero) in 
$\delta\bar\sigma = \bar\sigma - \bar\sigma_\smallqcp = 
\delta\sigma_{_H}^\smallqcp -i \delta\sigma_{_D}^\smallqcp$,
\begin{equation*}
\beta^\sigma \stackrel{\smallqcp}{\longrightarrow} g_\smallqcp \sum_{n=1}^\infty b_n \delta\bar\sigma^n
\approx y \delta\bar\sigma
= y^+\delta\sigma_{_H}^\smallqcp + i y^-\delta\sigma_{D}^\smallqcp.
\end{equation*}
For simplicity we have here assumed that the principal flow directions at $\qcp$ are aligned with the 
coordinate axes of $\mathcal M$, as in the IQHE.
Observe that these flow rates satisfy the \emph{antiholomorphic scaling relation}
$y^+ = - y^- = y  =  b_1 g_\smallqcp >0$, which is inherited by the delocalization exponents 
$\nu^+ = -\nu^- = \nu = 1/y>0$ \cite{LR:2006}.

Furthermore, since modular symmetry identifies all critical points, even without knowing the value
of $\nu$ modular symmetry makes a strong prediction: \emph{there is only one possible 
(absolute) value of the delocalization exponents in the QHE ($\nu^\pm = \pm\nu$), 
for any quantum phase transition.} This prediction is easy to falsify, but so far it appears to be 
consistent with experimental data \cite{LR:2006}.


\bibliography{PRB_BW13167_Biblio_100519}            


\end{document}